\documentclass{aa} 
\DeclareUnicodeCharacter{0327}{}

\usepackage{xcolor}
\usepackage{lscape}
\usepackage{multirow}
\usepackage{txfonts}

\newcommand{\msun}{M$_\odot$}
\newcommand{\rschw}{R$_\text{S}$}
\newcommand{\fcol}{$f_\text{col}$}
\newcommand{\rmag}{R$_\text{m}$}
\newcommand{\rco}{R$_\text{C}$}
\newcommand{\risco}{R$_\text{isco}$}
\newcommand{\rsph}{R$_{\text{sph}}$}

\newcommand{\nustar}{\textit{NuSTAR}}
\newcommand{\xmm}{\textit{XMM-Newton}}
\newcommand{\chandra}{\textit{Chandra}}
\newcommand{\swift}{\textit{Swift}}
\newcommand{\epicpn}{EPIC-pn}

\newcommand{\tbabs}{\texttt{tbabs}}
\newcommand{\diskbb}{\texttt{diskbb}}
\newcommand{\diskpbb}{\texttt{diskpbb}}
\newcommand{\cflux}{\texttt{cflux}}
\newcommand{\simpl}{\texttt{simpl}}

\newcommand{\cutoffpl}{\texttt{cutoffpl}}
\newcommand{\mekal}{\texttt{mekal}}
\newcommand{\xspec}{\textsc{XSPEC}}
\newcommand{\stingray}{\textsc{Stingray}}
\newcommand{\hendrics}{\textsc{HENDRICS}}
\newcommand{\chisq}{$\chi^2$}
\newcommand{\nh}{$n_{\text{H}}$}

\newcommand{\python}{PYTHON}
\newcommand{\fscatt}{$f_{scat}$}
\newcommand{\errors}[2]{$^{+#1}_{-#2}$}

\begin{document} 
   \title{Long-term X-ray spectral evolution of Ultraluminous X-ray sources: implications on the accretion flow geometry and the nature of the accretor}
   \titlerunning{Long term evolution of ULXs}


   \author{A. G\'urpide
      \inst{1}\and O.Godet\inst{1}\and F. Koliopanos\inst{1}\and N.Webb\inst{1}\and J.-F. Olive\inst{1}}

   \institute{Institut de Recherche en Astrophysique et Plan\'etologie, Universit\'e de Toulouse, UPS/CNRS/CNES, 9 Avenue du Colonel Roche, BP44346, F-31028 Toulouse Cedex 4, France\\}

   \date{}

 
  \abstract
   {The discovery of pulsations in several Ultraluminous X-ray sources (ULXs) has demonstrated that a fraction of ULXs are powered by super-Eddington accretion onto neutron stars (NS). This has opened the debate as to what is the neutron star to black hole (BH) ratio within the ULX population and what physical mechanism allows ULXs to reach luminosities well in excess of their Eddington luminosity: the presence of strong magnetic fields or rather the presence of strong outflows that collimate the emission towards the observer.}
   {In order to distinguish between these scenarios, namely, supercritically accreting BHs, weakly or strongly magnetised NSs, we study the long-term X-ray spectral evolution of a sample of 17 ULXs with good long-term coverage, 6 of which are known to host neutron stars. At the same time, this study serves as a baseline to identify potential new NS-ULX candidates.}
   {We combine archival data from \chandra, \xmm\ and \nustar\ observatories in order to sample a wide range of spectral states for each source. We track each source's evolution in a hardness-luminosity diagram (HLD) in order to identify spectral changes and show that these can be used to constrain the accretion flow geometry and in some cases, the nature of the accretor.}
   {We find NS-ULXs to be among the hardest sources in our sample with highly variable high-energy emission. On this basis, we identify M81 X-6 as a strong NS-ULX candidate, whose variability is shown to be akin to that seen in NGC 1313 X-2. For most softer sources with unknown accretor, we identify the presence of three markedly different spectral states that we interpret invoking changes in the mass-accretion rate and obscuration by the supercritical wind/funnel structure. Finally, we report on a lack of variability at high-energies ($\gtrsim$ 10 keV) in NGC 1313 X-1 and Holmberg IX X-1, which we argue may offer means to differentiate BH from NS-ULXs.}
   {We support a scenario in which the hardest sources in our sample might be powered by strongly magnetised neutron stars, so that the high-energy emission is dominated by the hard direct emission from the accretion column. Instead, softer sources may be explained by weakly magnetised neutron stars or black holes, in which the presence of outflows naturally explains their softer spectra through Compton down-scattering, their spectral transitions and the dilution of the pulsed-emission, should some of these sources contain neutron stars.}


   \maketitle
%

\section{Introduction} \label{sec:intro}
Ultraluminous X-ray sources are defined as extragalactic off-nuclear point-like sources with an X-ray luminosity in excess of $\sim$ 10$^{39}$ erg/s \citep[see][for a review]{kaaret_ultraluminous_2017}, albeit there is now evidence for a Galactic ULX \citep{wilson-hodge_nicer_2018}. Their nature remains for the most part unknown, and given this empirical definition they likely constitute a heterogeneous population of objects. It was initially proposed that ULXs could be powered by accreting intermediate-mass black holes \citep[IMBH:$\sim$ 100 -- 10$^5$ \msun; see][for a review]{mezcua_observational_2017} in the sub-Eddington regime \citep{colbert_nature_1999, matsumoto_discovery_2001}. The best ULX IMBH candidates may be those at the high-end of the High-Mass X-ray binary (HMXB) luminosity distribution \citep[e.g.][]{mineo_x-ray_2012}, reaching L$_\text{X}$ $\geq$ 10$^{41}$ erg/s in some cases. Some of these objects show transitions  or temporal properties that seem to be consistent with the expectation of a scaled-up version of Galactic Black Hole Binaries (GBHBs) \citep[e.g.][]{farrell_intermediate-mass_2009, godet_first_2009, pasham_400-solar-mass_2014} and/or show evidence for cooler accretion disks \citep[e.g.][]{feng_identification_2010, servillat_x-ray_2011, godet_investigating_2012, lin_multiwavelength_2020} which suggest masses in the IMBH regime. 

However, it was soon realised that the bulk of the ULX population (10$^{39}$ $<$ L$_\text{X}$ $<$ 10$^{41}$ erg/s) did not comply with the canonical states seen in GBHBs \citep[e.g.][]{stobbart_xmmnewton_2006, gladstone_ultraluminous_2009, grise_x-ray_2010}. It was thus suggested that stellar-mass black holes fed at super-Eddington mass-transfer rates could power these sources \citep{shakura_black_1973, king_black_2003, poutanen_supercritically_2007}, with geometrical beaming possibly further enhancing the observed luminosity \citep{king_ultraluminous_2001}. In this scenario, the intense radiation pressure blows-off some of the excess gas in the form of a wind or outflow, which can become optically thick to the hard radiation emitted in the inner parts of the disk \citep{poutanen_supercritically_2007}. The wind is expected to form a conic structure around the rotational axis of the compact object due to angular momentum conservation, producing highly anisotropic emission. It was thus proposed that the short-term variability and the differences in spectral shape in a broad ULX sample could be understood in terms of different viewing angles and mass-accretion rates \citep{sutton_ultraluminous_2013, middleton_spectral-timing_2015} with those ULXs with L$_\text{x}$ $\sim$ 10$^{39}$ erg/s showing a behaviour more consistent with stellar-mass black holes accreting close to or at their Eddington limit \citep[e.g.][]{middleton_x-ray_2011}. These studies were later extended to include ultraluminous supersoft sources (ULSs) \citep{urquhart_optically_2016}, which are interpreted as sources in which either the inclination of the system and/or the accretion rate are so high that the inner disk emission is completely reprocessed by the optically thick wind, with some sources possibly showing transitions between a ULX-like spectrum to a ULS state \citep{feng_nature_2016}. Indeed, numerical simulations of super-Eddington accretion onto black holes have shown extensively that the viewing angle has a strong impact on the observed spectrum \citep[e.g.][]{ohsuga_why_2007, sadowski_global_2015, ogawa_radiation_2017}. For this reason, it has been suggested that overall ULXs belong to a homogeneous class of objects, including stellar-mass black holes or lowly magnetised neutron stars, being fed at super-Eddington mass-transfer rates \citep{king_ulxs:_2016, king_pulsing_2017, pinto_thermal_2020}. Observational evidence supporting winds associated with super-Eddington accretion comes from high-resolution X-ray spectroscopy studies, that revealed absorption and emission features suggesting the presence of strong ionised winds colliding with the circumstellar medium \citep{pinto_resolved_2016, pinto_ultraluminous_2017, pinto_xmmnewton_2020}.

However, while it was generally accepted that accreting stellar-mass black holes were the engines behind ULXs below 10$^{41}$ erg/s \citep[e.g.][]{poutanen_supercritically_2007, gladstone_ultraluminous_2009, pintore_ultraluminous_2014}, the discovery of X-ray pulsations in one ULX \citep{bachetti_ultraluminous_2014} showed that NS can also attain super-Eddington luminosities. Motivated by the discovery of (now) 5 more pulsating ULXs (PULXs) \citep{furst_discovery_2016,israel_discovery_2017, israel_accreting_2017, carpano_discovery_2018, sathyaprakash_discovery_2019, rodriguez-castillo_discovery_2020} and the possible confirmation of another NS-ULX through the identification of a cyclotron line \citep{brightman_magnetic_2018}, several authors have highlighted the remarkable X-ray spectral similarity of the ULX sample with those with detected pulsations \citep{pintore_pulsator-like_2017, koliopanos_ulx_2017, walton_evidence_2018}, suggesting that NS-ULX may dominate the ULX population. Yet, there is still some debate as to what is the driving mechanism responsible for their extreme luminosities. 

\cite{pintore_pulsator-like_2017} suggested that the accretion column could be responsible for the hard emission in ULXs, as their X-ray emission could be described with a model commonly used to fit galactic X-ray pulsars. Instead, based on the theoretical calculations carried out by \cite{mushtukov_optically_2017}, \cite{koliopanos_ulx_2017} argued that the ULX spectra are consistent with accreting highly-magnetised NS (B>10$^{12}$ G), as the 0.3 -- 10 keV band can be described by two thermal components. In this scenario, the soft thermal component arises from a truncated accretion disk at the magnetospheric radius \citep{ghosh_disk_1978} while the hard emission is produced in an optically thick envelope as the accreting material is forced to follow the magnetic field lines, creating a closed envelope around the NS. The rotation of the envelope, due to the coupling with the NS through the magnetic field lines, and its latitudinal temperature gradient could explain the sinusoidal profiles seen in PULXs \citep[e.g][]{furst_tale_2018}, provided that the NS rotational axis is misaligned with the magnetic field axis. Alternatively, \cite{walton_evidence_2018} following the arguments from \cite{king_ulxs:_2016, king_pulsing_2017} showed that the interplay between the mass-accretion rate and magnetic field strength could explain the lack/presence of pulsations in ULXs, suggesting that ULXs might instead be powered by weakly magnetised neutron stars. This in turn may explain the different spectral shape of those ULXs for which broadband spectroscopy data is available, compared to the PULXs.

However, the long-term evolution of PULXs is marked by high levels of variability of 2 orders of magnitude or more in flux \citep{israel_accreting_2017, israel_discovery_2017} while other ULXs show variations of a factor 5 or less throughout year-time scales \citep{grise_long-term_2013, luangtip_x-ray_2016, pinto_xmmnewton_2020} with PULXs being also somewhat harder than the rest of the population \citep{pintore_pulsator-like_2017}. Whether these represent differences in the nature of the accretor or are due to a disfavourable viewing angle \citep[e.g.][]{king_pulsing_2020} is still not fully understood \citep[e.g.][]{walton_evidence_2018}. If beaming is a natural consequence of super-Eddington mass transfer \citep{king_masses_2009}, theoretical studies show that the fraction of observed BH-ULX could be higher than NS-ULX \citep{middleton_predicting_2017}. Similarly, binary population synthesis studies suggest that BH-ULX could even dominate the observed ULX population \citep{wiktorowicz_observed_2019}, as NSs need stronger beaming factors to reach L$_\text{X}$ > 10$^{39}$ erg/s. However, the evidence for supercritically accreting stellar-mass black holes remains elusive. The best observational evidence for such BH-ULX systems might be M101 ULX-1, which was estimated to host a $\sim$ 20 \msun\ BH, although its dynamical mass determination was challenged by \cite{laycock_revisiting_2015}. Constraining the BH to NS ratio in ULXs can lead to important clues about the formation path leading to such systems and their connection with HMXBs \citep[e.g.][]{mineo_x-ray_2012} and the formation of BH-BH and BH-NS systems.

Thus, while the spectral resemblance of the ULX population is undeniable, tracing their long-term evolution can be the key in understanding the geometry of the accretion flow and discriminating between super-Eddington accretion models and the nature of the accretor. Therefore, in this paper, we perform a comprehensive study of the long-term spectral evolution of a representative sample of 17 ULXs (including 6 NS-ULXs) in the 10$^{39}$ $<$ L$_x$ $<$ 10$^{41}$ erg/s range, in an effort to gain insights into the accretion flow geometry as well as the nature of the accretor. More specifically, we focus on studying each source spectral evolution in an attempt to assess which scenario best describes the variability observed: super-Eddington accretion onto a highly magnetised NS, a weakly magnetised NS or a black hole. This in turn gives clues about the nature of the accretor and by comparing its evolution with the known PULXs, we can identify potential strong NS-ULX candidates. To do so, we build a phenomenological model taking into account the new insights gained on the ULX broadband emission with \nustar\ \citep{bachetti_ultraluminous_2014,walton_nustar_2015, mukherjee_hard_2015, walton_unusual_2020} and track their spectral changes in a hardness-luminosity diagram. 

We describe the sample of sources selected for this work and the data reduction in Section \ref{sec:sample_data_reduction}. In Section \ref{sec:spectral_fitting} we describe the data analysis and results. We discuss our results in Section \ref{sec:discussion} and present our conclusions in Section \ref{sec:conclusions}.
\section{Sample selection and data reduction} \label{sec:sample_data_reduction}
\subsection{Sample selection} \label{sub:sample}
In order to have a representative sample of the possible range of spectral variability of each source, we searched in the literature for sources that have been observed on at least 5 occasions, well spaced in time, by either \xmm\ \citep{jansen_xmm-newton_2001} or \chandra\ \citep{weisskopf_chandra_2000}. We further required that they have either: high-quality \xmm\ data (with $\gtrsim$ 10000 total counts in pn) or simultaneous broadband coverage with \nustar\  \citep{harrison_nuclear_2013} in at least one epoch. We were less stringent with the data constraints on the PULXs, since they are the only sources for which the nature of the accretor is known and thus they will be crucial for our study. From the current PULXs sample (NGC 7793 P13, NGC 5907 ULX1, NGC 300 ULX1, M51 ULX--7, NGC 1313 X--2 and M82 X--2), we discarded M82 X--2 as this source is only resolved by \chandra\ and its spectra are frequently affected by pile-up. Its emission is also often contaminated with nearby sources and diffuse emission of unknown origin \citep{brightman_spectral_2016}, precluding a clean detailed spectral analysis of the source. Another source of interest is M51 ULX--8, which was identified as harbouring an accreting NS through the detection of a cyclotron resonance feature \citep{brightman_magnetic_2018}, although no pulsations have been reported to date. We therefore included this source in our sample to further investigate its spectral properties with respect to the PULXs.

For those ULXs for which the accretor is unknown, we included at least two sources from each of the different spectral regimes proposed by \cite{sutton_ultraluminous_2013} in order to have a representative characterisation of the ULX population. We also made sure to include those showing evidence for super-Eddington outflows with 3$\sigma$ detections \citep{pinto_resolved_2016, pinto_ultraluminous_2017}. Thus, we also included NGC 55 ULX1, even if it does not meet our criterion of having more than three epochs. The final sample selected for this study is presented in Table \ref{tab:observations_xmm}.

Finally, note that the luminosities reported for NGC 5907 ULX1 in this work are subject to an additional source of uncertainty, as there is a large discrepancy in the distance measurements to the host galaxy. The distance measurements range from $\sim$17 \citep{tully_cosmicflows-3_2016} to $\sim$ 12.9 Mpc \citep{crook_groups_2007} which can boost the inferred luminosity by a factor of $\sim$ 1.7. Here we adopted the most recent estimate of 17.06 by \citep{tully_cosmicflows-3_2016}.
\subsection{Data reduction} \label{sub:data_reduction}
\xmm\ data reduction was carried out using SAS version 17.0.0. We produced calibrated event files from \epicpn\ \citep{struder_european_2001} and MOS \citep{turner_european_2001} cameras with the latest calibration files as of March 2018 using the tasks \textit{epproc} and \textit{emproc} (version 2.24.1), respectively. We selected events from patterns 0 to 4 for pn and patterns $\leq$ 12 for the MOS cameras. The standard filters were used to remove pixels flagged as bad and those close to the CCD gaps. We created high-energy (10 $-$ 12 keV) lightcurves from single pattern events from the full field of view to assess the presence of high-background particle flaring periods that could contaminate our spectra. We filtered these periods by removing times where the count-rate was above a certain threshold by visually inspecting the lightcurves. These thresholds varied for each observation, and ranged from $\sim$ 0.3 cts s$^{-1}$ to 1.2 cts s$^{-1}$ and from $\sim$ 0.2 cts s$^{-1}$ to 0.5 cts s$^{-1}$ for the pn and MOS respectively. 

Generally, we extracted source events from circular regions with a radius of 40" and 30" for pn and MOS respectively owing to their different angular resolution, rejecting observations in which the source fell on a chip gap. We reduced the source regions to avoid contamination from nearby sources, chip-gaps, or in cases where the source was faint in order to increase the S/N, but always ensuring that at lest 50\% of the PSF was enclosed. We used elliptical regions in cases where the source was placed at large off-axis angles, resulting in a distorted PSF. This was the case, for example, for the pn observations 0657801601 and 0657802001 of Holmberg IX X-1, observations 0112521001, 0112521101, 0657801801, 0657802001, 0657802201, 0693850801, 0693850901, 0693851001, 0693851101 of M81 X-6 and observation 0656580601 for Circinus ULX 5. The background region was selected from a larger circular source-free region and on the same chip as the source when possible. For the pn we also tried to select the background region from a distance to the readout node as close as possible as for the source region. Finally, for M51 ULX-7 we used regions of $\sim$20" and $\sim$ 25" for pn and MOS detectors respectively, to reduce contamination from the diffuse emission the source is immersed in \citep{rodriguez-castillo_discovery_2020}. We discuss possible contamination by the diffuse emission and its treatment in Section \ref{sub:choice_model}.

We noted also that Holmberg IX X-1 and Holmberg II X-1 were bright enough in some occasions to cause pile-up in the \xmm\ detectors. We assessed the importance of pile-up using the tool \textit{epatplot} (version 1.22) when the source count rate was above the recommended values\footnote{\label{footnote:pileup}\url{https://xmm-tools.cosmos.esa.int/external/xmm_user_support/documentation/uhb/epicmode.html}}. In order to mitigate its effects, we excised the inner core of the PSF by using an annular extraction region. The inner excised radius was never below 10.25" and 2.75" for the pn and the MOS cameras respectively, to avoid introducing inaccuracies in the flux estimation, as recommended\footnote{\url{https://www.cosmos.esa.int/web/xmm-newton/sas-thread-epatplot}}. 

We finally used the tasks \textit{rmfgen} (version 2.8.1) and \textit{arfgen} (version 1.98.3) to generate redistribution matrices and auxiliary response files, respectively. We regrouped our spectra to have a minimum of 20 counts per bin to allow the use of \chisq\ minimisation and also avoiding oversampling the instrumental resolution by setting a minimum channel width of 1/3 of the FWHM energy resolution.

We reprocessed the \textit{Chandra} data using the script \textit{chandra\_repro} with calibration files from \textit{CALDB} 4.8.2. We used extraction regions given by the tool \textit{wavdetect} to extract source events. Background regions were selected from roughly 3 times larger, circular, nearby, source-free regions. The level of pile-up was assessed by inspecting the images created using the \textit{pileup\_map}\footnote{\url{https://cxc.harvard.edu/ciao/ahelp/pileup_map.html}} tool. We rejected observations with a pile-up fraction $\gtrsim$ 5\%. We only considered observations that registered $\geq$ 1000 counts, as we found that below this threshold data were of too poor quality to robustly discriminate between different models. All data were also rebinned to a minimum of 20 counts per bin.

\textit{NuSTAR} data was processed using the \textit{NuSTAR} Data Analysis Software version 1.8.0 with \textit{CALDB} version 1.0.2. We extracted source and background spectra using nuproducts with the standard filters. Source events were selected from a circular region of $\sim$ 60". The only exception was NGC 1313 X-1, for which we follow \cite{walton_iron_2016} and chose a region of 40 -- 50" to reduce contamination from a nearby source \citep[see][]{bachetti_ultraluminous_2013}. Background regions were selected from larger circular source-free regions and on the same chip as the source but as far away as possible to avoid contamination from the source itself. We regrouped the \textit{NuSTAR}\ spectra to 40 counts per bin, owing to the lower energy resolution compared to the \textit{EPIC} cameras. For this work, we only considered \nustar\ observations for which simultaneous soft X-ray coverage with \xmm\ was available. A summary of all observations considered can be found in Table \ref{tab:observations_xmm}.

\longtab{
\begin{longtable}{cccccc}   
 \caption{\label{tab:observations_xmm} Sample of sources selected for this study with their respective distance, Galactic absorption \citep{kalberla_leiden/argentine/bonn_2005, hi4pi_collaboration_hi4pi_2016} and log of observations considered. For each particular epoch, the number of Xs,Ns or Cs indicate the number \xmm, \nustar\ and \chandra\ observations respectively, that have been fitted together. }\\
 \hline\hline 
 \noalign{\smallskip}
Epoch & Telescope & Instrument & Date & Obs. ID & Good Exp \\ 
    &  &  &  &  & (ks) \\ 
\noalign{\smallskip}
 \hline \hline
 \noalign{\smallskip}
 \endfirsthead
\caption{Continued.}\\
\hline 
\noalign{\smallskip}
\endhead
\endfoot
\hline
\endlastfoot
\multicolumn{6}{c}{Holmberg II X$-$1 ($D$ = 3.27 Mpc\tablefootmark{a}, \nh\ = 3.7 $\times$ 10$^{20}$ cm$^{-2}$)} \\
 \noalign{\smallskip} 
 \multirow{1}{*}{1X} & \textit{XMM-Newton} &  pn/MOS1/MOS2 & 2002-04-10 & 0112520601 &  4.6/9.5/9.6 \\ 
 \multirow{1}{*}{2X} & \textit{XMM-Newton} &  pn/MOS1/MOS2 & 2002-04-16 & 0112520701 &  3.7/5.1/5.7 \\ 
 \multirow{1}{*}{3X} & \textit{XMM-Newton} &  pn/MOS1/MOS2 & 2002-09-18 & 0112520901 &  3.8/5.8/6.5 \\ 
 \multirow{1}{*}{4X} & \textit{XMM-Newton} &  pn/MOS1/MOS2 & 2004-04-15 & 0200470101 &  35.0/49.6/51.1 \\ 
 \multirow{1}{*}{5X} & \textit{XMM-Newton} &  pn/MOS1/MOS2 & 2010-03-26 & 0561580401 &  21.7/39.6/40.7 \\ 
 \multirow{3}{*}{6XNN} & \textit{XMM-Newton} &  pn/MOS1/MOS2 & 2013-09-09 & 0724810101 &  4.6/6.6/6.6 \\ 
& \textit{NuSTAR} & FPMA/FPMB & 2013-09-09 & 30001031002 & 31.4/31.3 \\
\hline 
 \end{longtable}
 \tablefoot{Observations for which an instrument was not available because it was not active or because the source was not in the field of view are marked with "-". (P) indicates a source for which pulsations have been identified. Distances from: \tablefoottext{a}{\cite{tully_cosmicflows-3_2016}}, \tablefoottext{b}{\cite{lelli_small_2015}}, \tablefoottext{c}{\cite{karachentsev_local_2017}}, \tablefoottext{d}{\cite{tully_our_2008}}, \tablefoottext{e}{\cite{karachentsev_m_2002}} and  \tablefoottext{f}{\cite{cappellari_atlas3d_2011}}. \tablefoottext{g}{No pulsations have been detected in this source but it was identified as a NS through the detection of a cyclotron line \citep{brightman_magnetic_2018}.}}
}
\section{Data analysis and results}\label{sec:spectral_fitting}
We used the X-ray spectral fitting package \xspec\ \citep{arnaud_xspec:_1996} version 12.10.1f for spectral fitting and quote uncertainties on spectral parameters at the 90\% confidence level for a single parameter of interest, unless stated otherwise. All fluxes were estimated using the pseudo-model \cflux\ in \xspec. 

In general, we fitted EPIC data and the ACIS data in the 0.3 -- 10 keV range. For the sources with the highest absorption columns (NGC 5907 ULX1, IC 342 X-1 and Circinus ULX5), we inspected the \textit{epatplot} to choose the most suitable energy range to perform spectral fitting on the EPIC data. For IC 342 X-1 and Circinus ULX5, we noticed strong deviations from the observed pattern distributions with respect to the epatplot model below $\sim$ 0.4 keV. This is to be expected since the low-energy part of these absorbed spectra is likely dominated by the charge redistribution tail (and possibly noise). We therefore restricted the lower energy range to 0.4 keV for these two sources. For \nustar, we typically considered the 3 -- 35 keV range, although to avoid including bins with negative number of counts in the $\chi^2$ minimisation we restricted the high-end of the energy range to those bins where the net number of counts was positive. 

When simultaneously fitting spectra from different instruments from the same epoch, we attempted to compensate for calibration uncertainties by introducing a multiplicative cross-normalisation factor that was allowed to vary between the different instruments. This factor was frozen to unity for the pn (or the two MOS detectors if no pn data was available). We used the same factor for the MOS detectors as we found them to be generally in good agreement, while FPMA and FPMB had each their own separate constant, as recommended\footnote{\url{https://heasarc.gsfc.nasa.gov/docs/nustar/nustar_faq.html}}. The agreement between the pn and the MOS cameras was usually within errors, with a few cases in M81 X--6 where the disagreement reached up to 10\%, due to the highly elliptical distorted PSF because of the off-axis position of the source on the detectors. The value of the cross-normalisation factor between \textit{EPIC} data and \nustar\ detectors was typically in agreement within the errors, reaching in some cases a 5-20\% disagreement, with the largest values found when the source was highly off-axis on the \nustar\ detectors.  

Finally, we fitted together i.e. we assumed the same spectral model for two or more datasets if they were close in time ($\sim$ few days) and, if after inspecting each observation separately, we found no significant variation in flux and spectral parameters. This is also noted in Table \ref{tab:observations_xmm} where we quote the different datasets that have been fitted together based on the above. Throughout this work, we use the word \textit{epoch} to refer to all datasets that have been fitted together assuming the same model. 

\subsection{Spectral modelling} \label{sub:model}
\subsubsection{Choice of model} \label{sub:choice_model}
Our first aim was to characterise the long-term spectral evolution of our sample in a simple and coherent manner, by studying variations of the spectral components (i.e. luminosity, radius, temperature, etc). As the latest studies have revealed that the 0.3 -- 10 keV band can be modelled by two thermal components \citep[e.g.][]{mukherjee_hard_2015, koliopanos_ulx_2017, koliopanos_investigating_2019, walton_unusual_2020} that can reproduce the curvature seen at high-energies, we first considered a phenomenological model  based on two multi-colour blackbody disks \citep[\diskbb\ in \xspec;][]{mitsuda_energy_1984} to fit the data in this band, taking into account interstellar absorption by neutral hydrogen with two absorbing components \tbabs\ in \xspec. One was frozen at the Galactic value along the source line of sight (see Table \ref{tab:observations_xmm}), and the other one was left free to vary to take into account possible absorption from the host galaxy and the system itself. We adopted abundances given by \citet{wilms_absorption_2000} and cross-sections given by \citet{verner_atomic_1996} as recommended. 

While other models have been commonly adopted to reproduce the hard emission ($\sim$ 2 keV -- 10 keV) in ULXs, like Comptonisation of soft disk photons in a warm optically thick corona or a more phenomenological power-law, we preferred the \diskbb\ over these for various reasons. First of all, a warm optically thick corona up-scattering photons from the disk is merely a proxy to reproduce the curvature seen at high-energies, as its physical interpretation is subject to several caveats \citep[see][and references therein]{koliopanos_ulx_2017}. Additionally, broadband spectroscopic studies have shown how this model fails to reproduce the spectral shape of ULXs \citep{walton_nustar_2015, mukherjee_hard_2015}. Therefore, for the purpose of reproducing the spectral curvature in (at least) the \xmm\ band, the \diskbb\ has been shown to be equally valid with only two parameters. The power-law (or a power-law with a cutoff for the same matter) unphysically diverges towards low-energies, thus taking up flux from the soft component and causing the absorption column to be overestimated. 

Another advantage of the dual thermal component is that thanks to its simplicity, it can be used as a proxy to represent more complex models. For instance, in the context of super-Eddington accretion, the soft black body has been frequently associated with an outflow, while the hard one has been associated with emission from the inner parts of the accretion flow \citep[e.g.][]{walton_broadband_2014}. Alternatively, \cite{koliopanos_ulx_2017} associated the soft component with an accretion disk truncated at the magnetospheric radius, and the hard component to the emission from the magnetospheric envelope \citep{mushtukov_optically_2017}. The \diskbb\ also allows to test easily theoretical predictions by studying the evolution of its temperature with its luminosity as we show in Sections \ref{sub:correlations_soft} and \ref{sub:correlations_hard}.

\begin{figure*}
   \centering
      \includegraphics[angle=-90,width=0.45\textwidth]{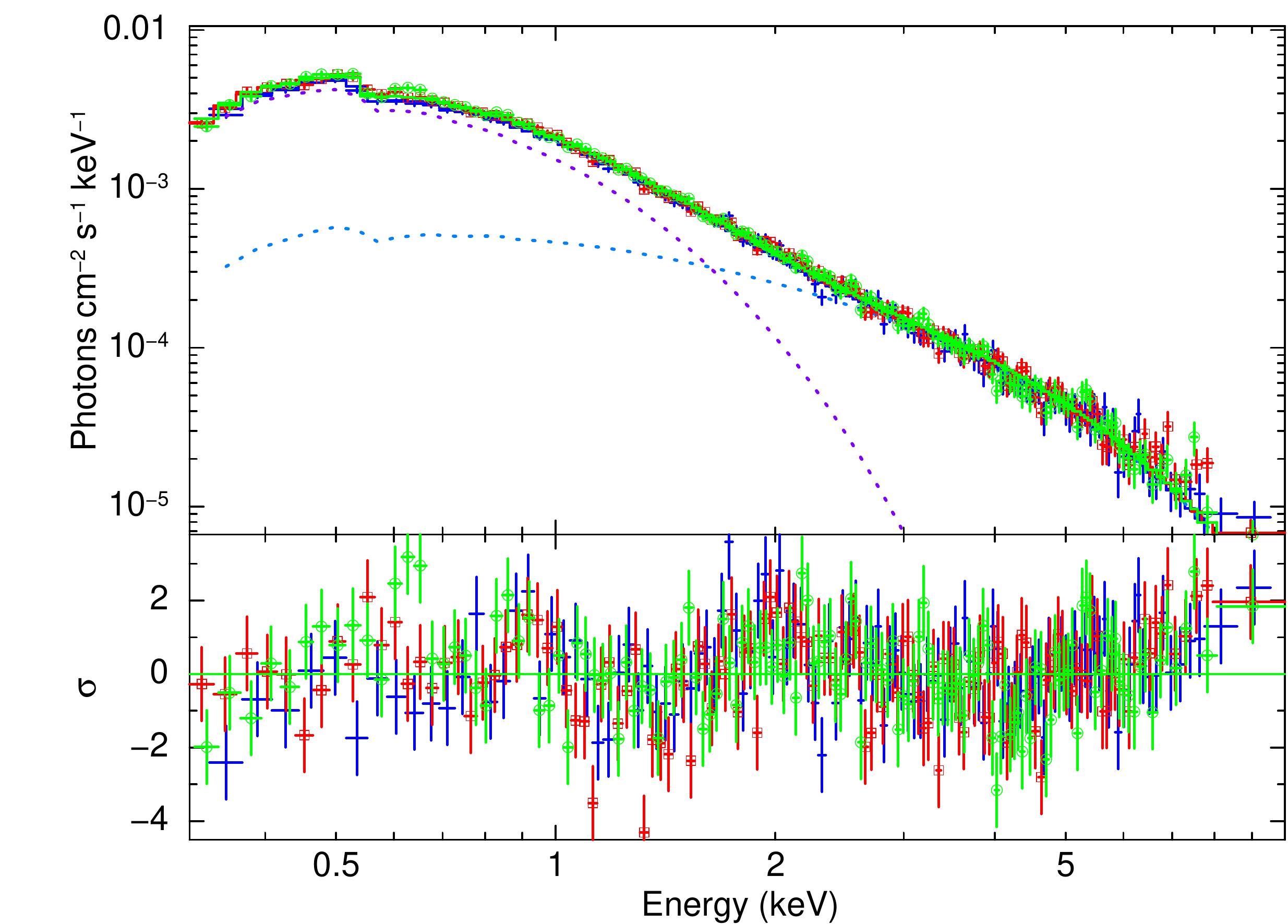}
 \includegraphics[angle=-90,width=0.45\textwidth]{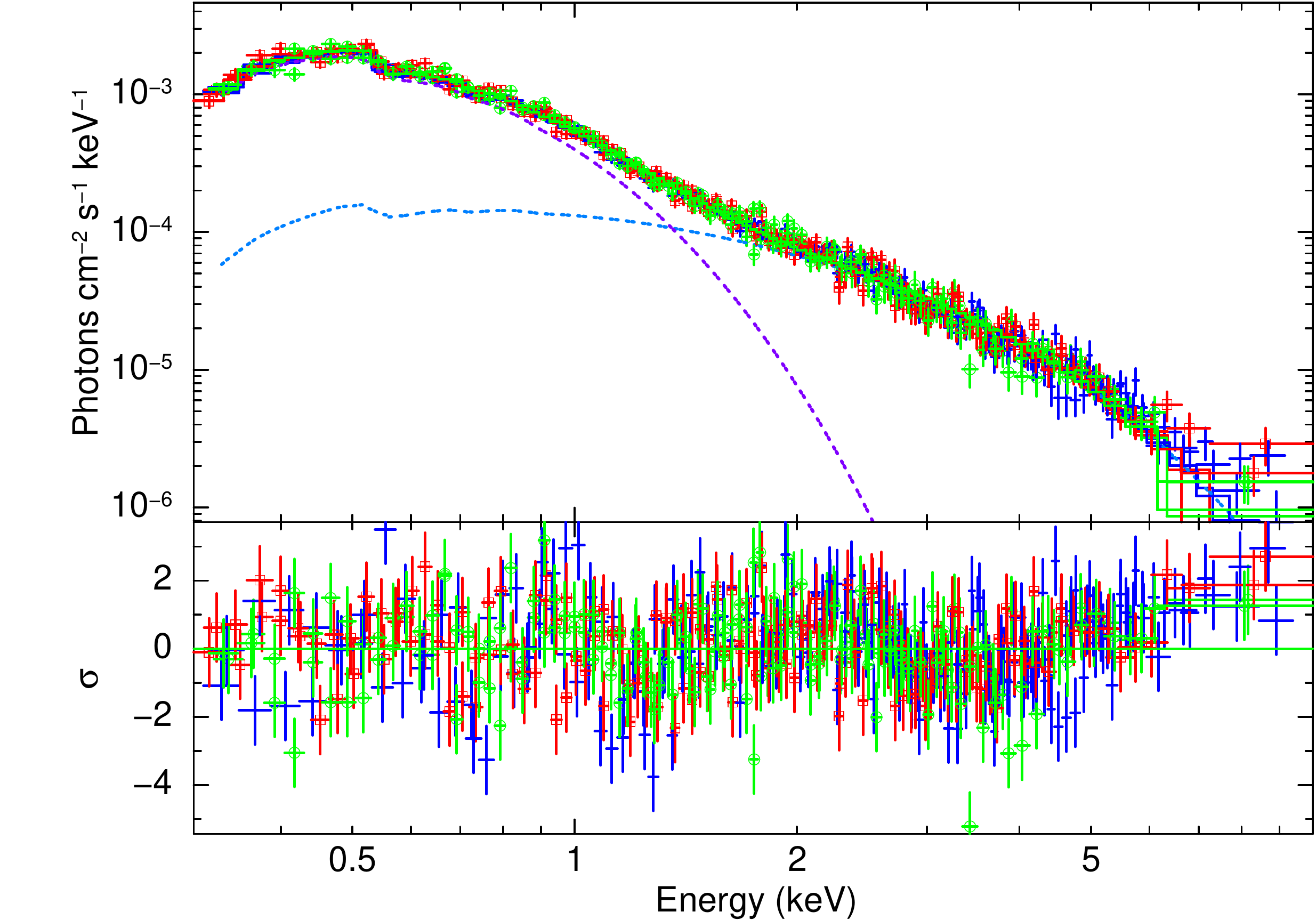}
   \caption{Unfolded pn (blue), MOS1(red) and MO2 (green) spectra fitted with an absorbed dual thermal component (\tbabs$\otimes$\tbabs$\otimes$(\diskbb + \diskbb) on \textit{left} observation 0200470101 of Holmberg II X-1 (\chisq$_r$ $\sim$ 1.34) and \textit{right} observations 0723130301 and 0723130401  of NGC 5408 X-1 (\chisq$_r$ $\sim$ 1.5), that we fit together given the lack of variability (see Section \ref{sec:spectral_fitting}). Data has been visually rebinned to have at least 3 $\sigma$ significance and a minimum of 35 counts per bin. Some clear residuals are seen at high energies indicating that the model is inadequate.}
    \label{fig:residuals}
    \end{figure*}

However, visual inspection of the fit residuals revealed strong residuals at high energies in the 0.3 -- 10 keV band in some epochs, indicating that our phenomenological model is not able to reproduce the emission at high energies. We show this for two high quality observations of Holmberg II X-1 and NGC 5408 X-1 in Fig \ref{fig:residuals}. This is perhaps not surprising, as broadband studies using \nustar\ data have revealed the presence of a faint hard power-law like excess dominating above $\sim$ 10 keV \citep[e.g.][]{mukherjee_hard_2015, walton_nustar_2015, walton_broadband_2017}. We found that for those epochs for which we had broadband coverage with \nustar, this excess can be well modelled as Comptonisation of the hard/hot thermal component as noted by previous studies. Since the nature of this Comptonisation component is still poorly understood \citep{walton_broadband_2017} and given the lack of broadband coverage for most of the observations considered here, we decided to use the \simpl\ model \citep{steiner_simple_2009} to reproduce it, as it does not assume any geometry but simply scatters a fraction of photons (\fscatt) of the seed component towards high energies emulating a power-law component with a certain $\Gamma$, with the advantage that it does not diverge towards low energies. 

In order to identify those epochs for which the \simpl\ model component is required, we would ideally rely on Monte-Carlo simulations. However, given the large number of datasets considered here this is not feasible. Instead, to have an estimate as to when the data quality allows to constrain this component we performed an F-test between the dual-thermal model described above and the model \tbabs$\otimes$\tbabs$\otimes$(\diskbb\ + \simpl$\otimes$\diskbb) and decided to include the \simpl\ model only if the probability of rejecting ($P_\text{rej}$) the simpler model was $\sim$ 3 $\sigma$. We also included the \simpl\ model for observations with $P_\text{rej}$ slightly below this value if we were able to constrain its parameters using near-in-time observations with $P_\text{rej}$ $>$ 3 $\sigma$ (see below). Similarly, we excluded the \simpl\ component in some cases with $P_\text{rej}$ $>$ 3 $\sigma$ if its parameters are largely unconstrained. We are aware that this is not an appropriate use of the F-test, however, the presence of this component was already shown to be significant by \citet{walton_evidence_2018} and indeed we found that this component was required to fit the highest quality datasets. We are not claiming whether this component is present or not, but instead we used these values to have a reference as to when the data quality allows to constrain this component. We present the result of this F-test in Table \ref{tab:ftest} where we also indicate for which observations we finally included this component.

For NGC 7793 P13 we found that this component is largely unconstrained even in the broadband datasets, resulting in no fit improvement when adding this component. The broadband datasets yield fits with $\chi^ 2_r$ $\sim$ 1.1 with the dual-thermal model alone, which are statistically acceptable, so we decided not to add this component.

Furthermore, as the \simpl\ model is mostly responsible for the emission above $\sim$ 8 keV in our modelling, its parameters are not well constrained when \chandra\ or \xmm\ data are considered alone, and only in long \xmm\ exposures with $>$ 40000 counts in pn we can constrain both \fscatt\ and $\Gamma$ simultaneously. We also found good consistency between the $\Gamma$-values in several epochs and that most sources are found recurrently at similar fluxes and hardness ratios (see Section \ref{fig:hid_diagram}). We thus tied $\Gamma$ across epochs where the source is found at similar flux and hardness ratio, while leaving \fscatt\ and the rest of the parameters free to vary. By doing so we were able to use the broadband information provided by \nustar\ to have some constraints on those epochs where this information is not available. For NGC 1313 X-1, after checking the consistency of $\Gamma$ across epochs and bearing in mind that its variability above 10 keV has been shown to be small \citep{walton_unusual_2020}, we tied $\Gamma$ for all observations where the \simpl\ model is used. For NGC 5408 X-1 and NGC 6946 X-1, we also found very little variability at high-energies and thus we again tied $\Gamma$ across all observations where we added the \simpl\ model. Finally, for NGC 1313 X-2, we also found that the same $\Gamma$-value can be tied for the four epochs for which we included the \simpl\ model, albeit in this case the source shows certain variability at high energies (see Section \ref{sub:spectral_transitions}).

For M81 X-6, M51 ULX-7, M51 ULX-8 and M83 ULX1, given the lack of broadband coverage and that the dual-thermal model gave an acceptable fit to the data, we found no need to try to improve the fit by adding the \simpl\ model and thus we did not explore this option. As stated in Section \ref{sec:sample_data_reduction}, the \xmm\ observations of M51 ULX-7 might be contaminated by some diffuse emission. To quantify the possible contamination from this diffuse emission, we extracted two MOS1 and MOS2 spectra (from obs id 0824450901) from two circular (25" in radius) regions located 34" from the source in two different directions. We fitted both emission spectra with a powerlaw and a \mekal\ model subject to Galactic absorption only. The spectral parameters of both spectra are consistent within errors, with $\Gamma$ = 2.9\errors{0.4}{0.6} and 2.8$\pm$0.3 and plasma temperature = 0.28$\pm$0.03 keV and 0.27\errors{0.04}{0.03} keV, for region 1 and 2 respectively. We thus ruled out strong spatial variations and computed the luminosity of region 1. We found a total unabsorbed luminosity of (4.9\errors{0.7}{0.3}) $\times$ 10$^{38}$ erg/s in the 0.3 -- 10 keV band, with the soft band (0.3 -- 1.5 keV) being $\sim$ 5 times brighter than the hard band. We thus estimated that the diffuse emission can make the source appear $\sim$ 25\% softer in the \xmm\ observations, given that M51 ULX-7 has a typical luminosity of 4 $\times$ 10$^{39}$ erg/s. We therefore added this diffuse emission model as a fixed additive component to the M51 ULX-7 continuum in all \xmm\ observations and decided to discard obsids 0677980701, 0677980801 and 0830191401 when the source was at its lowest.

Finally, some sources showed strong residuals at soft energies that have been associated with unresolved emission and absorption lines produced in an outflow colliding with the circumstellar gas \citep{pinto_resolved_2016, pinto_ultraluminous_2017, pinto_xmmnewton_2020}. In most cases, we ignored them, as we are interested in the continuum emission and this will not affect our results. There are two exceptions to this: for NGC 5408 X-1, including a Gaussian emission line at $\sim$ 1 keV gave a $\Delta \chi^2$ improvement that ranges from $\sim$ 50 to 122 (for 3 extra degrees of freedom) depending on the observation. We thus decided to also determine the parameters of this Gaussian in the joint fit of the high-quality datasets, by tying together all its parameter. We obtained E = 0.947$\pm$0.005 keV, $\sigma$ = 78$\pm$6 eV and normalisation = 2.7\errors{0.3}{0.2} $\times$ 10$^{-5}$ photons/cm$^{2}$/s. For NGC 5204 X-1, we also included a Gaussian emission line for the \chandra\ obsid 3933 following \citet{roberts_chandra_2006} as we again noted similar strong residuals. In this case we obtained E = 0.97$\pm$0.02 keV, $\sigma$ = 56\errors{24}{21} eV and normalisation = 2$\pm$1 $\times$ 10$^{-5}$ photons/cm$^{2}$/s, consistent with the values reported by \citet{roberts_chandra_2006}. 

\subsubsection{Treatment of the absorption column} \label{sub:absorption_column}
There is still no consensus as to whether the local absorption column in ULXs is variable \citep[e.g.][]{kajava_spectral_2009} or not \citep[e.g.][]{miller_revisiting_2013}. The main argument for variability of the absorption column in ULXs is the contribution from outflows \citep[e.g.][]{kajava_spectral_2009, middleton_diagnosing_2015}, which could imprint stochastic variability in \nh\ due to wind clumps crossing our line of sight \citep{takeuchi_clumpy_2013, middleton_spectral-timing_2015} depending on their column density and ionisation state. Alternatively, \citet{middleton_diagnosing_2015} argued that some expelled gas could cool down far from the source and contribute to the neutral absorption column. If this was the case, it is then not clear over which timescale \nh\ would react to instantaneous changes in the mass-accretion rate.

To make matters more complicated, \cite{miller_neutral_2009} showed by studying absorption edges at high spectral resolution in X-rays that the absorption column remains stable throughout spectral changes in a set of X-ray binaries, and that changes in the soft component must come from changes in the source spectrum and not from the absorption column itself. 

\begin{figure}
   \centering
   \includegraphics[width=0.49\textwidth]{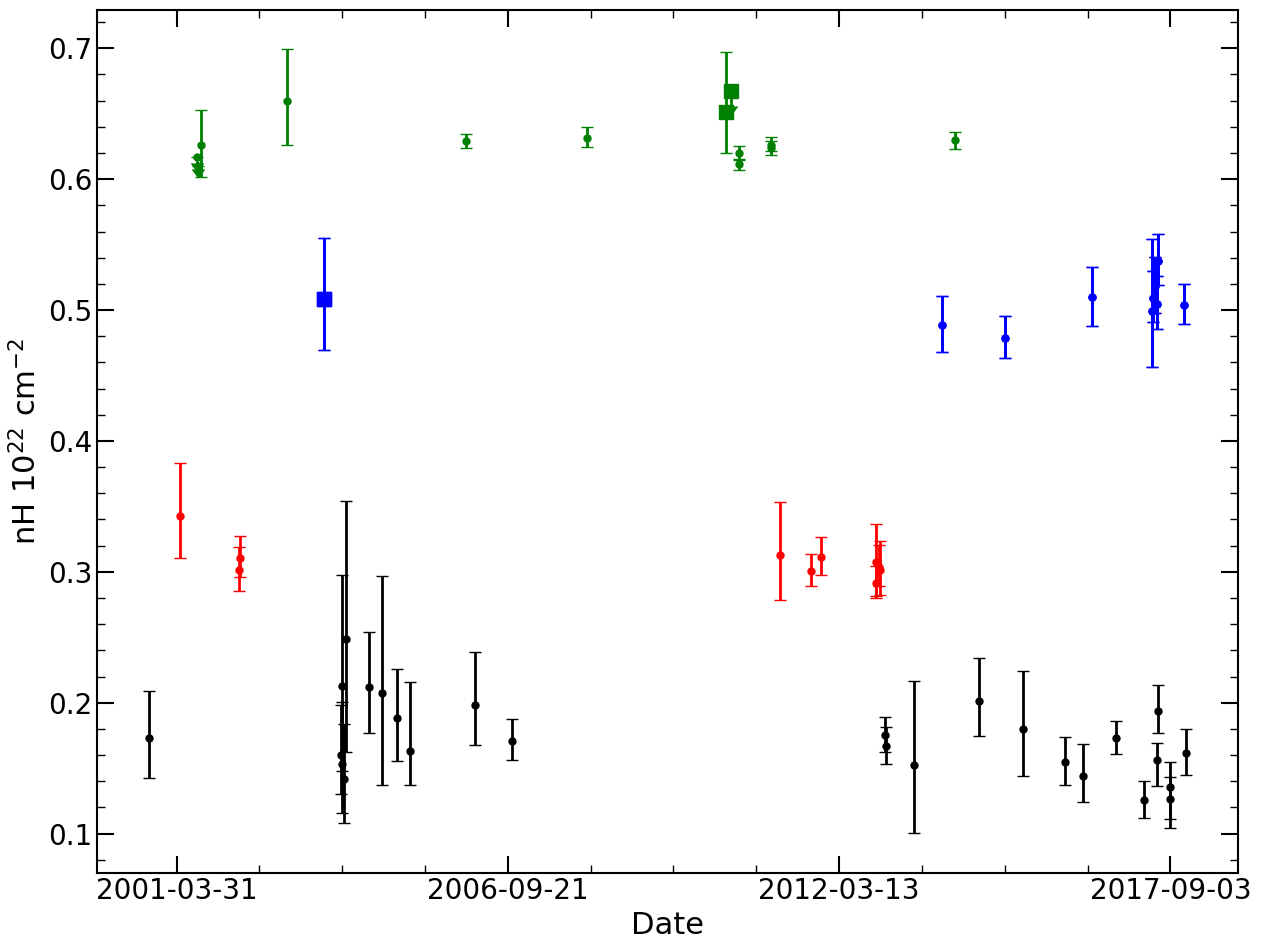}
   \caption{Evolution of the local absorption column over time for NGC 5408 X-1 (green), NGC 7793 P13 (blue), Holmberg IX X-1 (red) and NGC 1313 X-2 (black) shifted by 0.2 10$^{22}$ cm$^{-2}$ for clarity. All spectra have been fitted with a model consisting of an absorbed dual \diskbb\ model in the 0.3 -- 8 keV band. Errors are shown at 3$\sigma$ confidence level. Circles and squares correspond to \xmm\ and \chandra\ data respectively. }
    \label{fig:tbabs}
    \end{figure}

In view of these complications, we attempted to study variations of \nh\ by fitting individually all epochs for each source using the dual-thermal component over the 0.3 -- 8 keV band, so as to use the same model for all epochs and avoid introducing artificial changes in \nh\ due to the different energy ranges considered when using \nustar\ data. We found that in general, the values of the absorption column for a given source were consistent within 3$\sigma$ errors throughout epochs, which we show in Figure \ref{fig:tbabs} for four sources in our sample. Small discrepancies can be attributed to low data quality (e.g. short exposure \chandra\ observations and/or calibration uncertainties) rather than real physical-\nh\ changes. For NGC 5408 X-1, when fitting the 0.3 -- 8 keV band with the dual-thermal model we still saw strong residuals at high-energies, which may indicate that the dual-thermal model cannot adequately fit the 0.3 -- 8 keV band. We thus repeated the study of the \nh\ variations replacing the hard \diskbb\ with a power-law with a high-energy cutoff (\cutoffpl\ in \xspec) to ensure that the lack of mismodelling of the high-energy emission does not affect the \nh\ variations (or lack of) found before. With this model we again found all \nh\ values consistent at the 3 $\sigma$ level.

\begin{figure}
    \centering
     \includegraphics[width=0.49\textwidth]{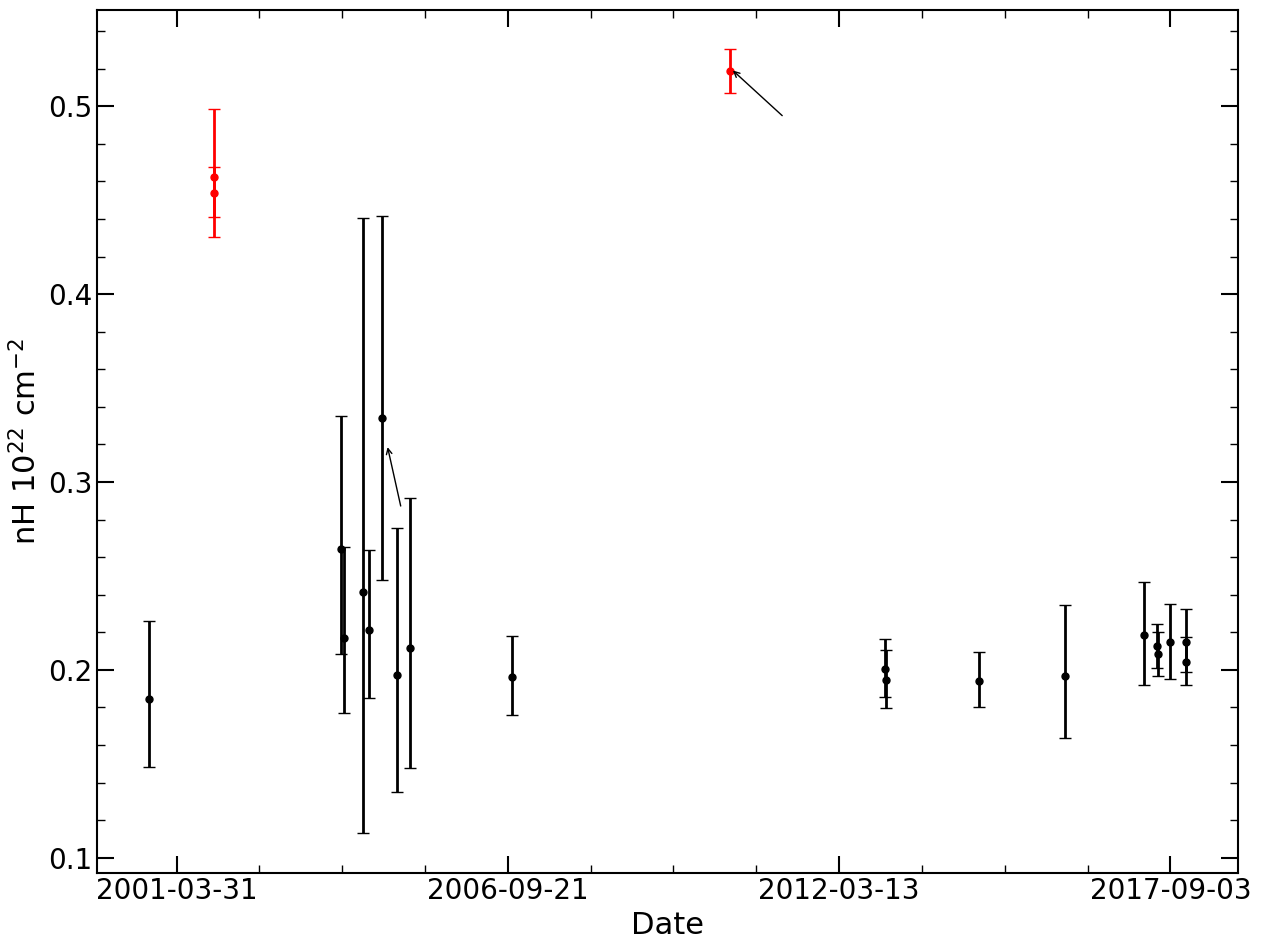}
    \caption{As for Figure \ref{fig:tbabs} but for NGC 1313 X-1 (black) and NGC 55 ULX1 (red). Arrows indicate epochs where a 3 $\sigma$ significant changes is seen with respect to some other epochs (see text for details). }\label{fig:tbabs_variable}
    \end{figure}

Conversely, we did find evidence for variability in the absorption column of NGC 1313 X-1 and NGC 55 ULX1 when following the same approach, where variations above the 3$\sigma$ confidence level were seen (see the arrows in Figure \ref{fig:tbabs_variable}). However, given the phenomenological nature of our model, we cannot rule out that these discrepancies are due to a change of the underlying continuum, changing therefore the parameter degeneracies. 

For the other sources not discussed, we did not find as strong evidence for variability in the absorption column as for the sources discussed above, although in some cases this could be due to poorly constrained model parameters. Thus, in view of the above considerations, we considered that the absorption column can be assumed to be constant for the most part, with the exception of epoch 2004-08-23 of NGC 1313 X-1 and epoch 2010-05-24 of NGC 55 ULX1. In these cases we allowed \nh\ to vary together with the continuum parameters as this is preferred by the fit. We found a $\Delta\chi^2$ improvement of $\sim$ 25 and 15 for NGC 1313 X-1 and NGC 55 ULX1 respectively when leaving \nh\ free, compared to the case where \nh\ is frozen to the average value (see next Section). We discuss this in more detail in Section \ref{sub:all_other_ulxs}.

\subsubsection{Spectral fitting approach}
Ideally, we would jointly fit all datasets for each source, tying the \nh\ and $\Gamma$ across certain datasets as explained above. However, such a procedure would be computationally prohibitive. Since the joint fit will mostly be driven by those datasets  with higher data quality, in order to reduce the computational burden of this approach we decided to do our spectral fitting in two steps: we first jointly fitted those datasets with better statistics, tying together $\Gamma$ across those epochs where the source is found at similar flux/hardness ratio and \nh\ across \textit{all} epochs, while the rest of the parameters are free to vary. We typically consider 3 to 8 datasets for each source that include those observations with the longest \xmm\ exposures when the three EPIC cameras are operational, those for which simultaneous \nustar\ coverage is available and in some rare cases, the longest \chandra\ observations. We next fitted individually the remaining observations of lower data quality with \nh\ and $\Gamma$ frozen at the values found in the joint fit. The results of the joint and the subsequent individual fits are reported in Table \ref{tab:individual_fits}. 

For Circinus ULX5, the joint fit approach resulted in largely unconstrained parameters for the low-energy components at soft energies. This is likely due to a combination of the calibration uncertainties at low energies (see Section \ref{sec:spectral_fitting}) and the high absorption column along the line of sight  \citep[$n_\text{HGal}$ = 50.6$\times$ 10$^{20}$ cm$^{-2}$;][]{hi4pi_collaboration_hi4pi_2016}. We thus considered only epoch 2013-02-03 to constrain \nh\ as it offers the best constrains on the broadband emission.

\longtab{
\begin{longtable}{ccccccccc}
 \caption{\label{tab:individual_fits}Best fit spectral fit parameters. Errors are quoted at a 90\% confidence level. The first epochs in each source correspond to the high-quality datasets, that were jointly fitted and from which we determine $\Gamma$ and \nh. The rest of the data were fitted individually by freezing \nh\ and/or $\Gamma$ to the best fit value found in the joint fit of the high-quality data.} \\
\hline\hline
\noalign{\smallskip}
 Epoch & \nh\ & kT$_\text{soft}$ & norm & kT$_\text{hard}$ & norm & $\Gamma$ & \fscatt\ & $\chi^2_r$/dof \\ 
& 10$^{20}$ cm$^{-2}$ & keV &  & keV & 10$^{-2}$ &  & \% &  \\	
\hline
\noalign{\smallskip}
\endfirsthead
\caption{Continued.}\\
\hline\hline
\noalign{\smallskip}
 Epoch & \nh\ & kT$_\text{soft}$ & norm & kT$_\text{hard}$ & norm & $\Gamma$ & \fscatt\ & $\chi^2_r$/dof \\ 
& 10$^{20}$ cm$^{-2}$ & keV &  & keV & 10$^{-2}$ &  & \% &  \\
\noalign{\smallskip}
\hline
\noalign{\smallskip}
\endhead
\noalign{\smallskip}
\hline
\endfoot
\multicolumn{9}{c}{Holmberg II X-1} \\ 
\noalign{\smallskip}
1X& \multirow{6}{*}{4.3$\pm$0.5} & 0.26 \errors{0.03}{0.08} & 39 \errors{18}{10} &0.47\errors{0.2}{0.3} & 333\errors{2592}{262} & 	\multirow{4}{*}{2.5 \errors{0.2}{0.5}}  & 62\errors{a}{30} & \multirow{6}{*}{1.01/ 2182}\\
2X& & 	 0.27$\pm$0.02 & 38\errors{11}{8} & 0.8\errors{0.4}{0.3} & 	 35\errors{199}{26}& & 70\errors{a}{37} & 	\\
4X& & 	 0.29$\pm$0.01 & 	 35\errors{6}{5} & 0.8\errors{0.2}{0.3} & 35\errors{92}{16} & & 47\errors{41}{21} & 	\\
5X& & 	 0.23$\pm$0.01 & 	 43\errors{7}{6} & 0.87\errors{0.09}{0.10} & 	 11\errors{7}{4}& 	 & 	 30$\pm$10 & 	 \\
6XNN&  & 	 0.27$\pm$0.01 & 	 31\errors{6}{5} & 	1.14\errors{0.22}{0.01} & 	 8\errors{2}{3} & \multirow{2}{*}{3.2$\pm$0.1} & 	 100\errors{a}{30} & \\ 
7XN&  &0.26$\pm$0.01& 36\errors{8}{7} &  1.07\errors{0.30}{0.07} & 10\errors{3}{5} & 	 & 100\errors{a}{50} &\\\\
\end{longtable} 
  \tablefoot{\tablefoottext{a} {Parameter frozen at the value determined in the joint fits.}\tablefoottext{b}{ Unbound parameter.}}
 }
 
\longtab{
\begin{longtable}{ccccccc}
 \caption{\label{tab:fluxes} Estimated unabsorbed luminosities in units of 10$^{39}$ erg/s for each epoch using the distances from Table \ref{tab:observations_xmm} and models from Table \ref{tab:individual_fits}. For jointly fit data, the error estimation takes into account the errors on the joint parameters. Uncertainties are quoted at 90\% confidence level. First column indicates the epoch given in Table \ref{tab:observations_xmm} sorted chronologically. The \simpl\ flux luminosity corresponds to the entire \diskbb$\otimes$\simpl\ component.}  \\
\hline\hline
\noalign{\smallskip}
 Epoch & L$_\text{soft}$& L$_\text{hard}$ & L$_\text{tot}$  & L$_\text{soft diskbb}$  & L$_\text{hard diskbb}$ & L$_\text{simpl}$  \\ 
 & (0.3-1.5 keV)  & (1.5-10 keV) & (0.3 -- 10 keV) & (0.01 -- 100 keV) & (0.01 -- 100 keV) & (0.01 -- 100 keV) \\
\hline
\noalign{\smallskip}
\endfirsthead
\caption{Continued.}\\
\hline\hline
\noalign{\smallskip}
 Epoch & L$_\text{soft}$& L$_\text{hard}$ & L$_\text{tot}$  & L$_\text{soft disk}$  & L$_\text{hard disk}$ & L$_\text{simpl}$  \\ 
 &  (0.3-1.5 keV)  & (1.5-10 keV) & (0.3 -- 10 keV) & (0.01 -- 100 keV) & (0.01 -- 100 keV) & (0.01 -- 100 keV) \\
\noalign{\smallskip}
\hline
\noalign{\smallskip}
\endhead
\noalign{\smallskip}
\hline
\endfoot
\multicolumn{7}{c}{Holmberg II X-1} \\ 
\noalign{\smallskip}
 1X & 6.1$\pm$0.2& 4.3$\pm$0.1& 10.4$\pm$0.2 &5\errors{1}{3} &5\errors{2}{1} &9\errors{4}{2} \\ 
2X & 5.5$\pm$0.2& 4.5$\pm$0.2& 10.0$\pm$0.2 &5.6\errors{0.4}{0.7} &3.8$\pm$0.9 &8$\pm$1 \\ 
3X & 1.87$\pm$0.05& 0.57$\pm$0.04& 2.44$\pm$0.06 &2.4$\pm$0.1 &1.0$\pm$0.1 & - \\ 
4X & 6.56$\pm$0.14& 4.52$\pm$0.06& 11.1$\pm$0.2 &7.0\errors{0.3}{0.4} &4.6\errors{0.3}{1.0} &8.0\errors{1.5}{0.6} \\ 
5X & 2.87\errors{0.08}{0.06}& 1.43$\pm$0.03& 4.30\errors{0.09}{0.04} &3.4$\pm$0.1 &1.79\errors{0.08}{0.09} &2.6\errors{0.4}{0.2} \\
6XNN & 4.17\errors{0.05}{0.11}& 4.23$\pm$0.06& 8.40\errors{0.07}{0.13} &4.7$\pm$0.2 &3.6\errors{0.6}{0.1} &6.4\errors{0.3}{0.2} \\ 
7XN & 3.98\errors{0.09}{0.11}& 4.02$\pm$0.06& 8.0$\pm$0.01 & 4.5$\pm$0.2 & 3.4\errors{0.8}{0.2} & 6.1$\pm$0.3 \\
\end{longtable} 
 }

\subsection{Hardness-luminosity diagram} \label{sub:hid_diagram}
As a first source classification and in order to highlight the differences and similarities between the sources in our sample, we started by building a hardness ratio luminosity diagram (HLD), similar to that often used for X-ray binaries  \citep{done_observing_2003} and also for ULXs \citep[e.g.][]{sutton_ultraluminous_2013}. We did this by computing unabsorbed fluxes rather than counts since given the different instruments employed for this work, relying on count rates is not feasible. Furthermore, fluxes have the advantage that can be corrected for absorption column. To do this, we retrieved the total unabsorbed luminosity in the 0.3 -- 10 keV band from the \tbabs$\otimes$\tbabs $\otimes$(\diskbb\ + \simpl$\otimes$\diskbb\ or \diskbb), depending on the epoch. The hardness ratio is computed as the ratio of unabsorbed fluxes in a soft band (0.3 -- 1.5 keV) and a hard band (1.5 -- 10 keV). This is motivated by the fact that the pulsating component in PULX has been shown to dominate at high energies \citep[e.g.][]{israel_accreting_2017, walton_super-eddington_2018} and thus we may expect to highlight the differences between pulsating and non-pulsating sources, while the soft component in ULXs usually stops dominating above $\sim$ 1 keV.

 \begin{figure*}
  \centering
  \includegraphics[width=\hsize]{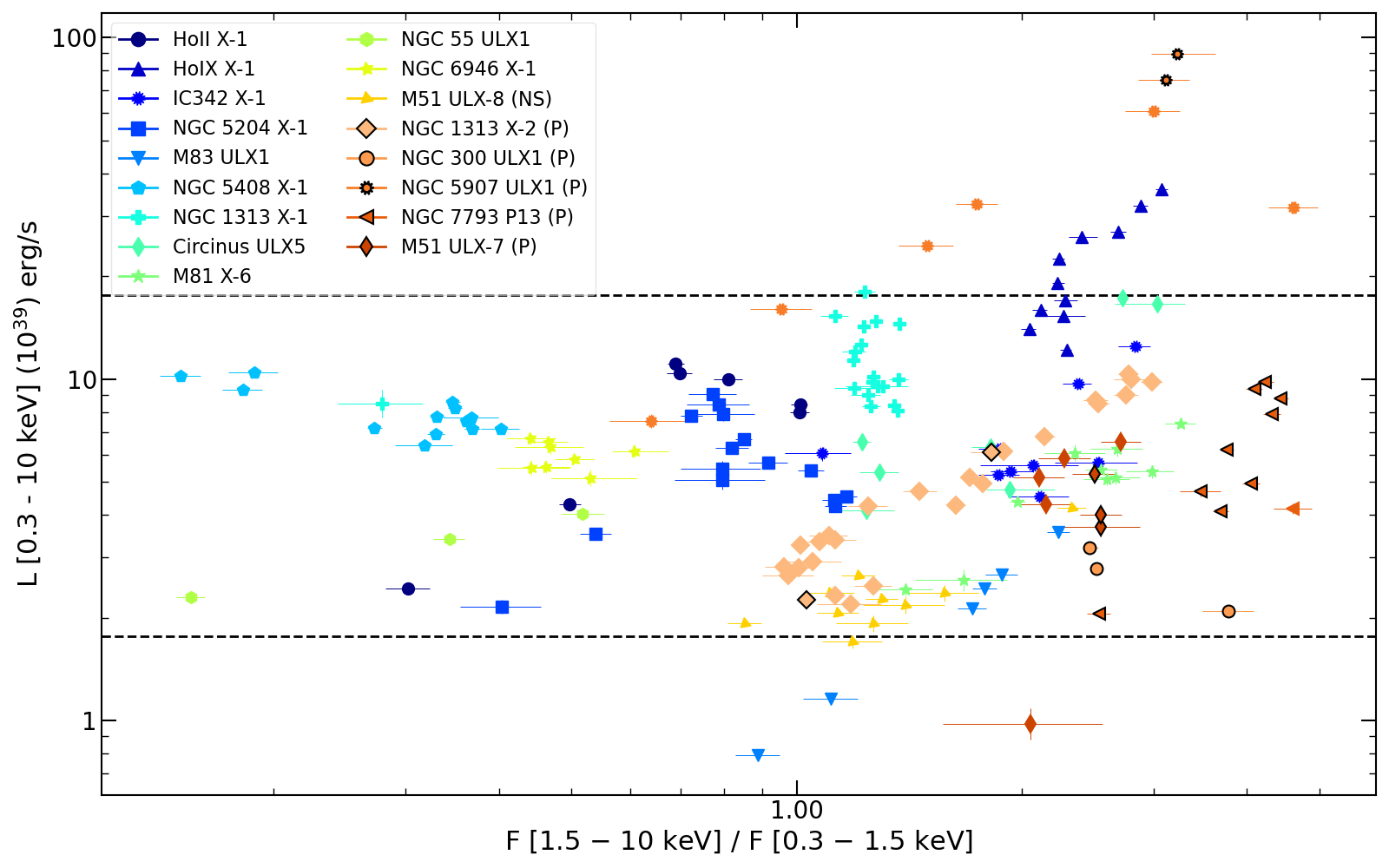}
  \caption{Hardness-luminosity diagram for the ULX sample selected for this study. All fluxes and luminosities are unabsorbed. Pulsating ULXs are shown in shades of orange and the epochs were pulsations have been reported in the literature \citep{israel_accreting_2017, furst_tale_2018,carpano_discovery_2018, vasilopoulos_ngc_2018, sathyaprakash_discovery_2019, rodriguez-castillo_discovery_2020} are highlighted by a black edge around the marker. The dashed black lines indicate respectively 10 and 100 times the Eddington limit for a neutron star ($\sim$ 2$\times$ 10$^{38}$ erg/s).}    
    \label{fig:hid_diagram}
    \end{figure*}

\begin{figure}
    \centering
    \includegraphics[width=0.45\textwidth]{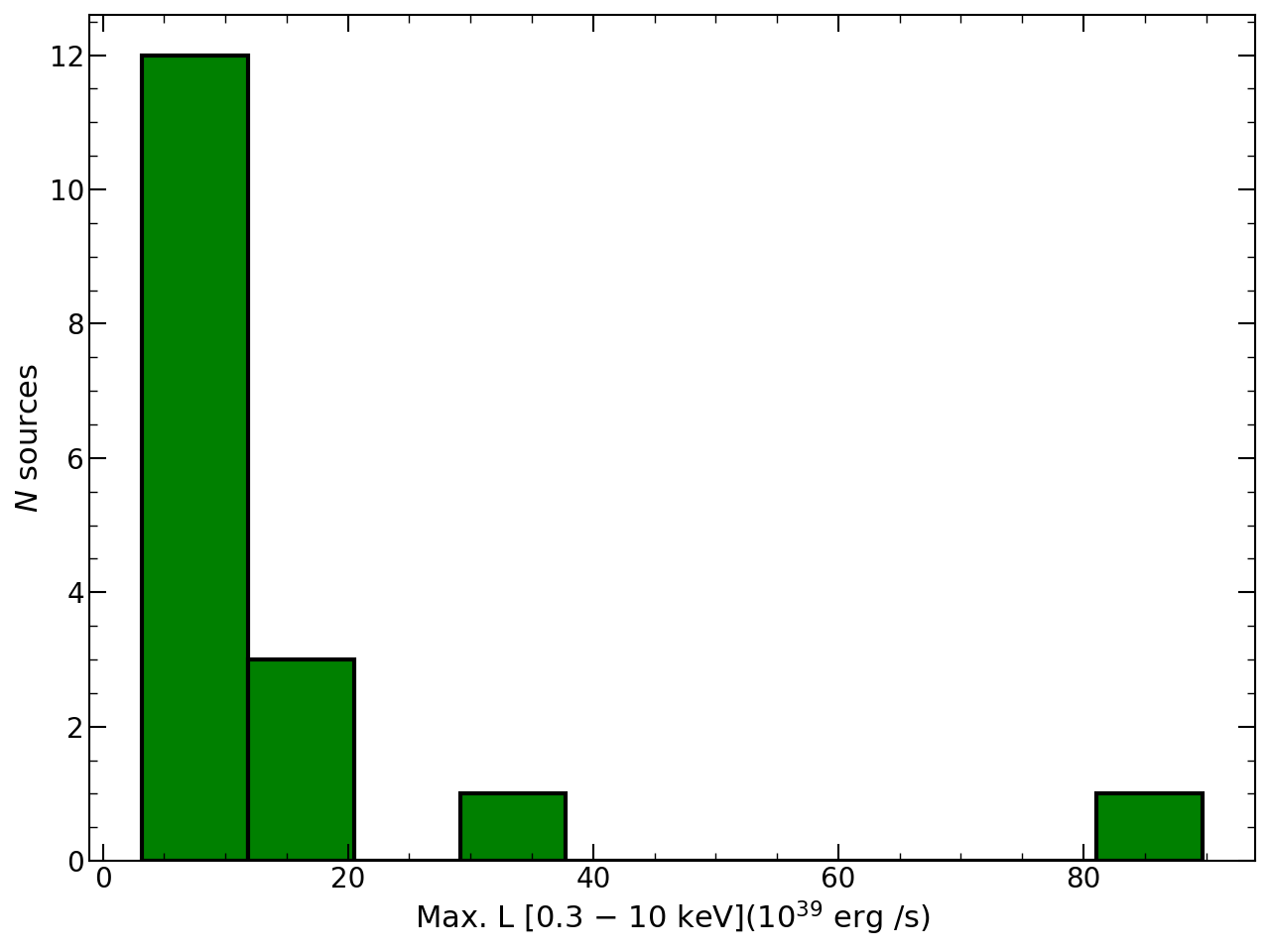}
    \caption{Histogram of the maximum unabsorbed luminosity in the 0.3 -- 10 keV band attained by each source considered in this work. A clear drop of sources with L$\geq$ 2 $\times$ 10$^{40}$ erg/s is seen.}
    \label{fig:l_histogram}
\end{figure}

Indeed, the results presented in Figure \ref{fig:hid_diagram} and Table \ref{tab:fluxes} show clearly that PULXs are harder than the rest of the sample, with some interesting exceptions. In general, PULXs harden in the 0.3--10 keV band with luminosity with high levels of flux variability. We found that most of the population reaches a maximum luminosity of $\sim$ 2 $\times$ 10$^{40}$ erg/s, with just Holmberg IX X-1 and NGC 5907 ULX1 above this value. This is also highlighted in Figure \ref{fig:l_histogram}, where a drop in sources reaching a maximum luminosity $\sim$ 2 $\times$ 10$^{40}$ erg/s is clearly seen.

Lastly, we note that as we have rejected observations below $\sim$ 1000 counts, transitions to the off states below $\sim$ 10$^{39}$ erg/s like those seen for NGC 5907 ULX-1 \citep{israel_accreting_2017}, NGC 7793 P13 \citep{israel_discovery_2017} or M51 ULX-7 \citep{brightman_swift_2019} are not reflected in this diagram.
\subsection{Spectral transitions} \label{sub:spectral_transitions}
We also present the temporal evolution of each individual source in the HLD in Figure \ref{fig:individual_hl}, along with the unfolded spectra of some selected epochs. For this, we selected 2 to 4 clearly distinct spectral states based on the HLD, in order to highlight the possible range of spectral variability of each source. When possible, we selected observations for which \nustar\ data is available so the broadband variability can be observed. We caution that given the sparse monitoring offered by \xmm\ and \chandra\ in some cases, care must be taken when looking at the source variations in the HLD. Arrows indicate the chronological order but in many cases we cannot guarantee that the source did not evolve differently between the epochs considered here. 

\begin{figure*}
    \centering
\includegraphics[width=0.48\textwidth]{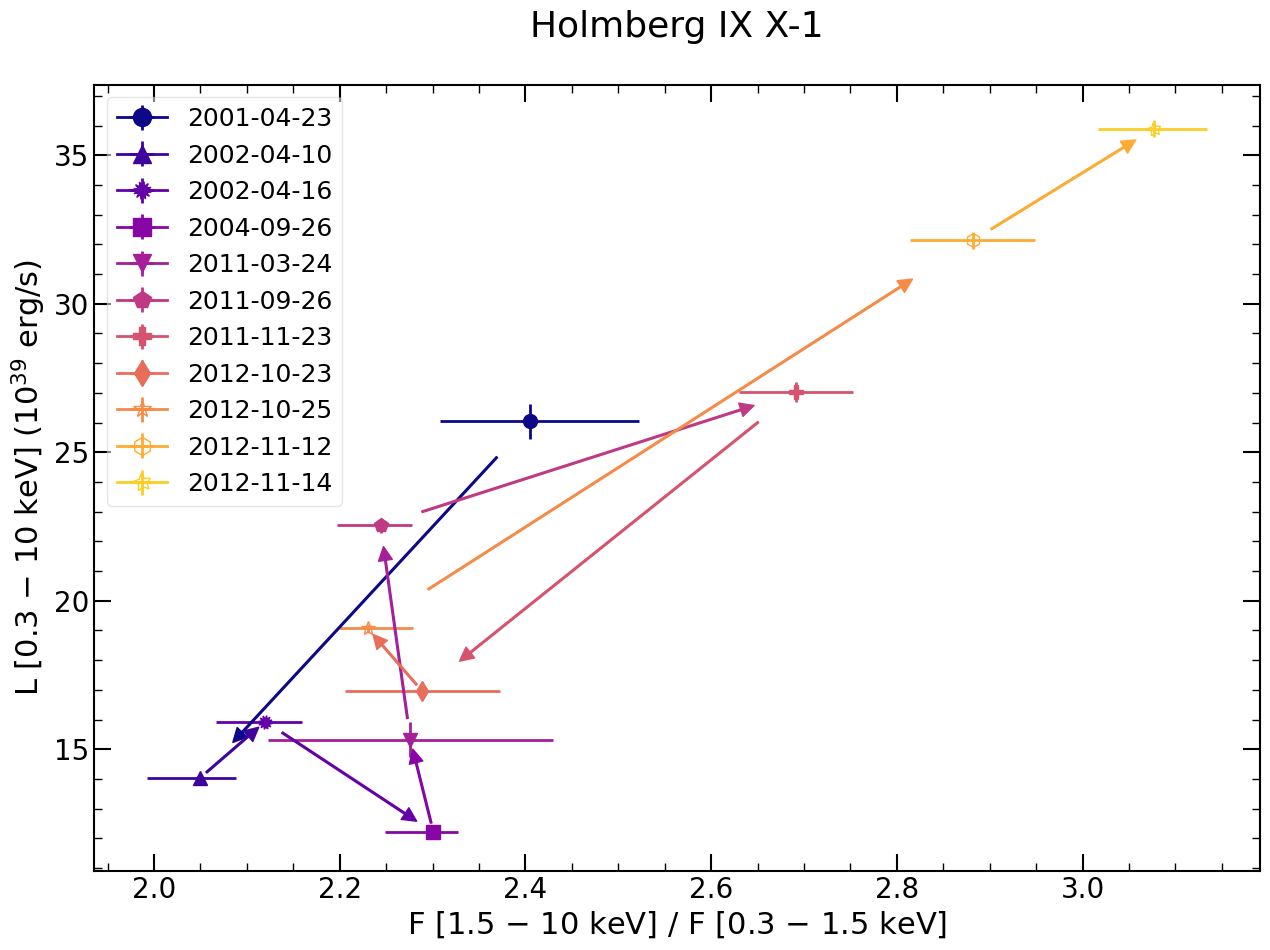}
\includegraphics[width=0.48\textwidth]{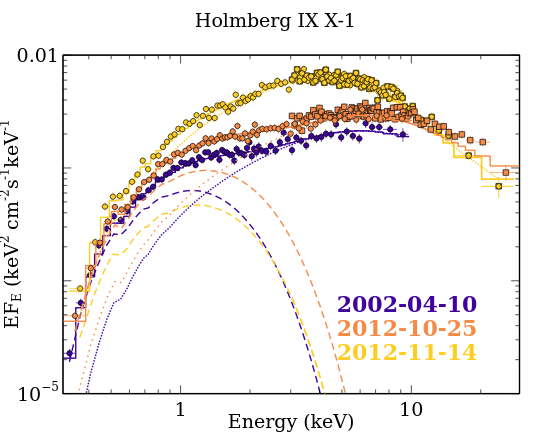}
\includegraphics[width=0.48\textwidth]{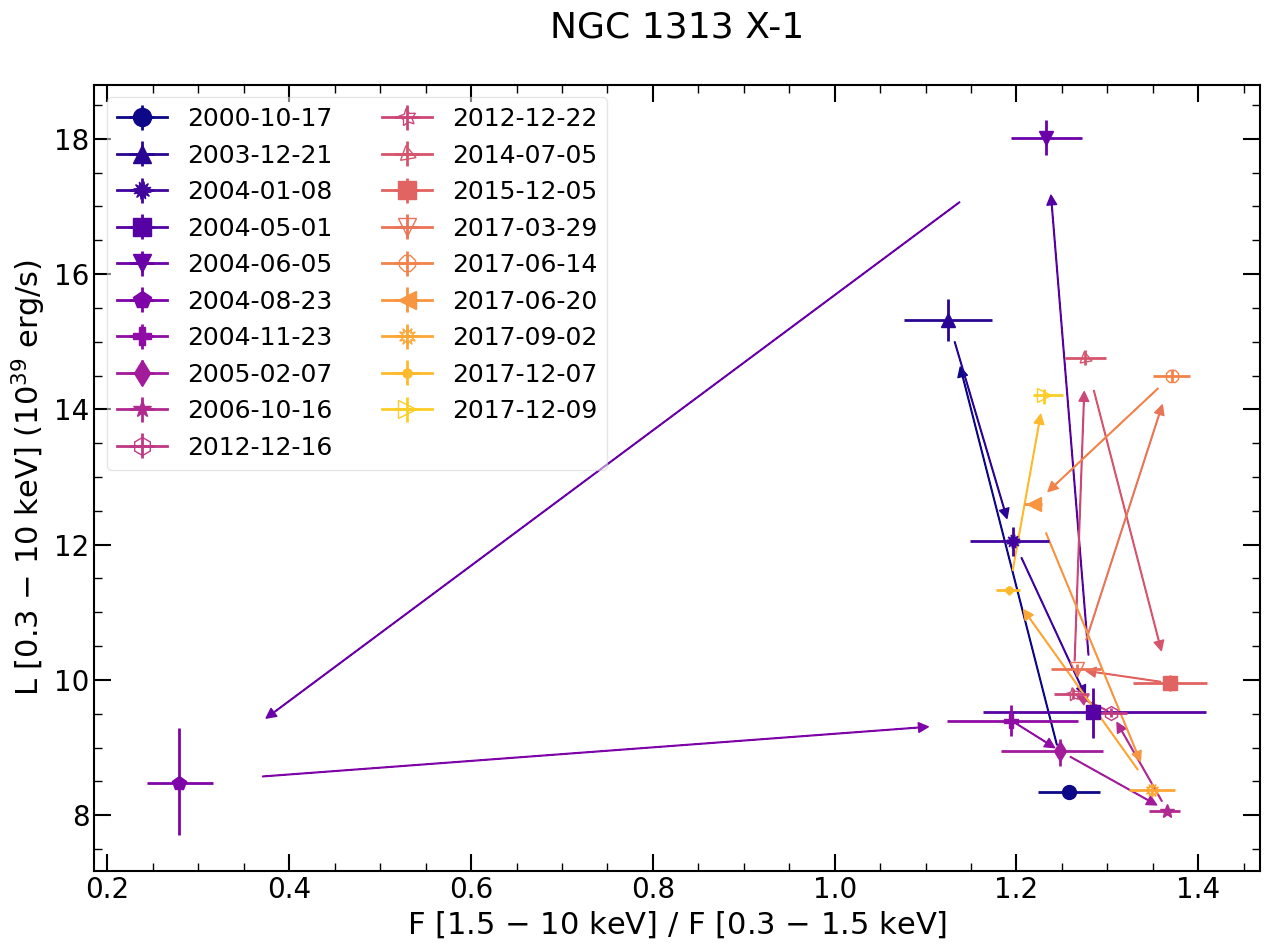}
\includegraphics[width=0.48\textwidth]{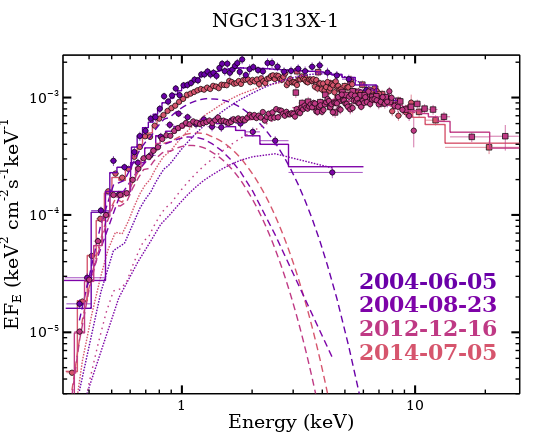}
    \includegraphics[width=0.48\textwidth]{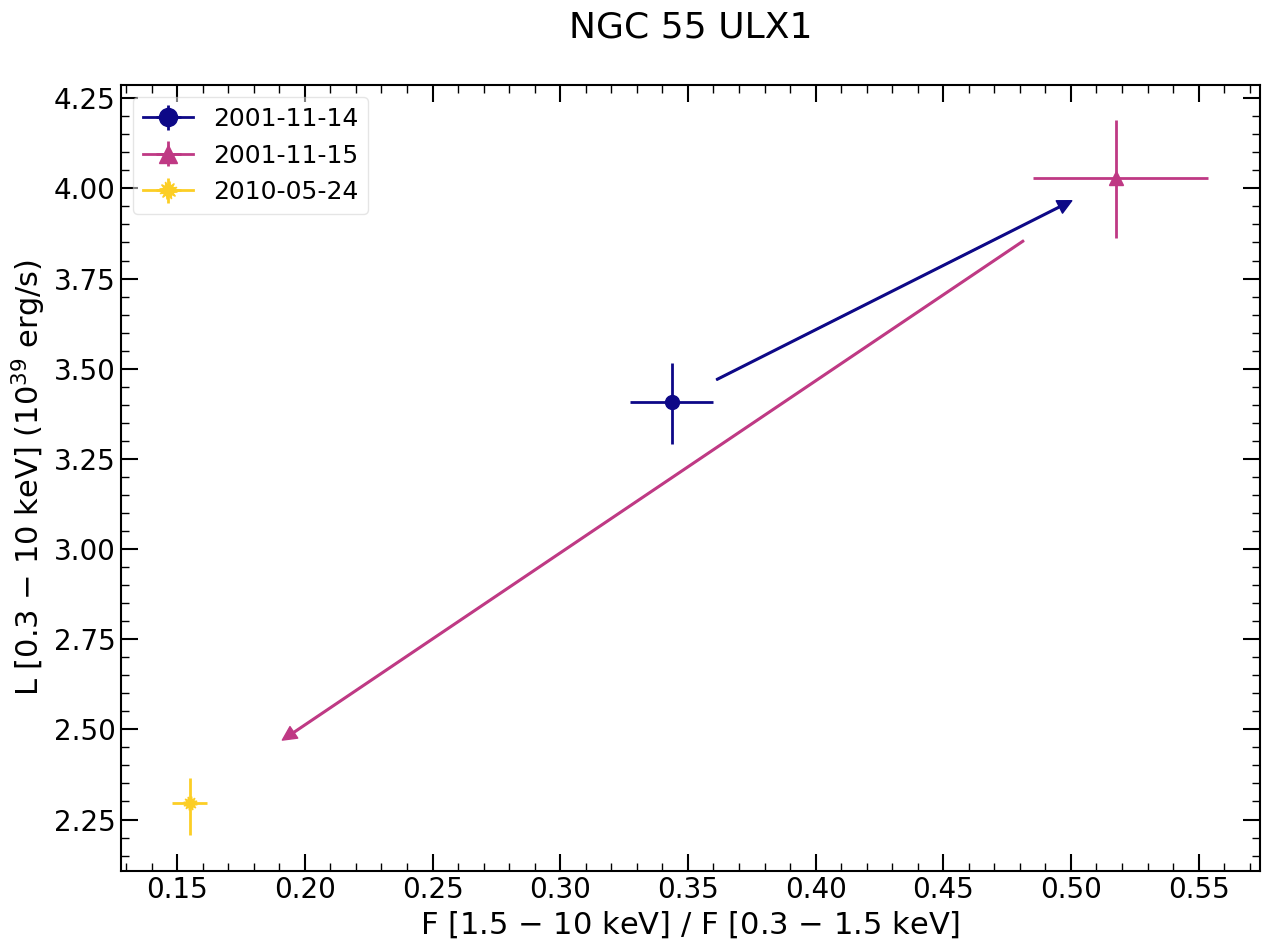}
\includegraphics[width=0.48\textwidth]{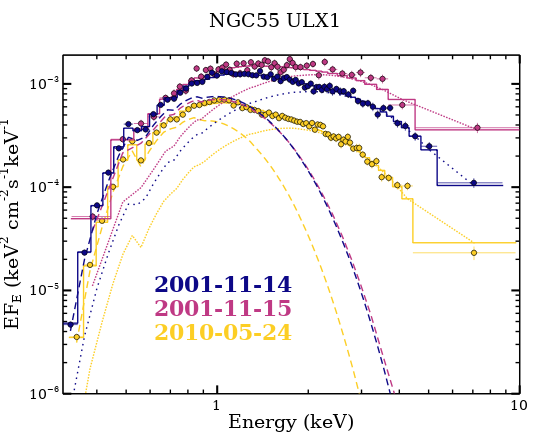}
 \caption{Left: Temporal tracks on the hardness-luminosity diagram. Filled coloured markers indicate fluxes obtained from \xmm\ data, unfilled markers indicate fluxes obtained from joint \xmm-\nustar\ data and black-filled markers indicate fluxes obtained from \chandra\ data. The legend indicates the date of the epoch in each case. Right: Unfolded spectra for selected epochs in which the source has experienced strong changes, following the same colour-code as for the left panels. Selected epochs are indicated in the legend. For the EPIC data, only pn is shown (circles) or MOS1 if pn is not available (as triangles up). In cases where we have used \nustar\ data, FPMA is shown, represented as squares and with the same colour as pn. \chandra\ data is shown with triangles down. The soft and hard \diskbb\ model components are shown with a dashed line and dotted line respectively, while the total model is shown with a solid line. For epochs where the \simpl\ model has been used see Table \ref{tab:individual_fits}. Data has been rebinned for clarity.} \label{fig:individual_hl}
 \end{figure*}
 \addtocounter{figure}{-1}
 \begin{figure*}
    \centering
    \includegraphics[width=0.48\textwidth]{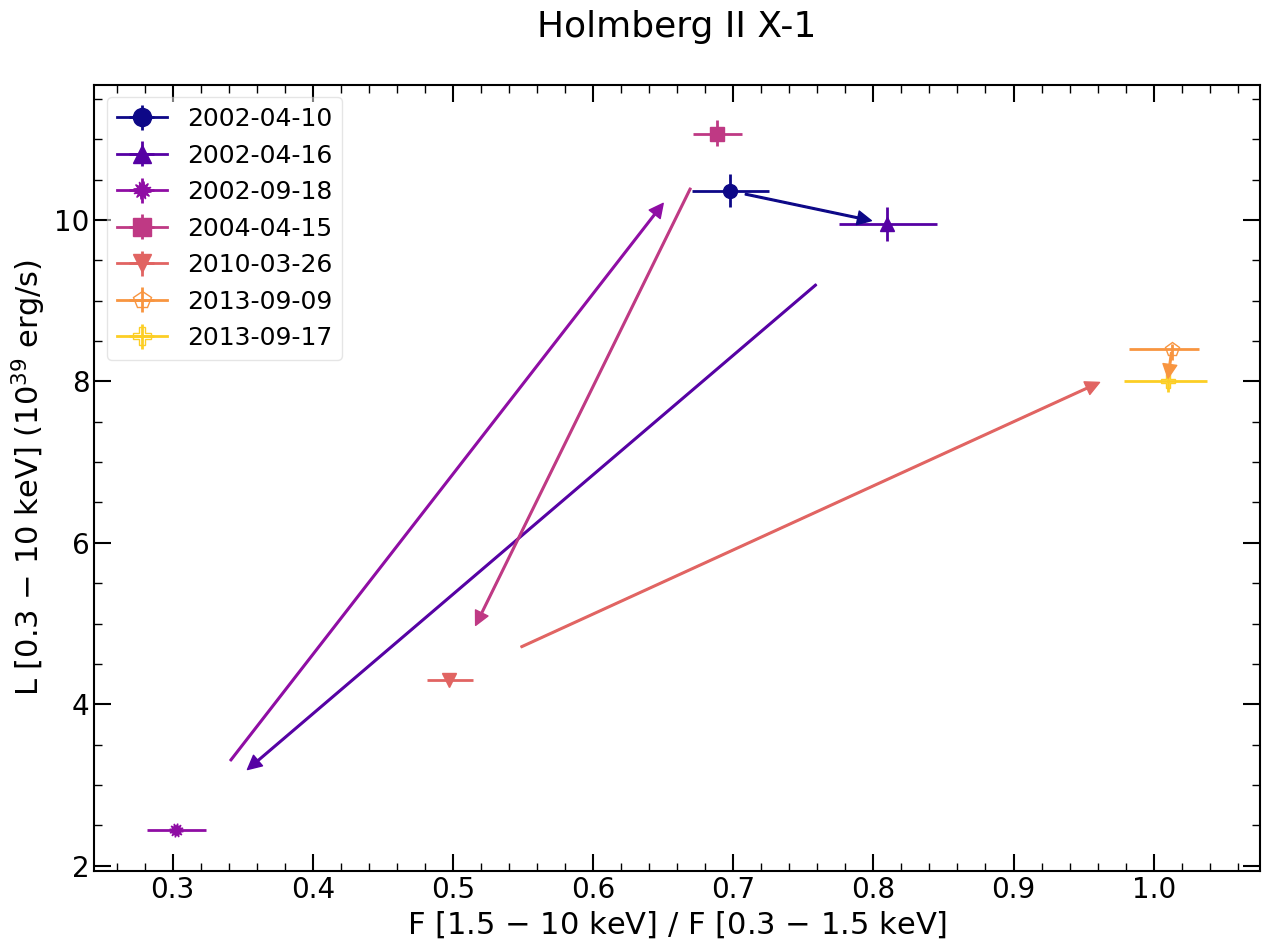}
\includegraphics[width=0.48\textwidth]{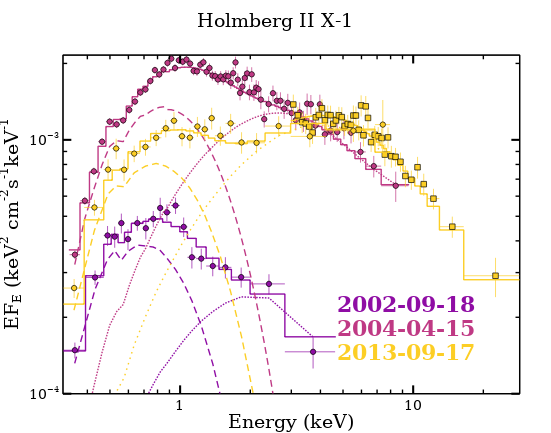}
   \includegraphics[width=0.48\textwidth]{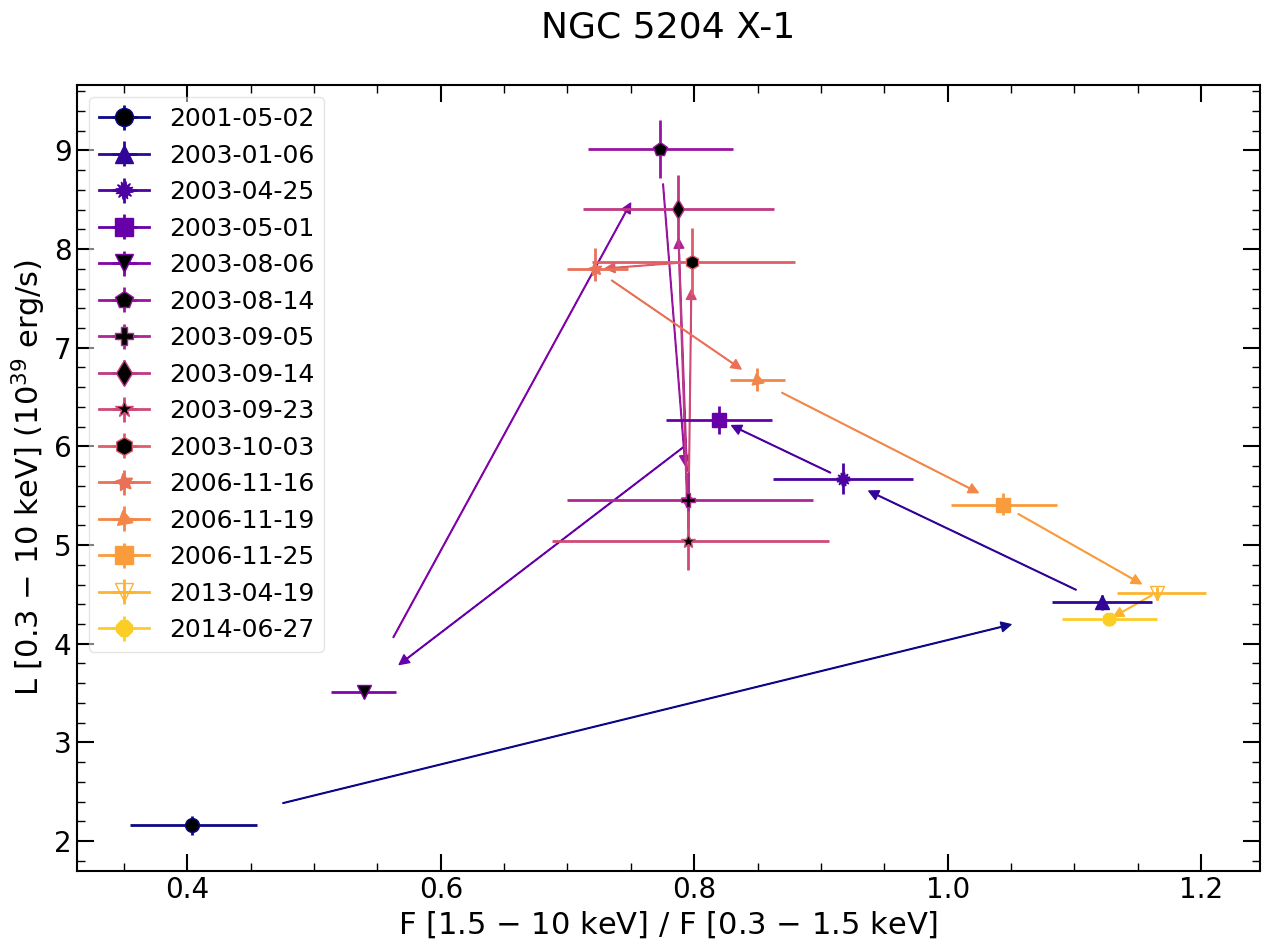}
\includegraphics[width=0.48\textwidth]{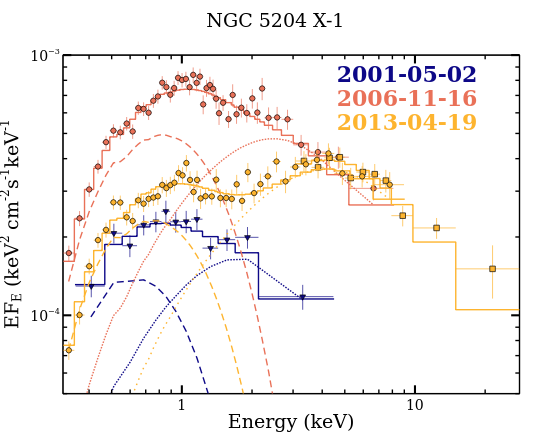}
    \caption{Continued}
\end{figure*}
 \addtocounter{figure}{-1}
 \begin{figure*}
    \centering
    \includegraphics[width=0.48\textwidth]{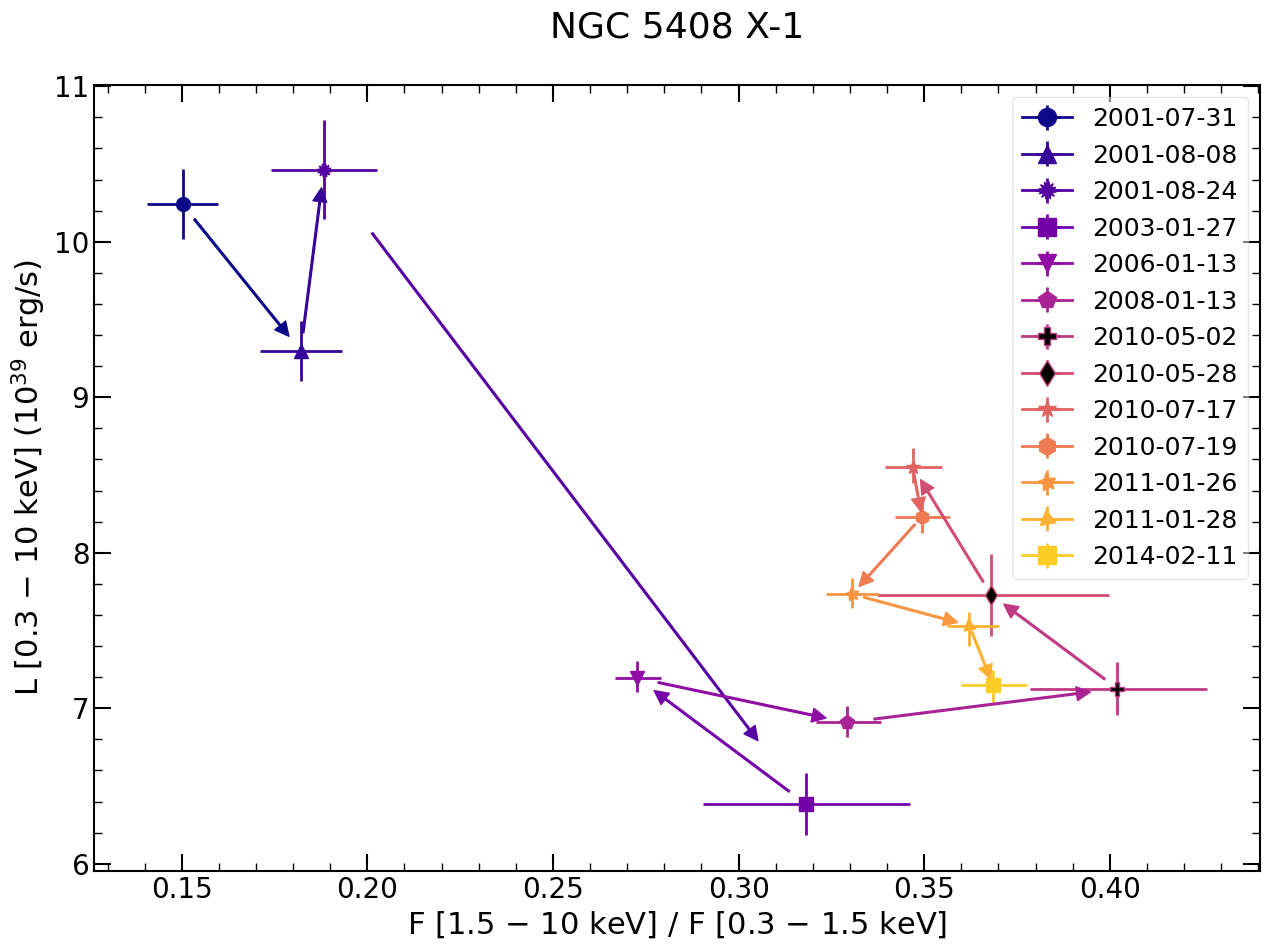}
\includegraphics[width=0.48\textwidth]{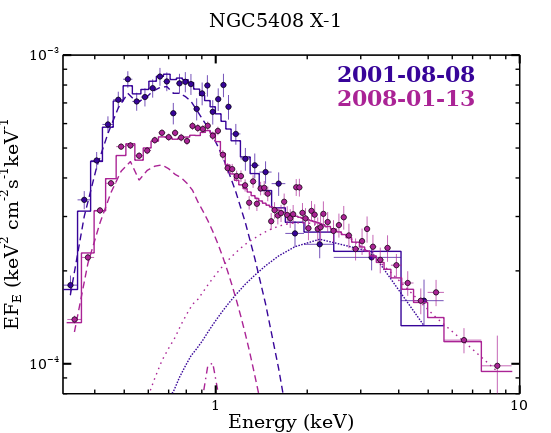}
\includegraphics[width=0.48\textwidth]{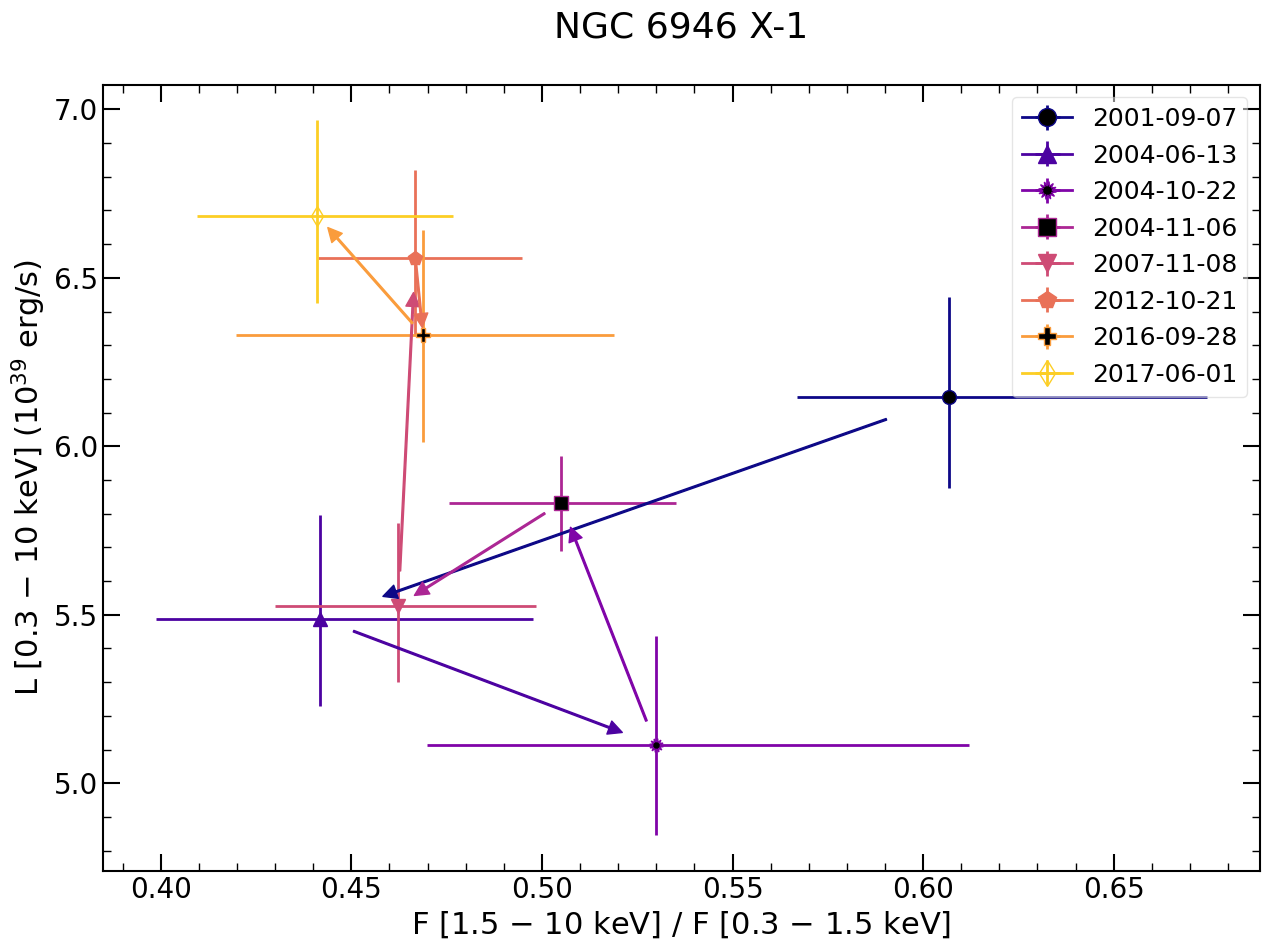}
\includegraphics[width=0.48\textwidth]{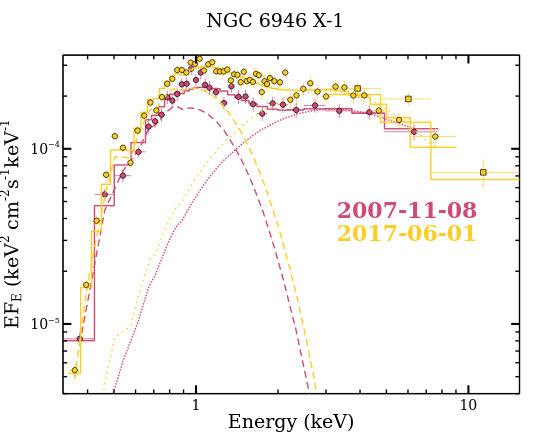}
    \includegraphics[width=0.48\textwidth]{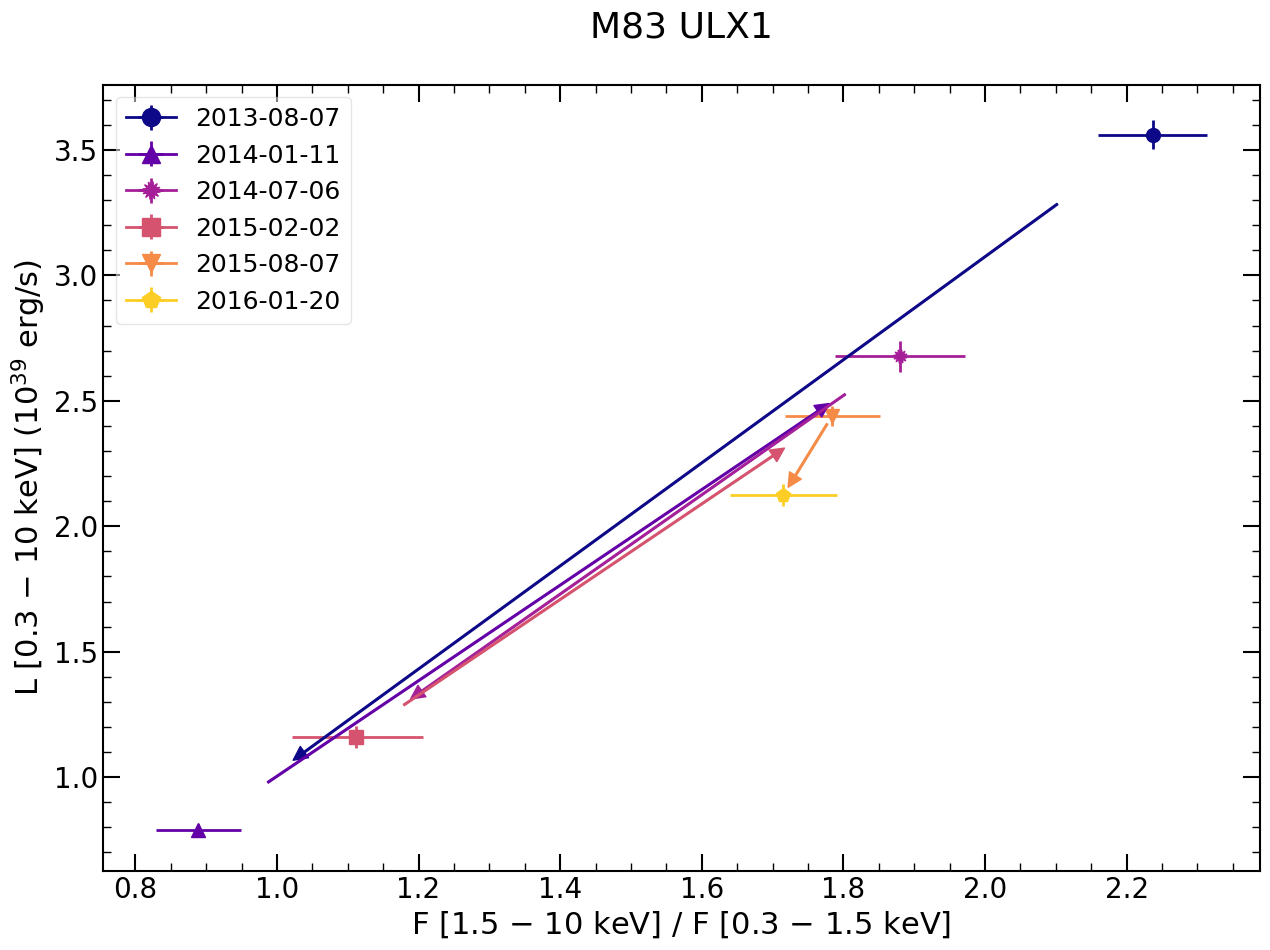}
\includegraphics[width=0.48\textwidth]{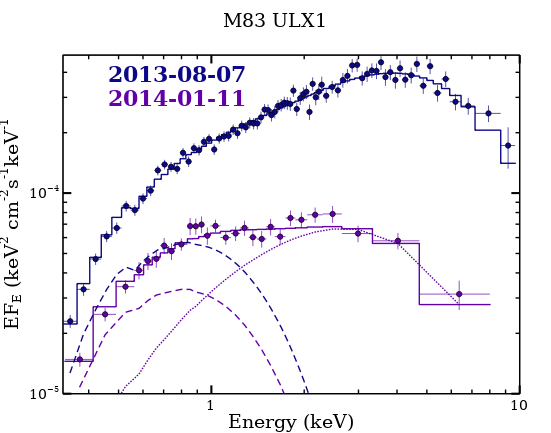}
    \caption{Continued}
\end{figure*}
 \addtocounter{figure}{-1}
 \begin{figure*}
    \centering
\includegraphics[width=0.48\textwidth]{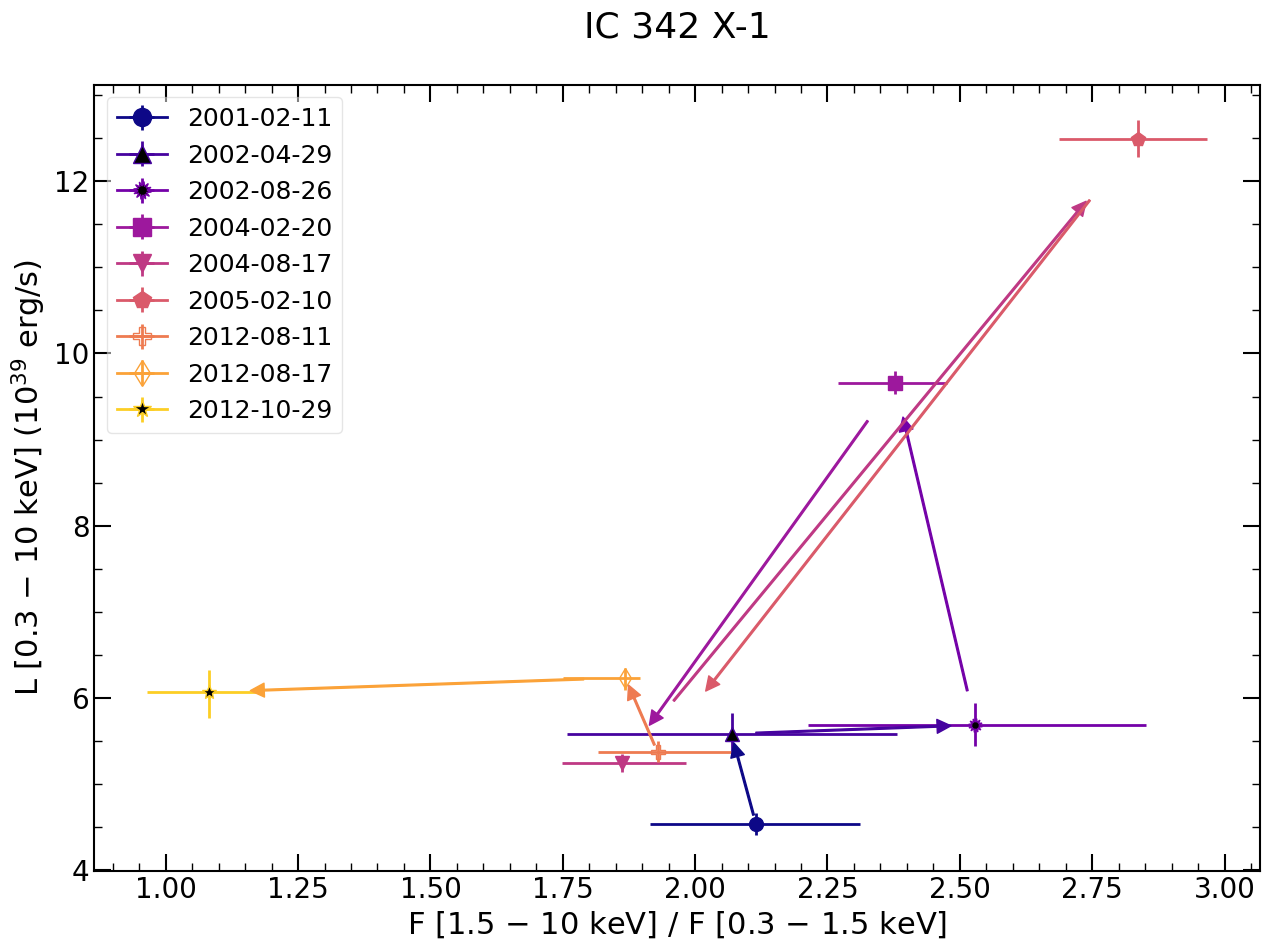}
\includegraphics[width=0.48\textwidth]{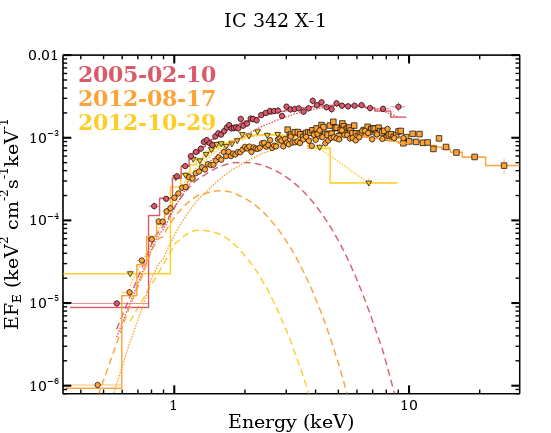}
\includegraphics[width=0.48\textwidth]{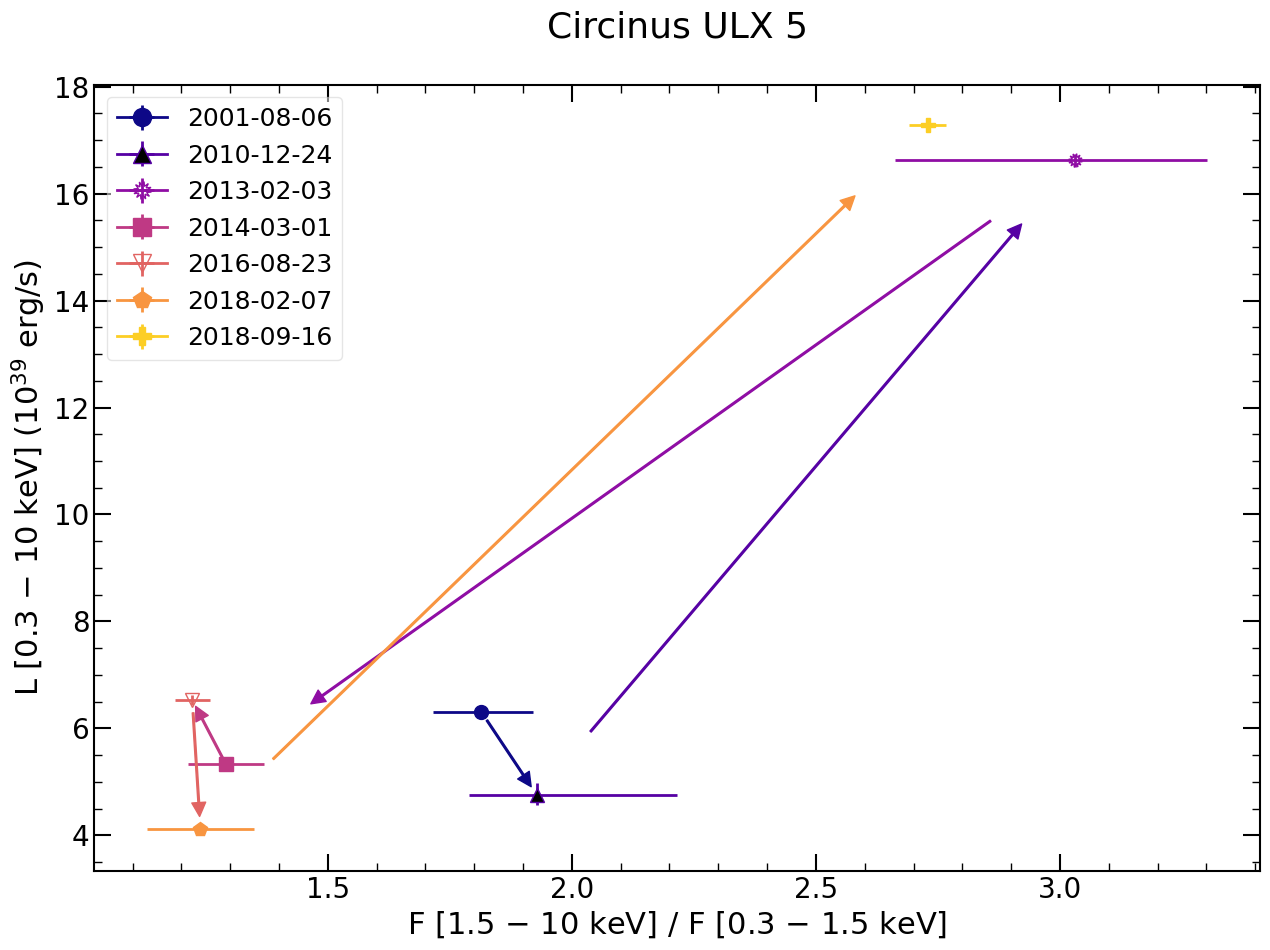}
\includegraphics[width=0.48\textwidth]{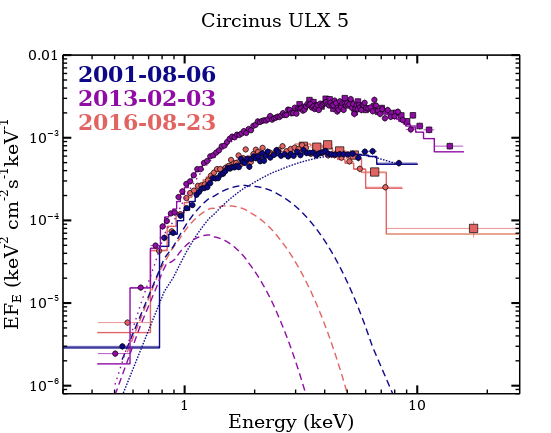}
  \caption{Continued}
\end{figure*}
 \addtocounter{figure}{-1}
 \begin{figure*}
    \centering
    \includegraphics[width=0.48\textwidth]{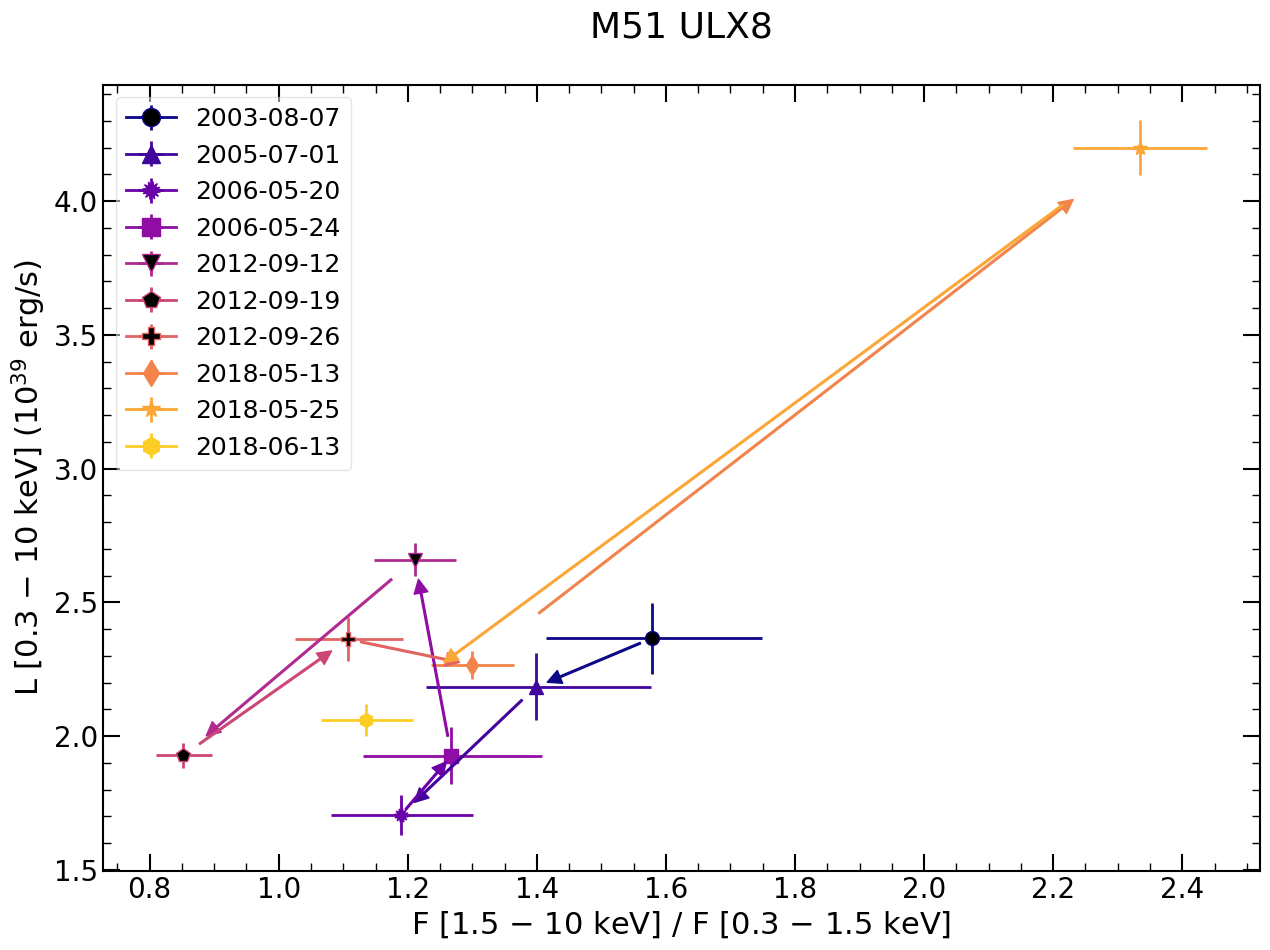}
\includegraphics[width=0.48\textwidth]{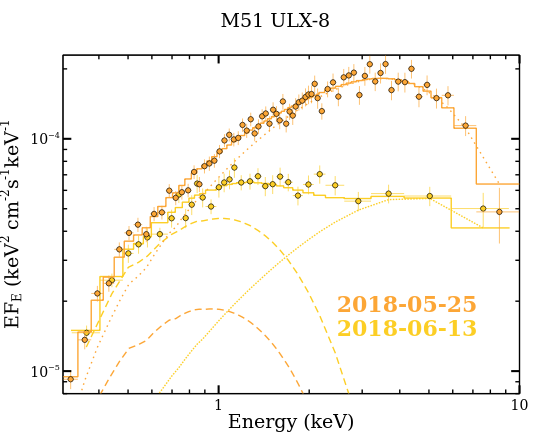}
    \includegraphics[width=0.48\textwidth]{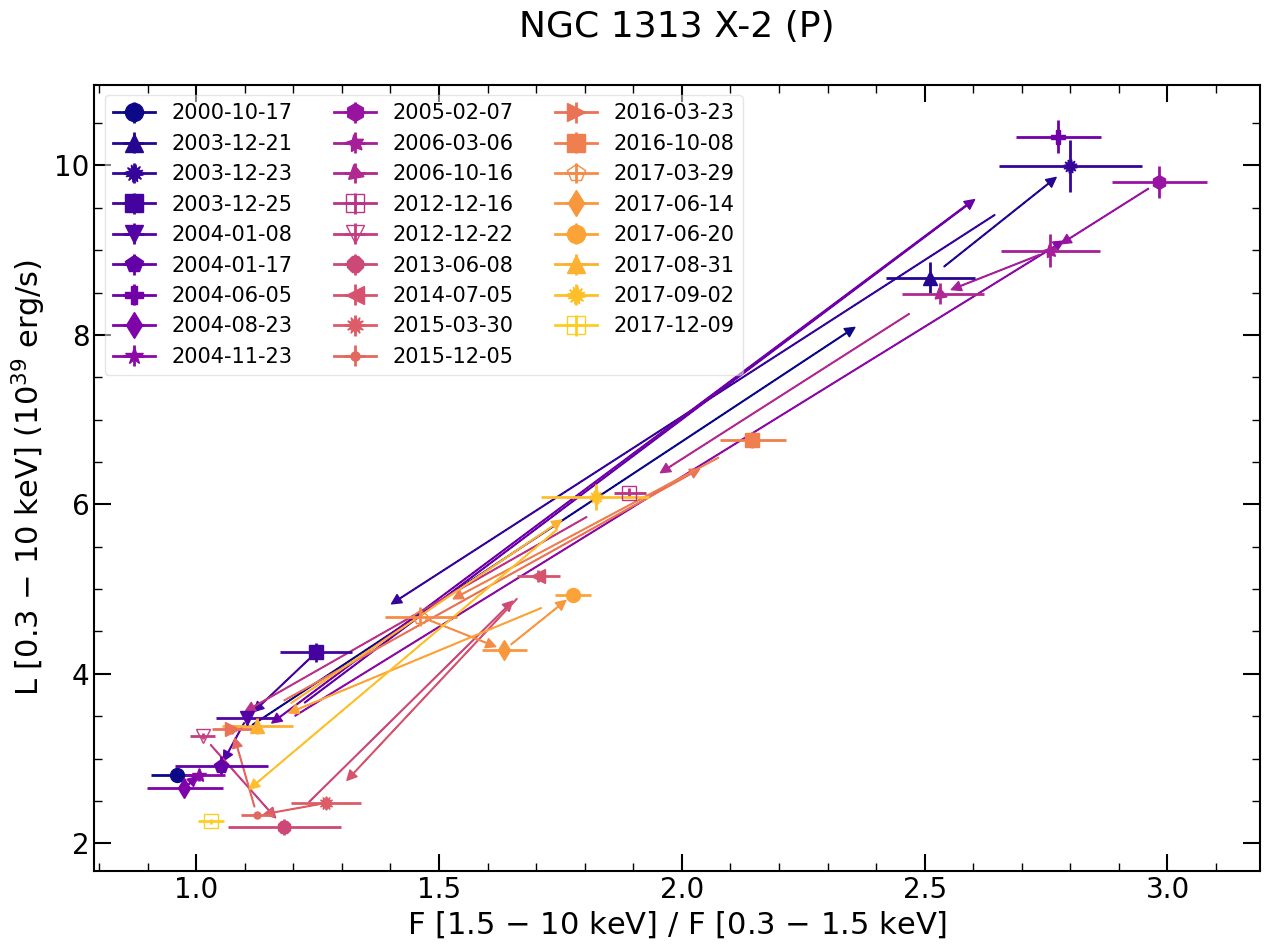}
\includegraphics[width=0.48\textwidth]{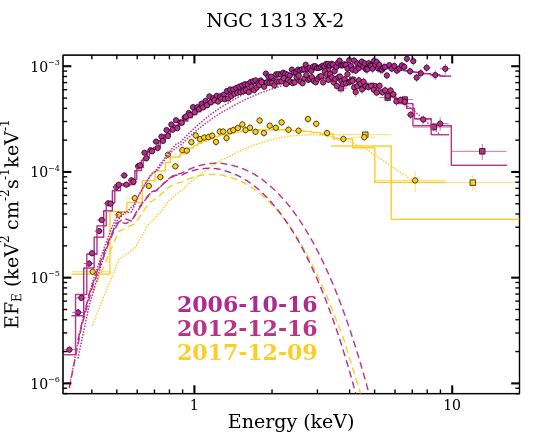}
    \includegraphics[width=0.48\textwidth]{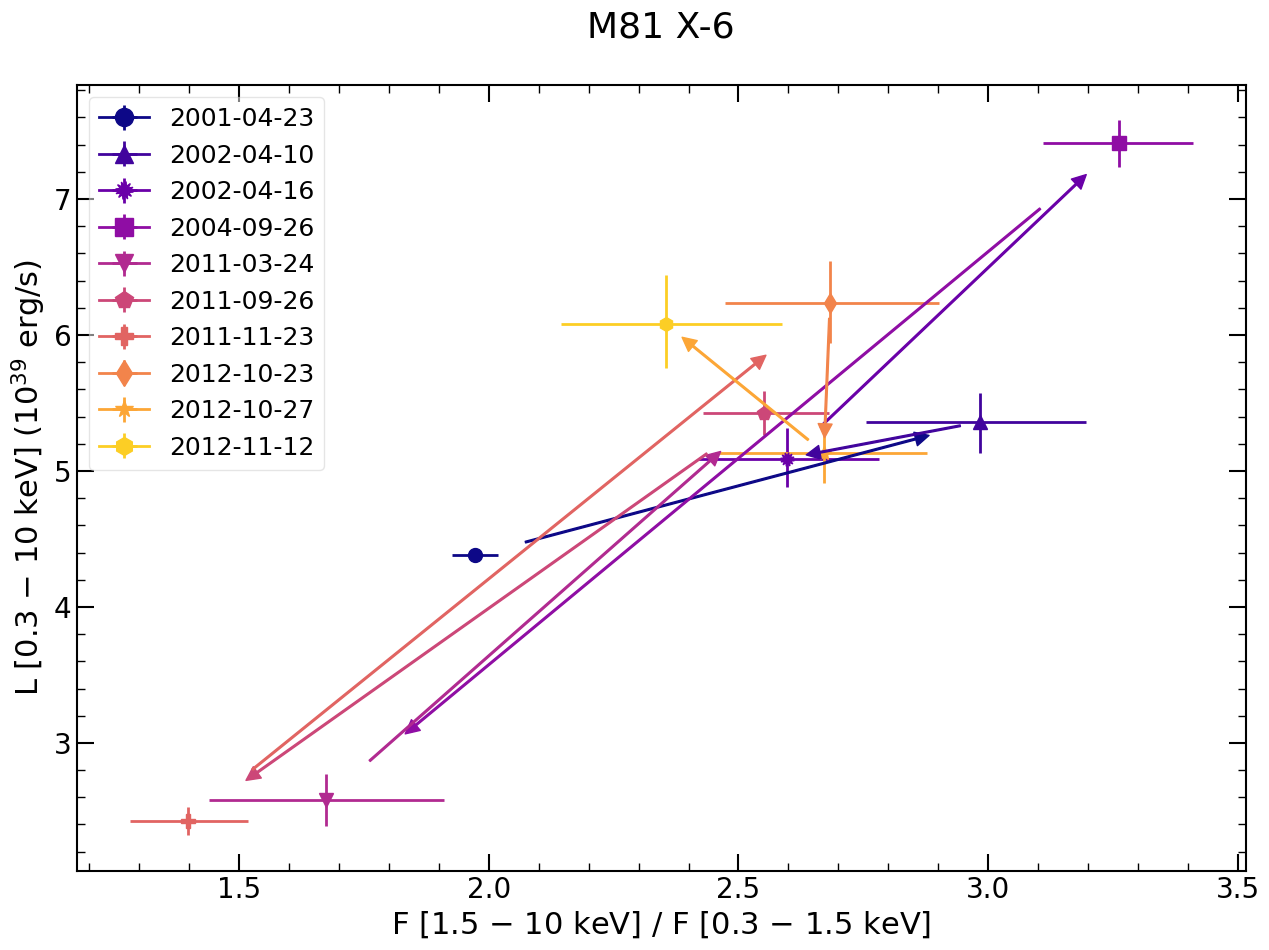}
\includegraphics[width=0.48\textwidth]{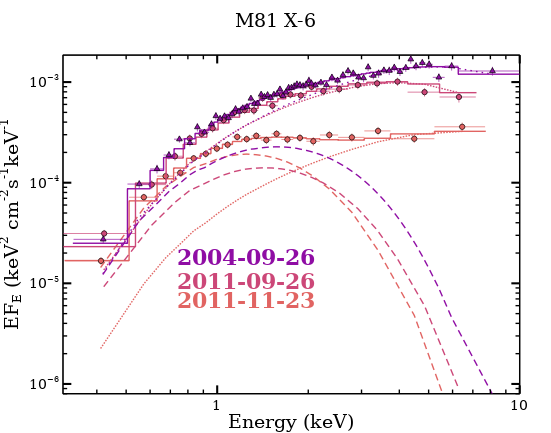}
    \caption{Continued}
\end{figure*}
 \addtocounter{figure}{-1}
 \begin{figure*}
    \centering
    \includegraphics[width=0.48\textwidth]{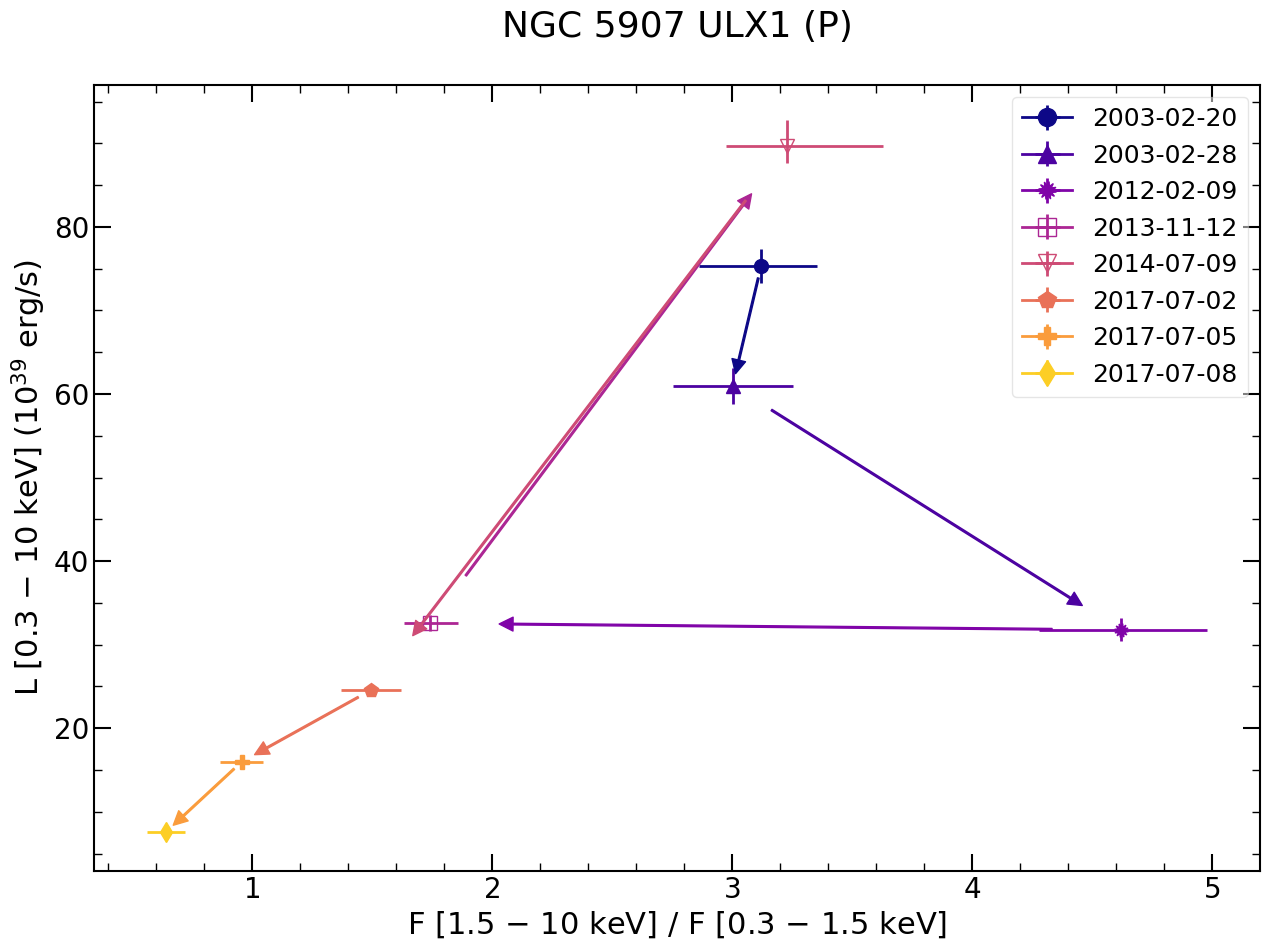}
\includegraphics[width=0.48\textwidth]{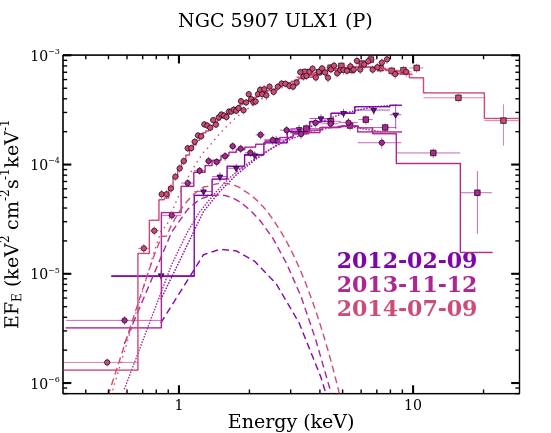}
\includegraphics[width=0.48\textwidth]{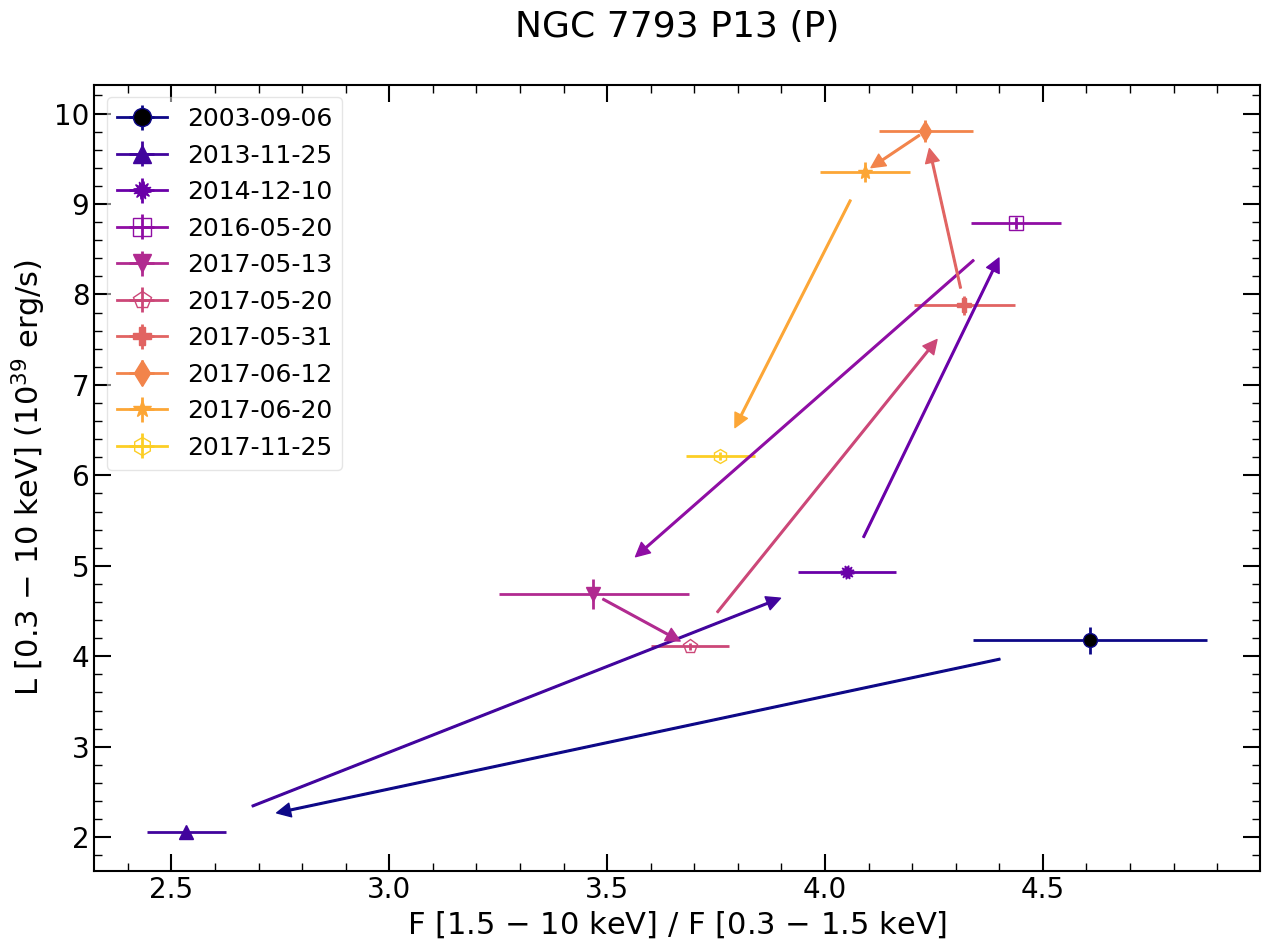}
\includegraphics[width=0.48\textwidth]{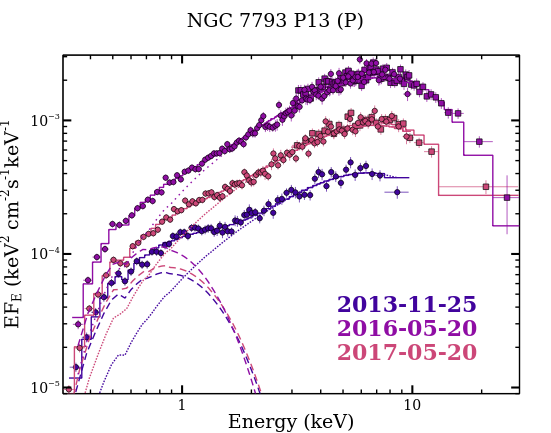}
    \includegraphics[width=0.48\textwidth]{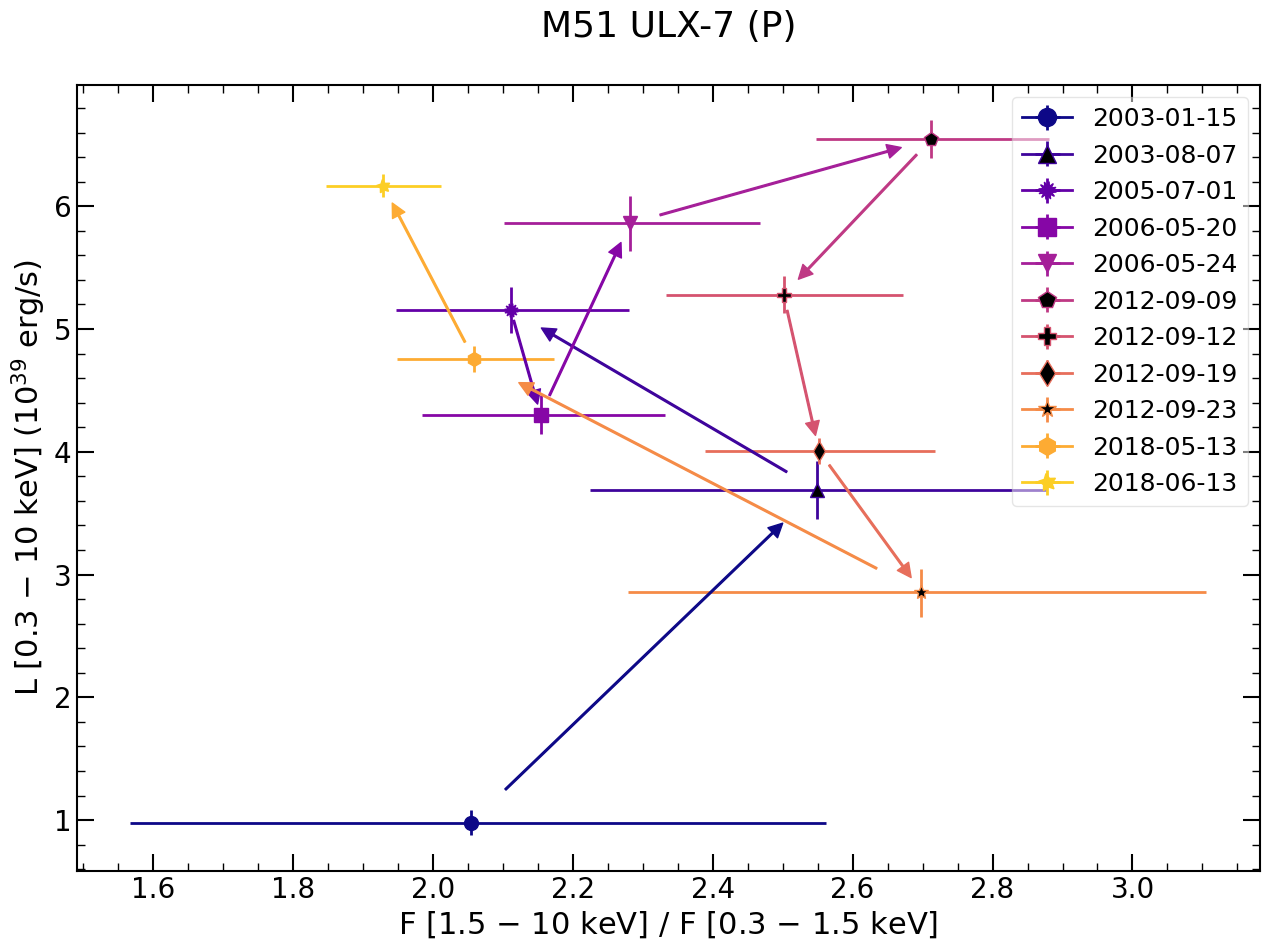}
\includegraphics[width=0.48\textwidth]{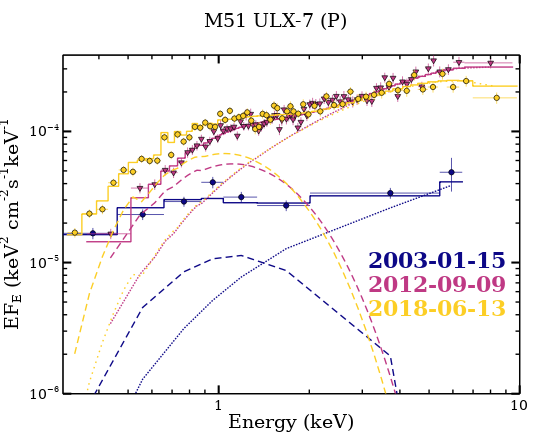}
    \caption{Continued}
\end{figure*}

\subsection{Spectral evolution of the soft thermal component} \label{sub:correlations_soft}
Several authors have attempted to study the nature of the soft component in ULX spectra by investigating its evolution on the luminosity--temperature (L--T) plane \citep[e.g][]{kajava_spectral_2009, miller_revisiting_2013}. However, these studies have frequently yielded contradictory results and thus there is still no consensus on its true nature.  We thus investigated the correlation of the bolometric luminosity of the cool \diskbb\ component with its temperature. We did this by retrieving unabsorbed luminosities of the \diskbb\ component in the 0.01 -- 100 keV\footnote{For the jointly fitted data, this flux calculation is done taking into account errors associated with all tied parameters.}. All fluxes are reported in Table \ref{tab:fluxes}. We did not attempt to derive any correlation for NGC 55 ULX1 and NGC 300 ULX1, due to the limited number of observations available for these sources. 

\begin{figure*}
 \centering
\includegraphics[width=0.49\textwidth]{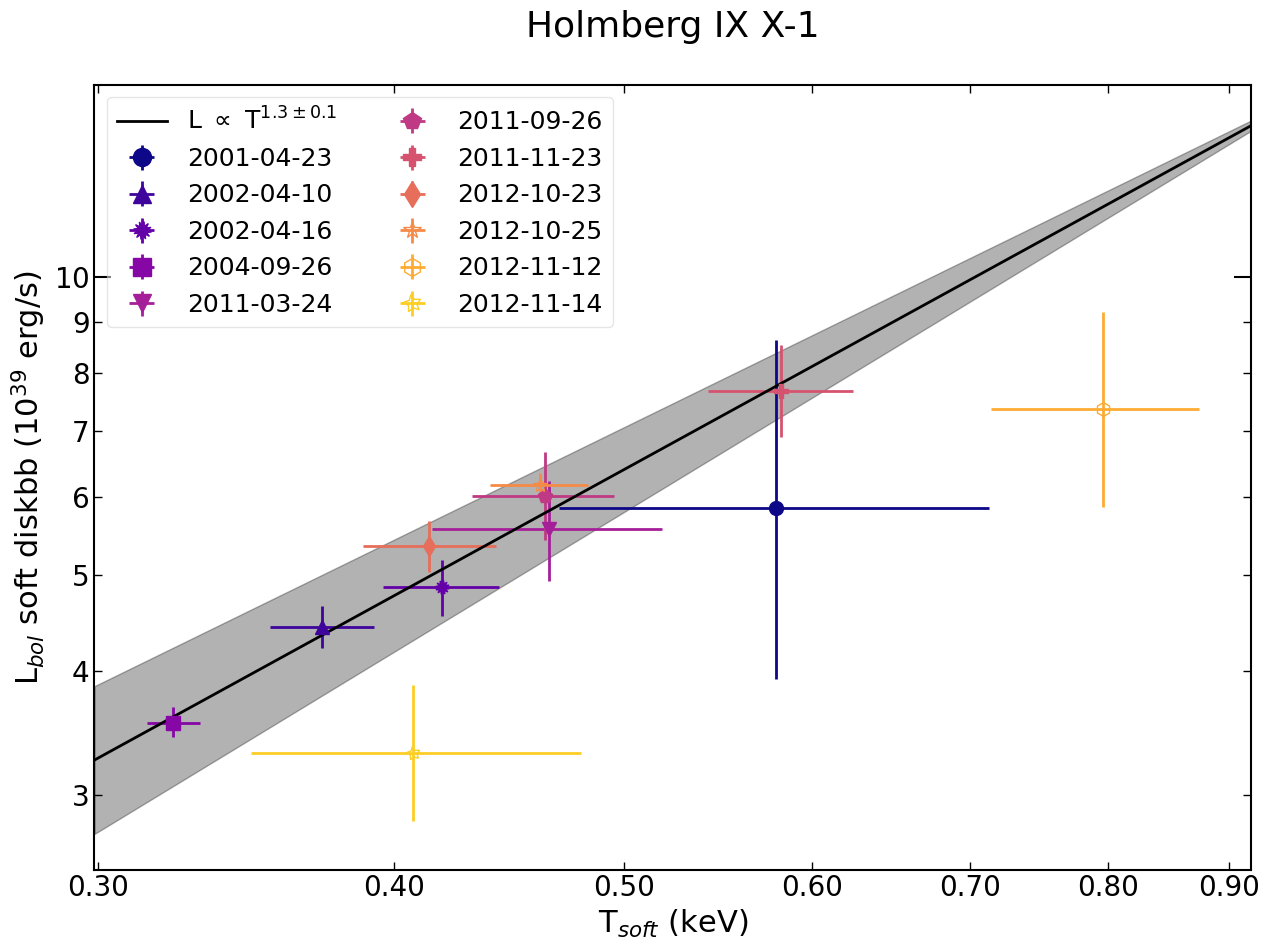} 
\includegraphics[width=0.49\textwidth]{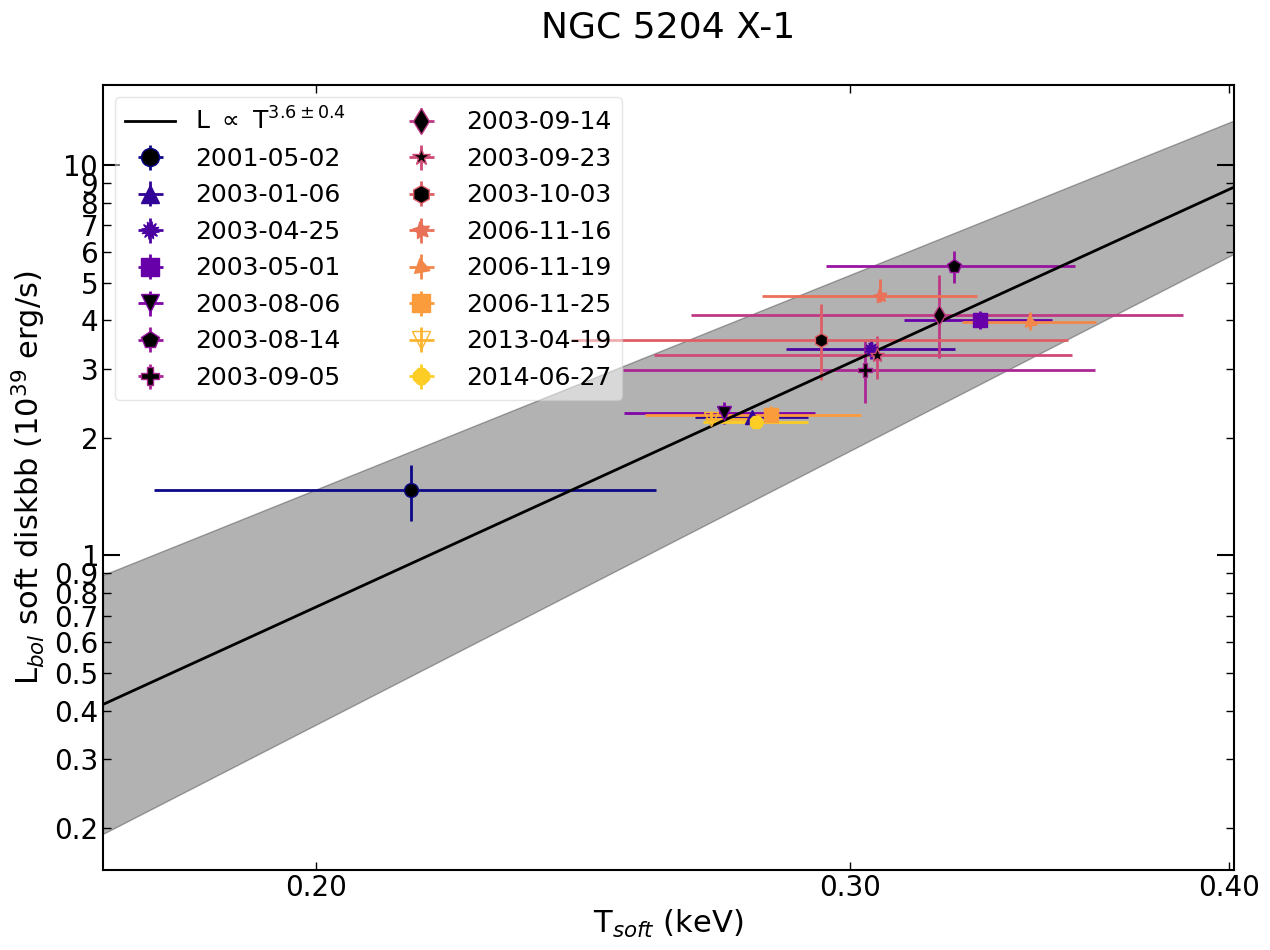} 
\includegraphics[width=0.49\textwidth]{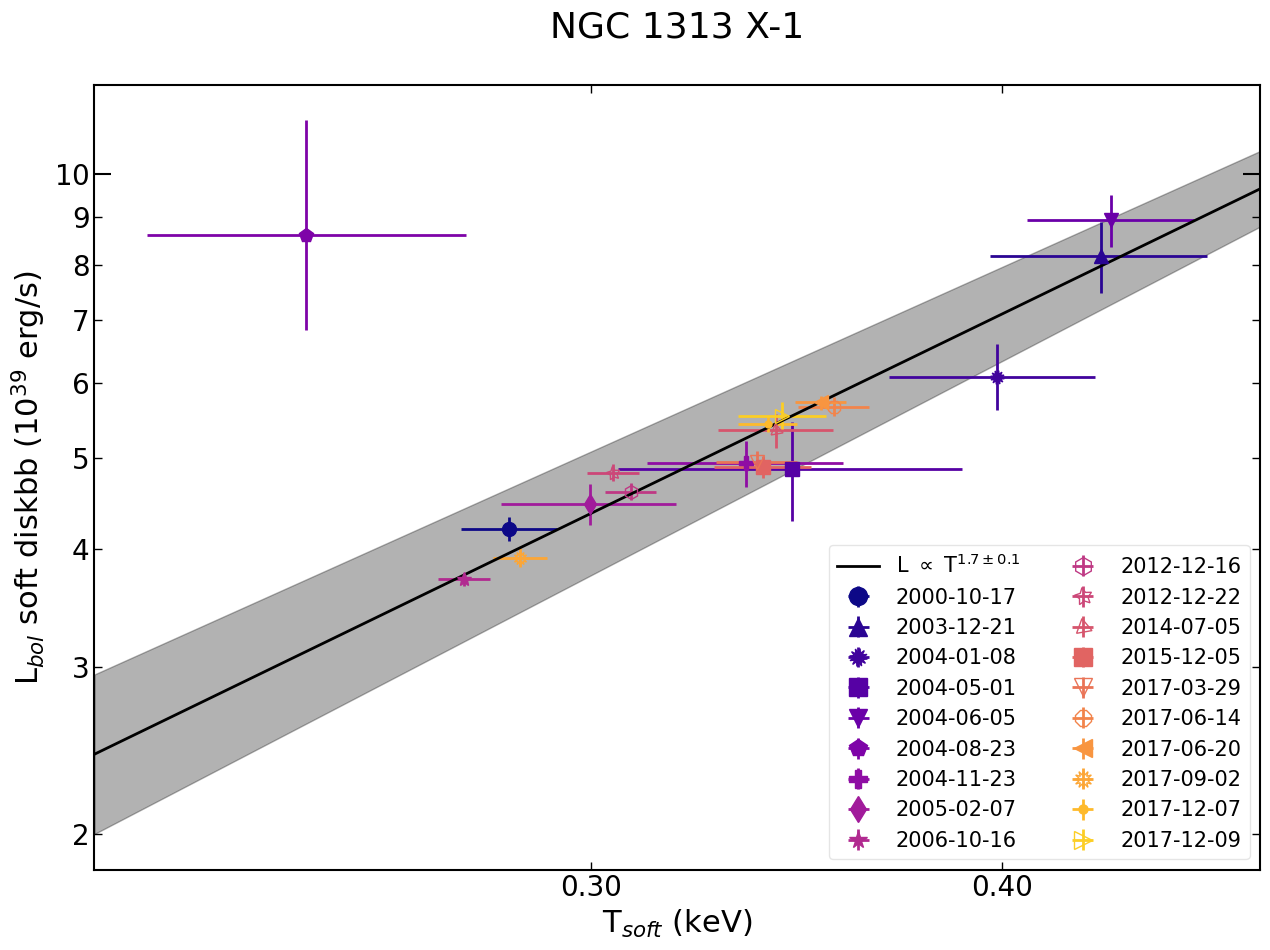}
\includegraphics[width=0.49\textwidth]{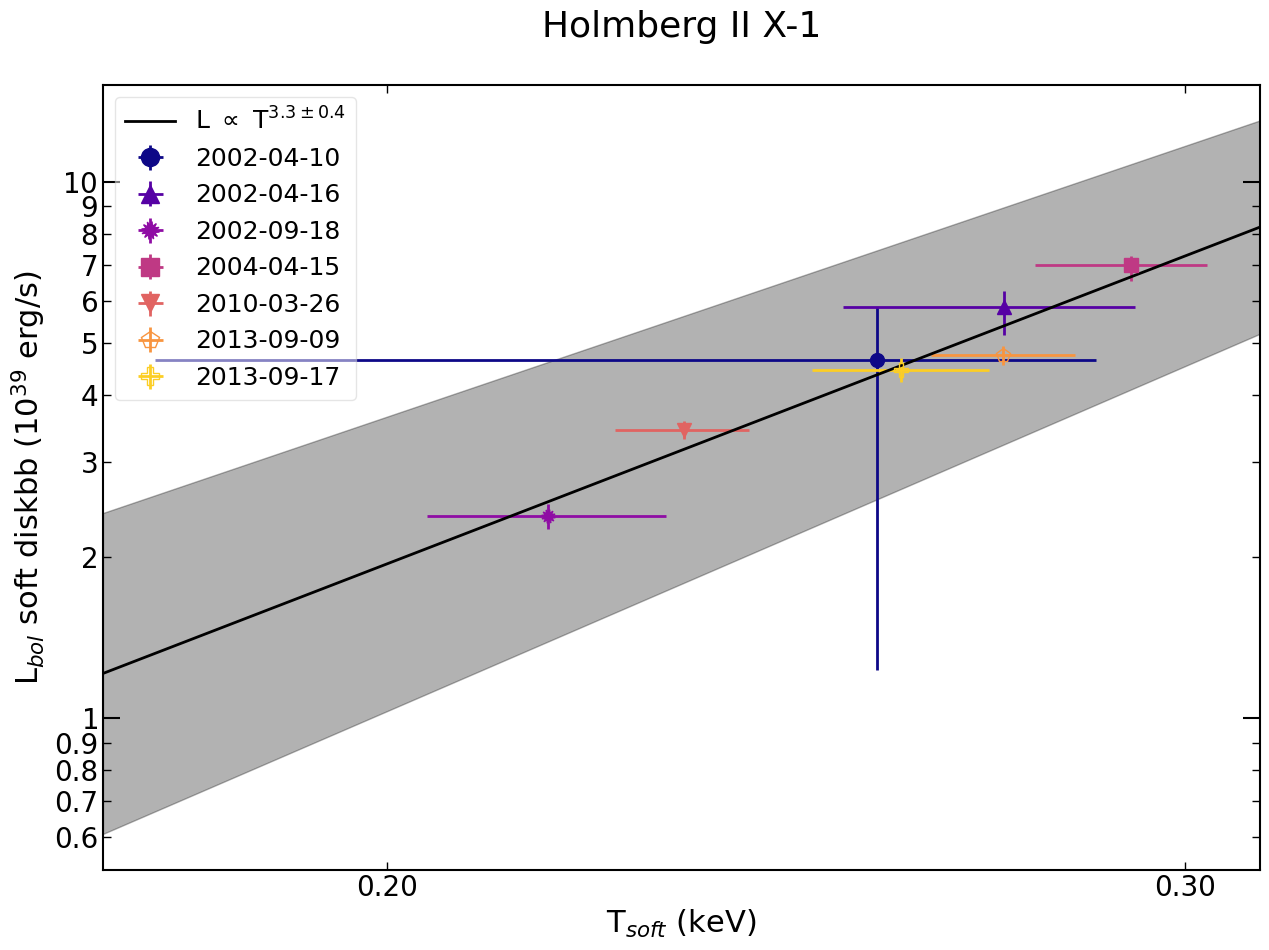}
 \caption{Examples of sources showing a positive correlation between the soft \diskbb\ component unabsorbed bolometric luminosity and its temperature. Black line shows the best fit power-law with grey shaded areas indicating the 90\% confidence interval on the exponent. Symbols are coloured as per Figure \ref{fig:individual_hl}. The data have been fitted with the model \tbabs $\otimes$ \tbabs $\otimes$ (\diskbb\ + \simpl\ $\otimes$\diskbb\ or \tbabs $\otimes$ \tbabs $\otimes$ (\diskbb\ + \diskbb\ ) (see text for details).} \label{fig:positive_sources}
\end{figure*}

\begin{figure*}
 \centering
\includegraphics[width=0.90\textwidth]{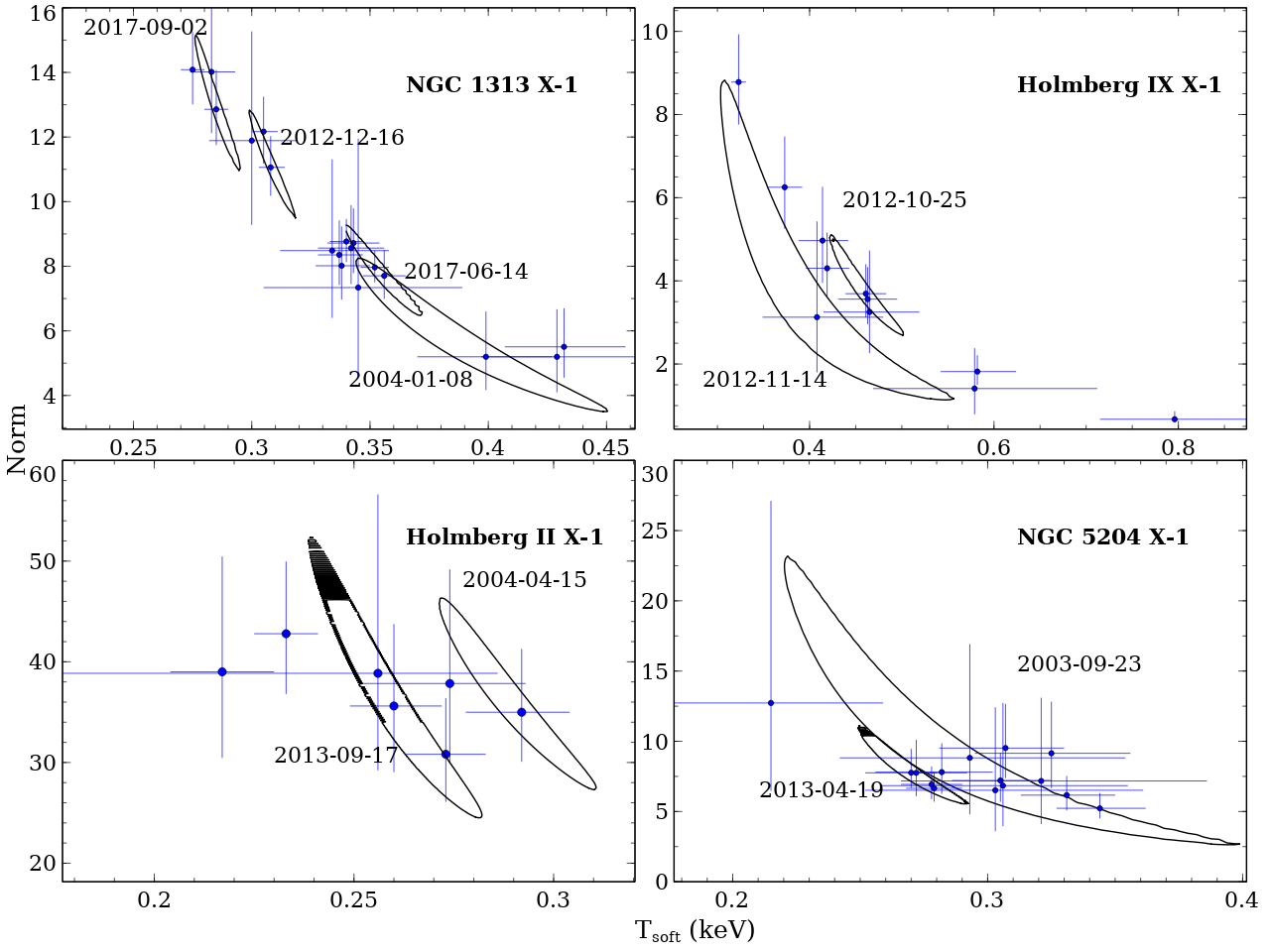} 
 \caption{99\% $\chi^2$ contours (solid black lines) between T$_\text{soft}$ and its normalisation for sources showing a positive L--T correlation. For each source, 2 to 4 different epochs are represented with the epoch indicated next to the contour. A clear anti-correlation is seen in all cases that closely follows the best fit values from all the epochs (blue datapoints, 90\% confidence level error bars), which indicates that the L--T positive correlation is due to the degeneracy between these two parameters. For the sake of readability, we have ignored epoch 2004-06-05 in the panel of NGC 1313 X-1 as it had a much higher normalisation ($\sim$ 51) compared to the other observations.} \label{fig:anti_correlation}
\end{figure*}

We next assessed whether L-T are correlated by running a Spearman correlation test\footnote{\url{https://docs.scipy.org/doc/scipy-0.14.0/reference/generated/scipy.stats.spearmanr.html}}. The results are reported in Table \ref{tab:correlation_softdiskbb}. Five of our sources show a strong positive L--T correlation (Spearman correlation $\geq$ 0.5 and a $p$-value of $\lesssim$ 0.05). For these, we fitted a power-law using the \texttt{python} routine \textit{odr}\footnote{https://docs.scipy.org/doc/scipy/reference/odr.html}, that takes into account both errors on x and y variables (see Figure \ref{fig:positive_sources} and Table \ref{tab:correlation_softdiskbb} for the results). In order to investigate whether these correlations were driven by the degeneracy between T$_\text{soft}$ and its normalisation, we derived 99\% $\chi^2$ confidence contours around the best-fit T$_\text{soft}$ and its normalisation for those sources showing a positive L--T correlation. Given the extensive computational time required by the \texttt{steppar} command in \xspec, we did this only for some selected epochs, ensuring that at least one was from the joint fit. The results are shown in Figure \ref{fig:anti_correlation} for NGC 1313 X-1, Holmberg IX X-1, Holmberg II X-1 and NGC 5204 X-1, where we also overlaid the best-fit T$_\text{soft}$ and normalisation from all epochs. In \textit{all} cases where a positive L--T correlation is observed, including IC 342 X-1 that we did not show for brevity, a strong degeneracy between the $T_\text{soft}$ and its normalisation is observed, highly correlated with the datapoints from the individual observations, indicating that the L--T correlations are simply due to a degeneracy between these two parameters. 

\begin{figure*}
 \centering
\includegraphics[width=0.49\textwidth]{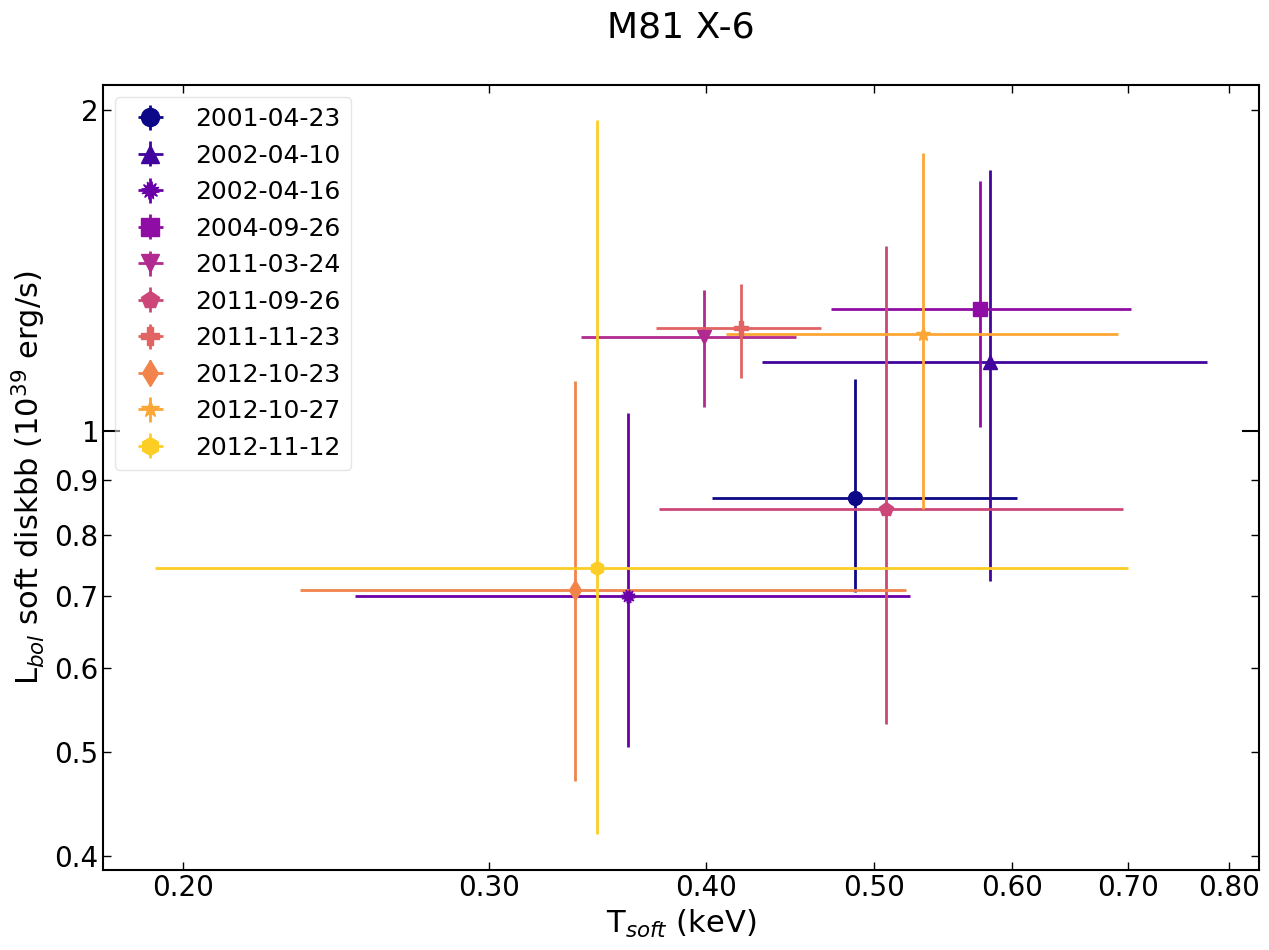} 
\includegraphics[width=0.49\textwidth]{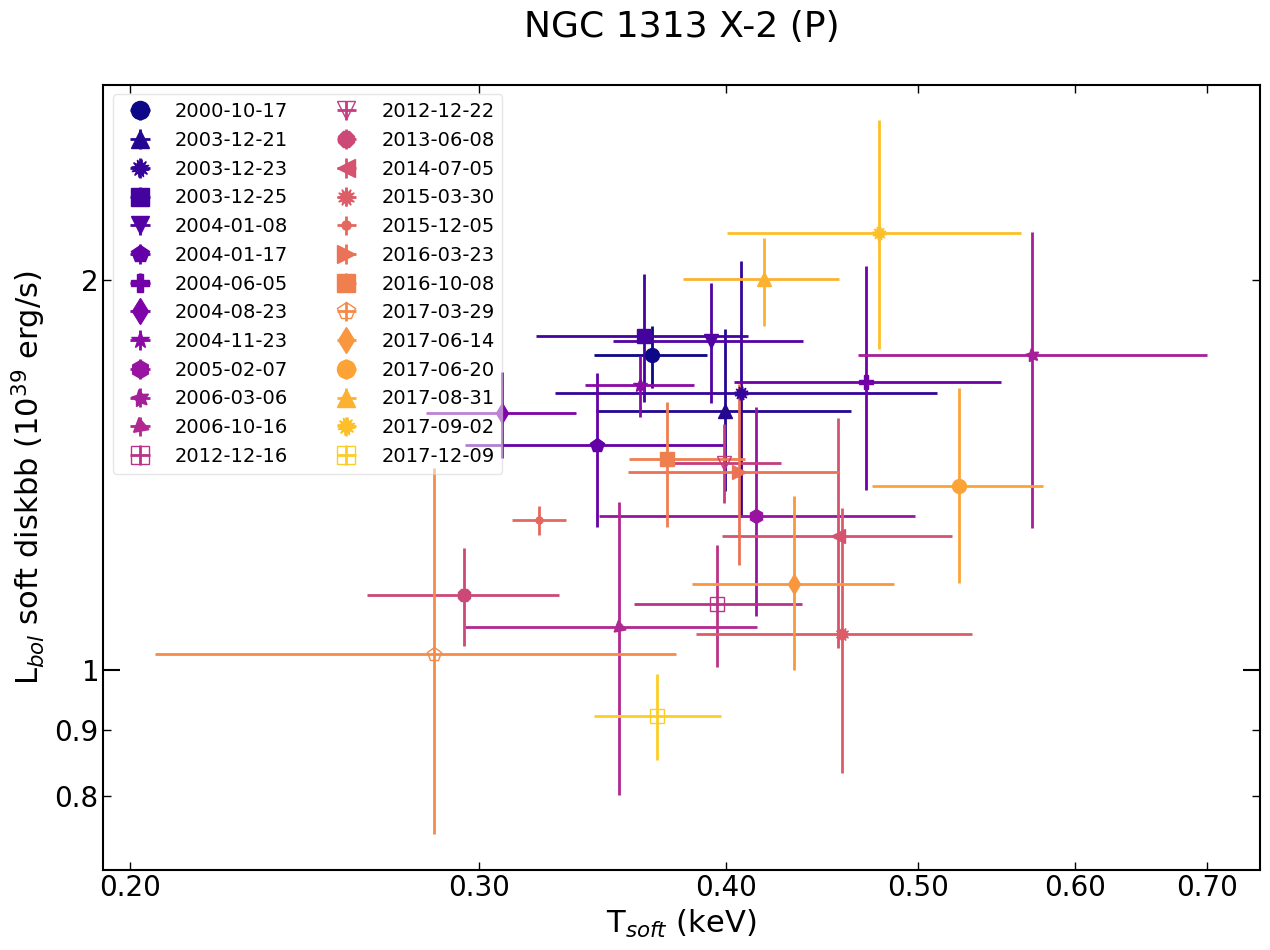} 
 \caption{As for Figure \ref{fig:positive_sources} but for sources showing no correlation between the soft \diskbb\ bolometric luminosity and its temperature. } \label{fig:uncorrelated_sources}
\end{figure*}

 For M81 X-6, while the correlation test could indicate a positive correlation, visual inspection of the data clearly reveals that there is no apparent trend (see Figure \ref{fig:uncorrelated_sources}). We recall that the Spearman's test coefficient does not take into account errors on the parameters and hence these values need to be treated with caution. In fact, the fit with the powerlaw yields a large error on the index $\alpha$ = 2.5 $\pm$ 2.0 (90\% confidence level), while the normalisation is consistent with 0.

\begin{table} 
 \centering 
 \caption{Results of the L-T correlation for the soft thermal component. We quote the Spearman's rank correlation coefficient, the false alarm probability ($p$-value) and the index ($\alpha$) of the L$\propto$ $T^\alpha$ relationship with its 90\% confidence level uncertainty.} 
 \label{tab:correlation_softdiskbb}
 \begin{tabular}{lccc} 
 \hline
 \noalign{\smallskip}
Source & Spearman & $p$-value & $\alpha$  \\ 
\noalign{\smallskip}
\hline 
NGC 7793 P13 &--0.13	&0.73 & -\tablefootmark{a}  \\ 
NGC 5907 ULX1 & 0.57 & 0.14 & -\tablefootmark{a} \\
NGC 5408 X-1 & --0.14 &0.65 & -\tablefootmark{a}  \\ 
Circinus ULX5 &0.32& 0.5 & -\tablefootmark{a}   \\ 
HoIX X-1 & 0.87 & 0.0005 &  1.31 $\pm$ 0.14 \\ 
M81 X-6 & 0.65 &0.04 &  -\tablefootmark{b}  \\ 
IC342 X-1 & 0.67 & 0.05 & 1.27$\pm$0.28 \\ 
NGC 1313 X-1 & 0.68 & 0.001 &  1.69 $\pm$ 0.13 \\
NGC 5204 X-1 &0.80	&0.00001 &  3.55 $\pm$ 0.43 \\ 
Ho II X-1 &0.96	&0.0005 &  3.26$\pm$ 0.39  \\ 
NGC 1313 X-2 & 0.25 & 0.21 &  -\tablefootmark{a} \\
NGC 6946 X-1 & 0.45 & 0.26 &  -\tablefootmark{a} \\
M83 ULX1 & 0.71	&0.11 & -\tablefootmark{a} \\ 
M51 ULX-7 & --0.22 &0.52 &  -\tablefootmark{a}\\ 
M51 ULX-8 & 0.54 &0.108 &  -\tablefootmark{a}\\ 
\hline 
 \end{tabular} 
 \tablefoot{
 \tablefoottext{a}{Non-correlated source.}
 \tablefoottext{b}{Not considered correlated due to uncertainties.}}
 \end{table}
\subsection{Spectral evolution of the hard thermal component} \label{sub:correlations_hard}
Similarly, we studied the hard thermal component evolution in the L--T plane. In cases where we have employed the \simpl\ model on this component, we retrieved the \textit{intrinsic} flux of the hard \diskbb\ by using \cflux\ as \simpl$\otimes$\cflux$\otimes$\diskbb\ and freezing the normalisation of the \diskbb. Derived fluxes are presented in Table \ref{tab:fluxes} and the results from the L--T correlation in Table \ref{tab:hard_correlations}.

\begin{figure*}
    \centering
    \includegraphics[width=0.49\textwidth]{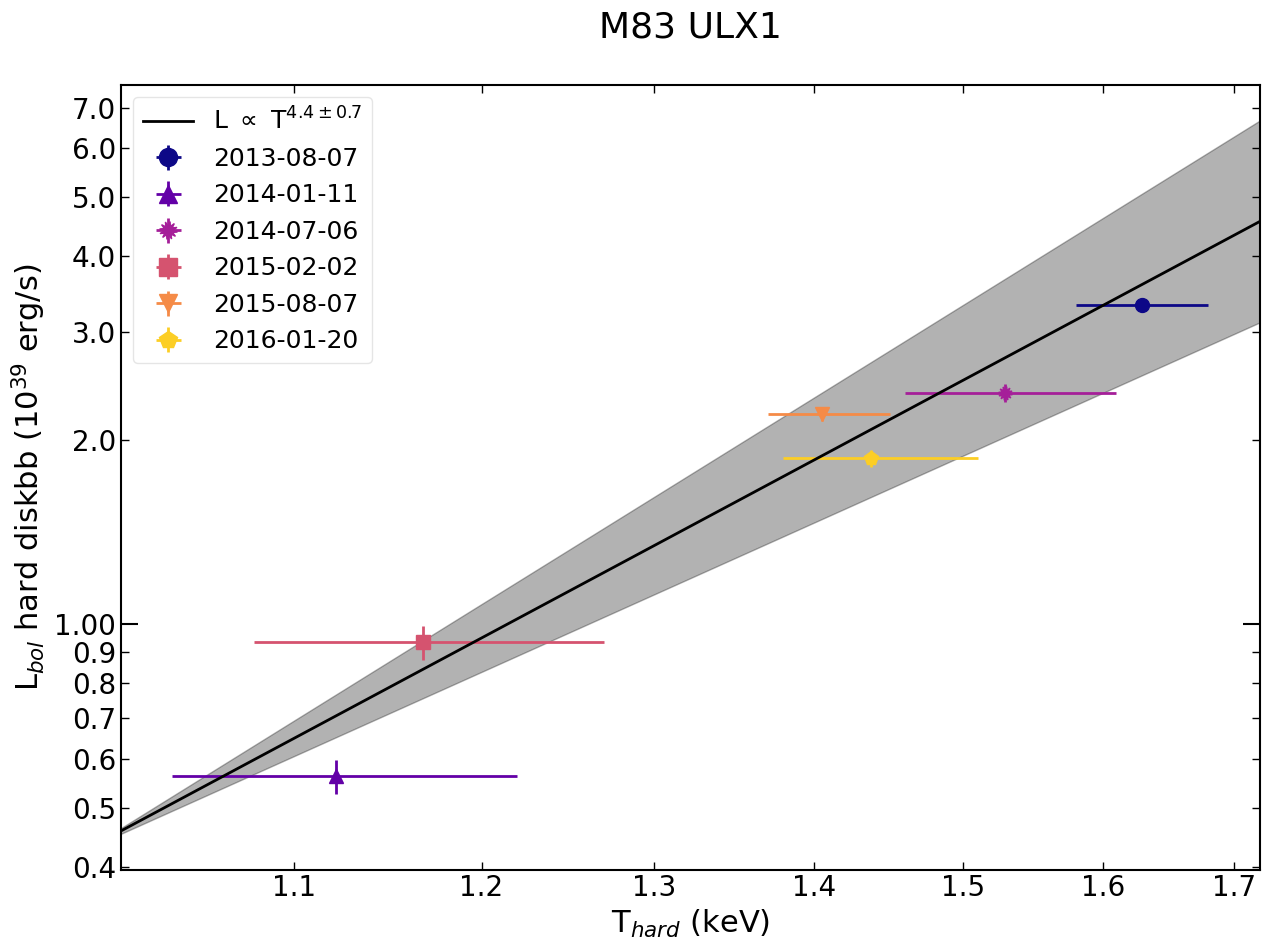}
    \includegraphics[width=0.49\textwidth]{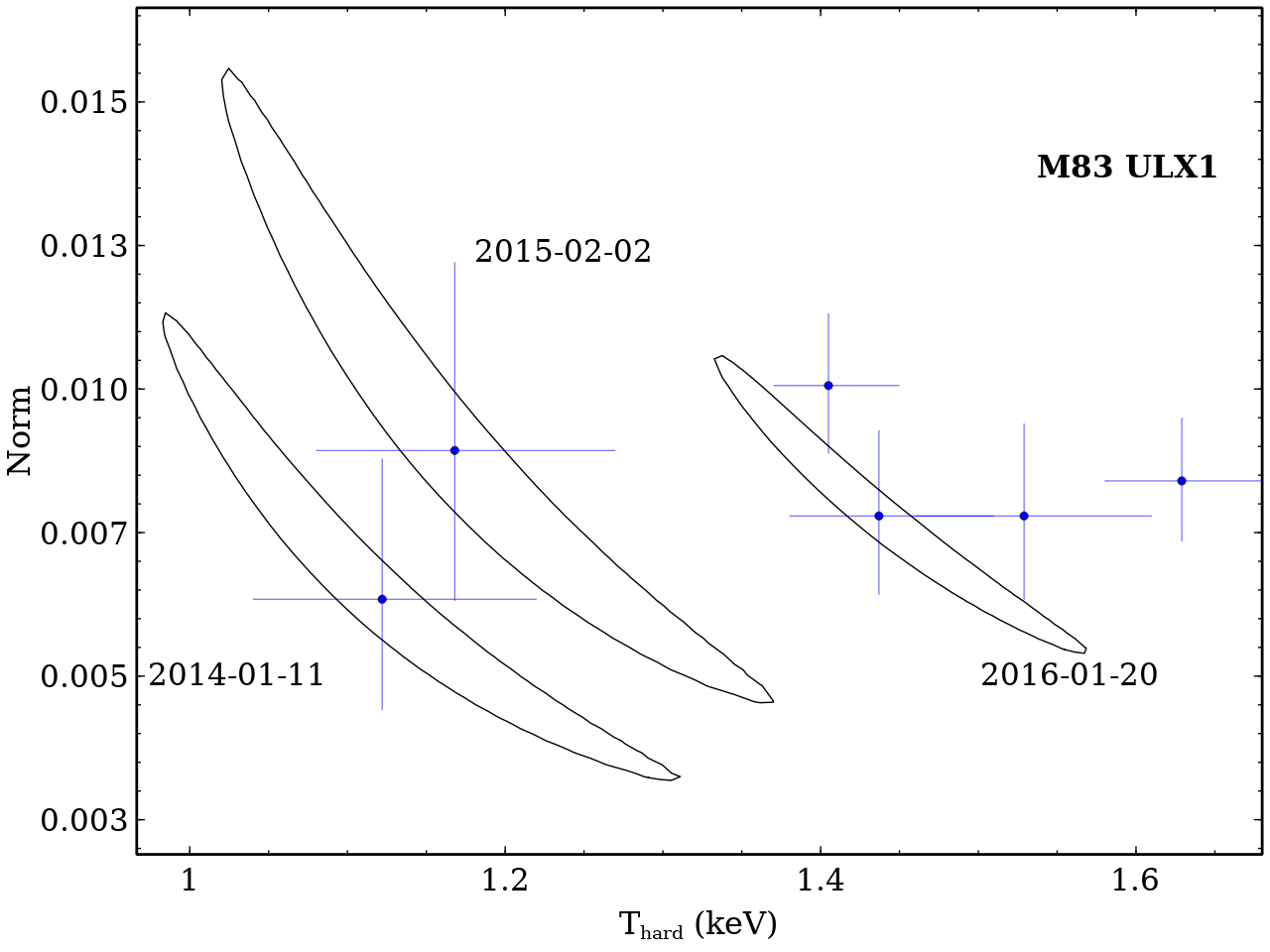}
     \caption{Left: As for Figure \ref{fig:positive_sources} for M83 ULX1 for the hard \diskbb. Right: As for Figure \ref{fig:anti_correlation} for the countours around the best-fit T$_\text{hard}$ and its normalisation for three selected epochs of M83 ULX1. The degeneracy between these two parameters can be ruled out as the cause of the positive L--T correlation.} \label{fig:degeneracy_hard}
\end{figure*}

Most of the sources show no clear correlation in the hard thermal component. The only exception is M83 ULX1, which shows a strong positive correlation (Spearman correlation of 0.94 with a $p$-value = 0.005, see Figure \ref{fig:degeneracy_hard}). We again computed $\chi^2$ contours around the best-fit temperature and normalisation for some selected epochs and overlaid the results from the spectral fitting on them as shown in Figure \ref{fig:degeneracy_hard}. In this case, it seems that the degeneracy between these two parameters is not driving the existing correlation.

\begin{figure}
    \centering
    \includegraphics[width=0.45\textwidth]{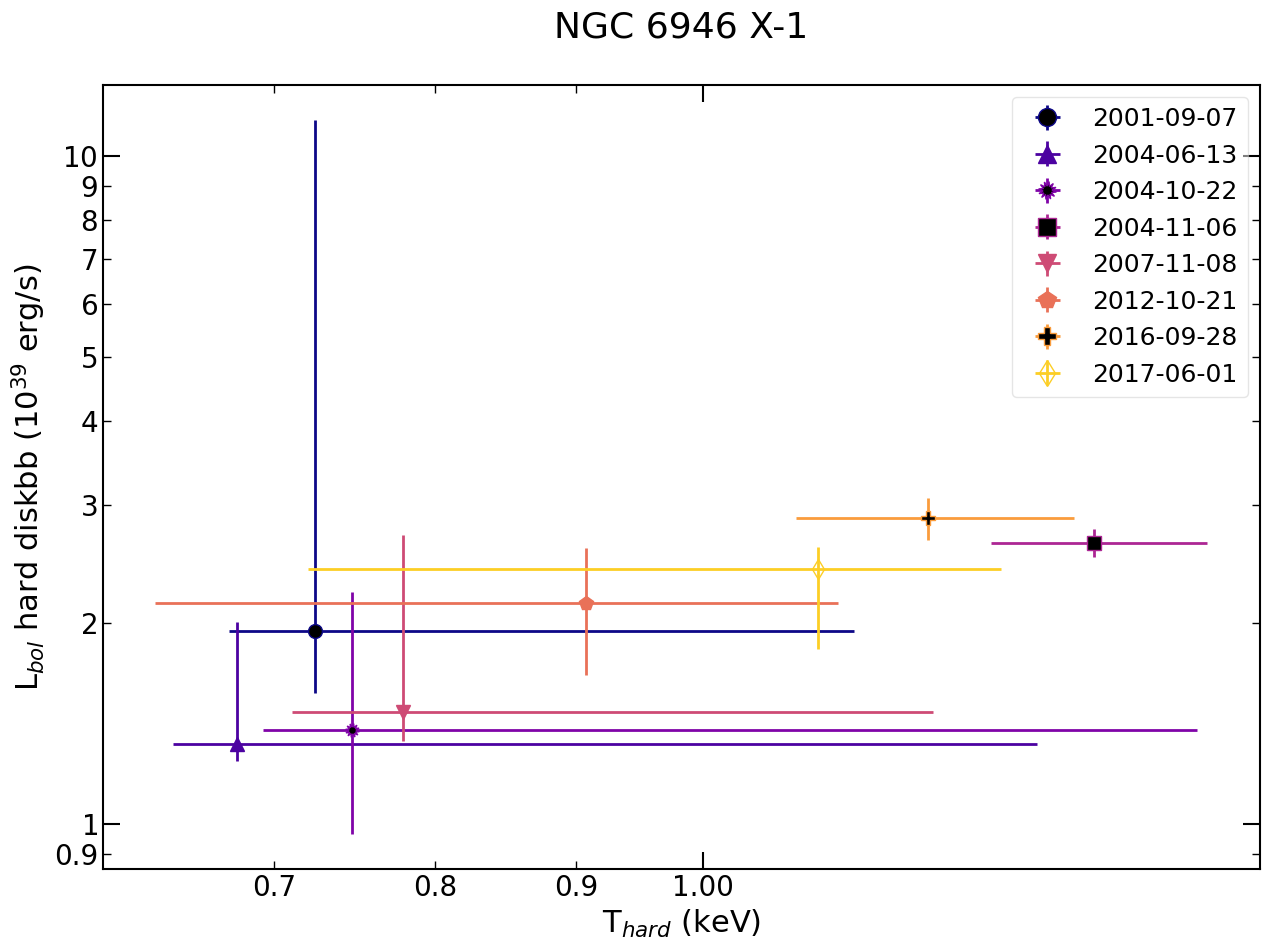}
    \caption{Bolometric luminosity of the hard \diskbb\ component as a function of its temperature for NGC 6946 X-1. Due to the uncertainties associated with the parameters, we do not consider these two quantities to be correlated (see text for details). Symbols as per Figure \ref{fig:individual_hl}.}
    \label{fig:ngc6946_hard}
\end{figure}

While NGC 6946 X-1 might seem positively correlated based on the Spearman test alone, examination of the data clearly revealed that there is an overall lack of strong variation of the hard component in most observations (see Figure \ref{fig:ngc6946_hard}). Furthermore, we noted a certain bias in those observations for which we have not included the \simpl\ model, as the temperature of the hard \diskbb\ appears to be systematically higher, as the \diskbb\ is pushed towards high-energies due to the lack of an additional high-energy component.  

\begin{table} 
 \centering 
 \caption{Results of the L--T correlation for the hard thermal component. Columns as per Table \ref{tab:correlation_softdiskbb}. }
 \label{tab:hard_correlations}
 \begin{tabular}{lccc} 
 \hline
 \noalign{\smallskip}
Source & Spearman & p-value & $\alpha$ \\ 
\noalign{\smallskip}
\hline 
 \noalign{\smallskip}
NGC 7793 P13 & 0.45	&0.19 & -\tablefootmark{a}\\ 
NGC 5907 ULX1 & 0.52 & 0.18 &-\tablefootmark{a}\\
NGC 5408 X-1 & 0.49 &0.01& -\tablefootmark{a}\\ 
Circinus ULX5 & --0.46& 0.29 & -\tablefootmark{a} \\ 
Ho IX X-1 & 0.31&0.38 & -\tablefootmark{a}\\ 
M81 X-6 & --0.08&0.83 &  -\tablefootmark{a}  \\ 
IC342 X-1 &0.05 & 0.91 & -\tablefootmark{a} \\ 
NGC 1313 X-1 & 0.3 & 0.2 & -\tablefootmark{a} \\
NGC 5204 X-1 &0.15	&0.58& -\tablefootmark{a} \\ 
Ho II X-1 & --0.37 & 0.47 &  -\tablefootmark{a} \\
NGC 1313 X-2 &--0.07 & 0.73 &  -\tablefootmark{a}\\
NGC 6946 X-1 & 0.90 & 0.002&  -\tablefootmark{b} \\
M83 ULX1 & 0.94	&0.005 & 4.36 $\pm$ 0.70 \\ 
M51 ULX-7 & --0.12&0.75 &  -\tablefootmark{a} \\ 
M51 ULX-8 & --0.17&0.67 &  -\tablefootmark{a} \\ 
\hline 
 \end{tabular} 
\tablefoot{\tablefoottext{a}{Non-correlated source.}
 \tablefoottext{b}{Not considered correlated due to uncertainties (see text for details)}}
 \end{table}

\section{Discussion} \label{sec:discussion}
We have examined the long-term spectral evolution of a sample of sources containing known pulsating and ULXs for which the accretor is unknown, in order to gain insights on their nature and the accretion processes driving their extreme luminosities. While our spectral fitting approach is phenomenological, we have attempted to understand the nature of the soft and the hard components by means of L--T correlations. The positive L--T correlations for the soft component found in Section \ref{sub:correlations_soft} are broadly consistent with previous results reported by \cite{miller_revisiting_2013}, who studied a similar sample of sources and also assumed a constant absorption column, albeit using a different model based on an accretion disk whose photons are Compton-up scattered by an optically thick corona of hot ($T_\text{e}$ $>$ 2 keV) electrons. Similar correlations were also reported for NGC 5204 X-1 and Holmberg II X-1 by \cite{feng_spectral_2009} using an absorbed \diskbb\ and a powerlaw. These correlations have frequently been used to argue in favour of an accretion disk as the nature of the soft component in ULXs. However, our analysis reveals that these are driven by the existing degeneracy (see Figure \ref{fig:anti_correlation}) between the temperature of the soft component and its normalisation and thus these relationships cannot be used reliably to support this scenario. Indeed, it has been shown that depending on the assumption of the underlying model, the correlation can disappear entirely \citep[see e.g.][]{luangtip_x-ray_2016,walton_unusual_2020} (see also \cite{goncalves_weakness_2006} for a discussion on the weaknesses of associating the soft component with an accretion disk). Nevertheless, these correlations might be useful to identify sources evolving in a similar fashion.

Conversely, the positive L--T correlation found for the hard thermal component of M83 ULX1 might have a physical origin as we argue in Section \ref{sub:m83}, given that this correlation is not driven by the existing degeneracy between the parameters (see Figure \ref{fig:degeneracy_hard}).

Similarly, \cite{kajava_spectral_2009} used the found negative L--T correlation for the soft component to argue in favour of an outflow as the nature of the soft component. However, this result was likely artificially created by employing a powerlaw for the high-energy emission while leaving the absorption column free to vary. We thus suggest that the beaming formalism derived by \cite{king_masses_2009} using the results from \cite{kajava_spectral_2009}, linking the mass-transfer rate ($\dot{m}_\text{0}$) with the beaming factor ($b$) for super-Eddington accretion onto BHs or weakly magnetised NSs, should be revisited in view of these new results. Overall, our study suggests that given the limitation from using phenomenological models to describe the ULX continuum, the observed changes in the best-fit spectral parameters may not be the most appropriate way to build a physical picture for our sources. For this reason, most of our discussion will be based on the variability observed in terms of luminosity and hardness ratios (see Figures \ref{fig:hid_diagram} and \ref{fig:individual_hl}).

Figures \ref{fig:hid_diagram} and \ref{fig:l_histogram} show that most sources seem to have a maximum luminosity of roughly 2 $\times$ 10$^{40}$ erg/s. M51 ULX-7 and M82 X-2 (the other PULX not analysed in this work) also reach luminosities of a few $\times$ 10$^{40}$ erg/s \citep{brightman_spectral_2016, brightman_swift_2019}. While the sample presented here is rather small in a statistical sense, it has the advantage that we have taken into account the long-term variability of each source, it is also free of contaminants and the absorption column has been carefully estimated. On this basis, our work seems to support larger sample studies \citep{swartz_complete_2011}, where a possible cutoff at around 2 $\times$ 10$^{40}$ erg/s was observed in the luminosity function of a sample of 107 ULX candidates. \cite{mineo_x-ray_2012} showed that ULXs seem to extend the X-ray luminosity function of HMXBs up to a possible energy cutoff also at around 2 $\times$ 10$^{40}$ erg/s, supporting ULXs as being an evolutionary stage of HMXBs. Thus, ULXs might be related to systems where the companion star expands and starts to fill its Roche Lobe as suggested by \cite{king_pulsing_2020}. As NS accretors are more numerous among HMXBs \citep[e.g.][]{casares_x-ray_2017}, it is possible that a substantial fraction of ULXs with L$_\text{x}$ $\sim$ 10$^{39-41}$ erg/s could host NS. However, we stress that given the limited sample studied here, we may be running into small numbers when distinguishing a cutoff distribution from a powerlaw one \citep{walton_2xmm_2011}.

A luminosity cutoff around 2 $\times$ 10$^{40}$ erg/s would be consistent with the maximum accretion luminosity NSs can attain according to \cite{mushtukov_maximum_2015}. The authors considered the accretion column model proposed by \cite{basko_limiting_1976} and the reduction of the electron scattering cross-section in the presence of high-magnetic fields, that allows to overcome the Eddington limit. \cite{mushtukov_maximum_2015} suggest that NSs could reach $\sim$ 10$^{40}$ erg/s for a reasonable parameter space of magnetic fields and spin periods. Indeed, the hard component (the \diskbb\ or the more complex \simpl$\otimes$\diskbb\ when applicable), that we would associate with the accretion column \citep[e.g.][]{walton_potential_2018}, reaches a maximum luminosity of $\sim$ 10$^{40}$ erg/s in NGC 7793 P13, M51 ULX-7 and NGC 1313 X-2. However, the fact that we find a PULX (NGC 5907 ULX1) also above the break is problematic. The hard component is a factor $\sim$ 8 above this theoretical value. It is possible that by assuming isotropic emission in our luminosity calculations, we may be overestimating the luminosity of the accretion column. We discuss this in more detail in Section \ref{sub:pulxs}.

On the other hand, general relativistic radiation magnetohydrodynamic (GRRMHD) numerical simulations of super-Eddington accretion onto black holes also predict saturation of the maximum luminosity with the mass-accretion rate \citep{narayan_spectra_2017}, as their models start to become radiatively inefficient above $\sim$ 10 $\dot{M}_\text{Edd}$. Their simulations show that the observed luminosity saturates at 2 $\times$ 10$^{40}$ erg/s for a 10 \msun\ black hole viewed close to face-on, which might be supported by our observations.

The mass-transfer rate ($\dot{m}_0$) in ULXs is generally accepted to be super-Eddington, even if the sustainability of such process at such extremes over long timescales (sometimes over several decades) has still to be understood. Models put forward to explain the emission of super-Eddington accretion onto black holes predict that as the mass-transfer rate increases beyond the classical Eddington limit, a radiatively driven outflow will be launched from within the spherisation radius (\rsph), the radius at which the disk reaches the local Eddington limit \citep{shakura_black_1973, poutanen_supercritically_2007}. The outflow leaves an optically thin funnel around the rotational axis of the compact object, so that at low inclinations we see a hard spectrum dominated by the inner parts of the disk  \citep[Hard ULXs:][]{sutton_ultraluminous_2013}. At higher inclinations the outflow becomes optically thick to Thomson scattering \citep{poutanen_supercritically_2007} and thus an observer at such inclinations should see a softer and fainter spectrum \citep[soft ULXs:][]{sutton_ultraluminous_2013} as most of the emission will be Compton down-scattered in the wind before reaching the observer \citep{middleton_spectral-timing_2015, kawashima_comptonized_2012}. A corollary of this scenario is that as the wind funnel narrows as the mass-accretion rate increases \citep{king_masses_2009, kawashima_comptonized_2012} and the wind starts to enter our line of sight, the contribution from the soft component will dominate the emission, as the wind starts to down-scatter out of the line of sight part of the hard emission. Therefore, a source with a hard ULX aspect could shift to a soft ULX aspect under certain conditions. A similar effect can occur if the source precesses \citep{abolmasov_optically_2009,middleton_spectral-timing_2015}, but we may see differences in the long-term variability between these two scenarios. The increase in the mass-accretion is also expected to increase the Thomson optical thickness of the gas within the funnel, preventing high-energy photons from escaping to the observer without being scattered \citep{narayan_spectra_2017,kawashima_comptonized_2012}. It is argued that, in extreme cases, either due to a high-mass accretion rate or/and higher inclination angle, a source may appear as a super-soft ultraluminous source (ULS) \citep{urquhart_optically_2016}, where most of the emission comes from the wind photosphere.

One key observational property of the presence of the funnel-like structure created by strong outflows is highly anisotropic emission. While a relationship between anisotropy and super-Eddington mass-transfer rates may be a natural consequence of super-Eddington accretion onto black holes \citep{king_masses_2009}, NS may have means to circumnavigate this relationship. Crucially, in the presence of a strong magnetic field, the disk might be truncated before radiation pressure starts to be significant to inflate the disk and drive strong outflows \citep[e.g.][]{chashkina_super-eddington_2019}. This occurs when the magnetospheric radius, given by:
\begin{equation} \label{eq:magnetospheric_radius}
R_\text{m} = \xi \left (\frac{R_{\text{NS}}^{12}B^4}{2GM_{\text{NS}}\dot{M}^2}\right )^{1/7}
\end{equation}
where R$_\text{NS}$ is the radius of the neutron star, $B$ is its magnetic field, M$_\text{NS}$ is the mass of the neutron star, $\dot{M}$ is the mass accretion rate at R$_\text{m}$ and $\xi$ is a dimensionless parameter that takes into account the geometry of the accretion flow and is usually assumed to be 0.5 for an accretion disk \citep{ghosh_accretion_1977}, is larger than the spherisation radius (\rsph). \rsph\ in turn has a linear dependence on $\dot{m}_\text{0}$. This offers means for a NS to be fed at high-mass transfer rates, while largely reducing anisotropy. On the other hand, we may expect that for weakly magnetised NSs, the disk will become super-critical given the dependency of \rmag\ on $B$. In this case the emission will be collimated by the outflow in a similar fashion as for super-critically accreting black holes \citep[e.g.][]{king_pulsing_2017, takahashi_general_2017}. Additionally, we may expect a NS with strong outflows to appear softer, as outflows will Compton down-scatter the emission from the accretion column.

We therefore can make use of the long-term variability observed in the HLD to discuss which scenario best describes the variability observed in each source: super-Eddington accretion onto BHs, weakly magnetised NSs or highly magnetised NSs. At the same time, the wealth of data analysed in this work allows us to identify groups of sources showing common evolution and similar spectral states, while identifying new NS-ULX candidates based on their similarity with the PULXs. Given that our analysis focuses on discussing the spectral transitions observed in the HLD (see Figure \ref{fig:individual_hl}), our study is less constraining for NGC 6946 X-1 and NGC 5408 X-1\footnote{Note that the increase of luminosity of NGC 5408 X-1 in 2001 is less certain given the short exposure of the 2001 observations and our assumption of constant \nh.}, given the lack of marked spectral states. We therefore do not attempt to offer any interpretation for these two sources.

Finally, a common characteristic of the PULXs seems to be a hard spectrum accompanied with high-levels of variability at high-energies ($\gtrsim$ 6 keV) (see last four sources in Figure \ref{fig:individual_hl} and also Figure 1 from \cite{pintore_pulsator-like_2017}). On this basis, we argue that M81 X-6 constitutes the best NS-ULX candidate from our sample, given its harder spectrum \textit{and} its high spectral variability at high energies as we discuss in Section \ref{sub:sub_pulxcandidates}. This high-energy component is likely associated with emission from the accretion column in the case of PULXs \citep{walton_super-eddington_2018, walton_evidence_2018} and may offer means to distinguish NS- and BH-ULXs (which we discuss in Section \ref{sub:hard_component}). While IC 342 X-1 and Circinus ULX5 also show a hard spectra with high-levels of variability at high-energies, they both show a soft and dim state that we argue is akin to that seen in NGC 1313 X-1, a much softer source with relatively stable high-energy emission (see Figure \ref{fig:individual_hl}) and thus we discuss them in Section \ref{sub:opticall_thick}. Nevertheless, we stress that the spectral similarity across the sample is undeniable as also noticed in previous studies \citep[e.g.][]{pintore_pulsator-like_2017, walton_evidence_2018}.

\subsection{PULXs} \label{sub:pulxs}
Remarkably, our analysis shows (Figure \ref{fig:hid_diagram}) that PULXs are among the hardest sources in our sample, something also noted by \cite{pintore_pulsator-like_2017} employing a colour-colour diagram. Interestingly, PULXs with higher pulse-fractions, e.g. NGC 7793 P13 \citep[PF$\sim$20\%;][]{israel_discovery_2017}, M51 ULX-7 \citep[PF$\sim$12\%;][]{rodriguez-castillo_discovery_2020} and NGC 300 ULX1 \citep[PF$\sim$50\%;][]{carpano_discovery_2018} tend to appear harder (see Figure \ref{fig:hid_diagram}) while the softest PULX in our sample, NGC 1313 X-2, has the lowest pulse-fraction \citep[PF$\sim$5\%][]{sathyaprakash_discovery_2019}. Figure \ref{fig:individual_hl} shows that harder PULXs show little variability in the HLD (see the case of NGC 7793 P13 and M51 ULX-7\footnote{Note that epochs 2003-01-15 and 2013-11-25 of NGC 7793 P13 and M51 ULX-7, where the sources are found in a softer and dimmer state, are likely associated with the sources resuming from the propeller regime.}), while in contrast NGC 1313 X-2 shows strong changes in hardness by a factor of $\sim$3, without undergoing to the propeller regime. Below we discuss whether these differences in the long-term evolution could be explained due to the different interplay between \rmag\ and \rsph. We note that in the super-Eddington regime, the interaction between the magnetic field and the disk is likely to be more complex than as predicted by Equation \ref{eq:magnetospheric_radius} as radiation pressure can dominate over the gas ram pressure \citep{takahashi_general_2017, chashkina_super-eddington_2019} but for the qualitative picture discussed here we neglect these facts. For NGC 300 ULX1, given the limited number of observations we cannot offer a detailed discussion as for the other sources and thus we will not consider this source further.

\textit{NGC 1313 X-2}: \rmag $<$ \rsph\ -- This source shows a strong bi-modal behaviour in the HLD \citep[see Figure \ref{fig:individual_hl} and also][]{feng_spectral_2006, pintore_x-ray_2012} that is likely associated with the 158 day quasi-periodicity reported by \cite{weng_evidence_2018} using \swift-XRT data, although an association is not straightforward since several periodicities are found in the periodogram presented by the authors. Our analysis reveals that this bi-modal behaviour is driven by a highly variable hard component while the soft emission is rather stable (the soft \diskbb\ varies by a factor $\lesssim$ 2 in luminosity while the hard component luminosity varies by factor $\lesssim$ 5, see Figure \ref{fig:uncorrelated_sources}). This bi-modal behaviour is confirmed by \swift-XRT long-term monitoring \citep{weng_evidence_2018}. 

This variability is unlikely to be produced by the propeller effect, where the centrifugal barrier of the magnetosphere prevents further infalling of gas onto the NS \citep{illarionov_why_1975}, as for a period of $\sim$ 1.47 s \citep{sathyaprakash_discovery_2019}, we should expect a drop in luminosity by a factor $\sim$ 220 \citep[see for instance equation 5 from][]{tsygankov_propeller_2016} assuming a NS radius of 10 km and a mass of 1.4 \msun, whereas the observed drop in luminosity from brightest to dimmest is only about a factor $\sim$ 5 (using the luminosities in the 0.3 -- 10 keV band from epochs 2006-03-06 and 2013-06-08). 

Instead, the absence of transitions to the propeller regime can be used to set constraints on the maximum magnetic field of the source, since we require that the magnetospheric radius is \textit{always} smaller than the co-rotation radius, the radius at which the Keplerian disk velocity equals the rotation of the NS. We thus require that at the minimum luminosity observed in the source, \rmag $<$ \rco. We can rearrange equation (37) from \cite{mushtukov_maximum_2015} to derive an upper limit on the magnetic field strength:
\begin{equation} \label{eq:limiting_l}
B_{12} \lesssim 3.8 \times \xi ^{-7/4} m ^{1/3}  R_\text{6}^{-5/2} P^{7/6} L_{intr}^{1/2}
\end{equation}
where L$_{intr}$ is the minimum \textit{intrinsic} luminosity in units of 10$^{39}$ erg/s, B$_{12}$ is the magnetic field in units of 10$^{12}$ G, R$_\text{6}$ is the NS radius in units of 10$^6$ cm, $m$ is the NS mass in \msun, $P$ is the period in seconds and $\xi$ is the same dimensionless parameter as for Equation \ref{eq:magnetospheric_radius} which we take again to be 0.5. We can express the intrinsic luminosity in terms of the observed luminosity taking into account a beaming factor ($b$) \citep[e.g.][]{king_pulsing_2020} so that $L_\text{intr}$ = $b\,L_\text{observed}$ with $b$ $\leq$ 1.
 Setting L$_\text{observed}$ = 2.2$\pm$0.1 (from epoch 2013-06-08), R$_\text{6}$ = 1, $m$ = 1.4 and $P$ = 1.47 s we obtain an upper limit on the magnetic field of $B$ = ($b^{1/2}$33.0$\pm$0.7) $\times$ 10$^{12}$ G. Thus, this suggests that we can rule out extreme magnetic fields ($B$ $\geq$10$^{13}$ G) for moderate values of $b$ ($\lesssim$ 0.1).

\begin{figure}
    \centering
    \includegraphics[angle=-90, width=0.49\textwidth]{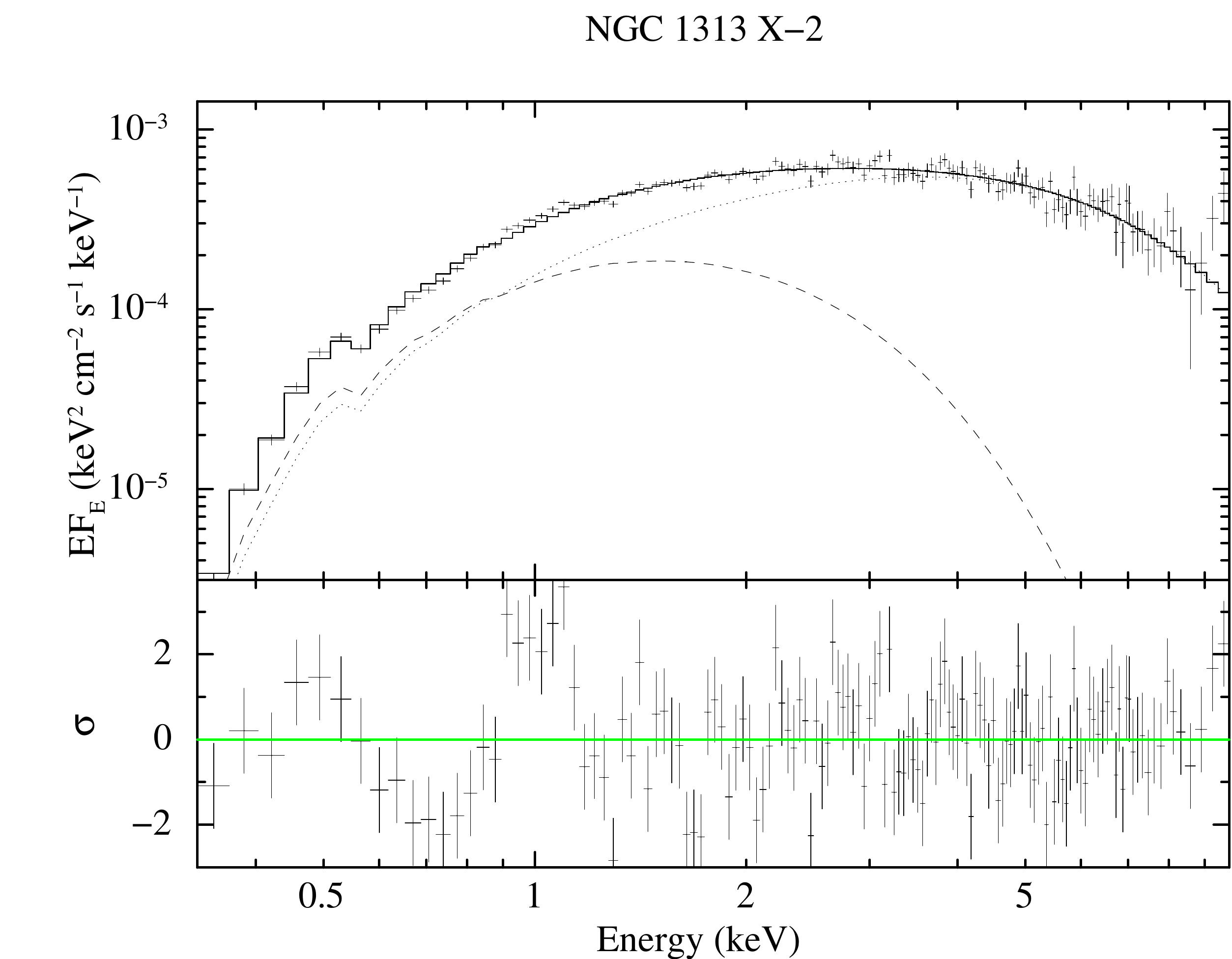}
    \caption{Unfolded spectra of NGC 1313 X-2 of epoch 2017-06-20 (only pn is shown for clarity) fitted with an absorbed dual thermal-component. The soft and hard \diskbb\ components are shown with dashed line and the dotted line respectively, while the total model is shown with a solid line. Strong residuals are seen at soft energies at around 1 keV.}
    \label{fig:ngc1313x-2_residuals}
\end{figure}

Therefore the hardening of the source with luminosity might instead support a scenario in which a conical outflow from a supercritical disk imprints highly anisotropic emission. Changes in the viewing angle due to the source precession and a varying degree of down-scattering in the wind could thus explain the source variability. For a discussion on possible mechanisms for precession in ULXs we refer the reader to \cite{vasilopoulos_m51_2020}. The stability of the soft component and the fact that the changes in HR and luminosity are likely driven by a super-orbital period support that changes in the mass-accretion rate are not the main source of variability. Additionally, the presence of strong outflows may be supported by the residuals observed at soft energies around 1 keV (see Figure \ref{fig:ngc1313x-2_residuals}) in the epoch with the longest exposure (2017-06-20), reminiscent of those seen in NGC 5408 X-1 and NGC 55 ULX 1 \citep{middleton_broad_2014, pinto_resolved_2016, pinto_ultraluminous_2017}. 

Overall the variability of the source and its low pulse-fraction ($\sim$ 5\%) support a scenario in which NGC 1313 X-2 is a weakly magnetised NS, in which \rsph $>$ \rmag\ and therefore outflows and precession cause the emission to be highly anisotropic. This is also supported by the recent ray tracing Monte-Carlo simulations of \cite{mushtukov_pulsating_2021}, showing that larger scale-height flows lead to lower pulse-fraction due to the increased number of scatterings.

\textit{NGC 7793 P13} and \textit{M51 ULX-7}: \rmag > \rsph\ -- Conversely, both the long-term evolution of NGC 7793 P13 and M51 ULX-7 show little variability in terms of hardness ratio, albeit also being associated with super-orbital periodicities, of 66.9 days \citep[see][Figure 5]{furst_tale_2018} in the case of NGC 7793P13 and of 39 days in the case of M51 ULX-7 \citep{vasilopoulos_m51_2020}. Indeed, long-term \swift-XRT monitoring shows no clear bi-modal behaviour \citep{weng_evidence_2018} in a hardness-intensity diagram in the case of NGC 7793 P13. Both sources also show very hot hard component (T$_\text{hard}$ $\sim$ 3 keV), albeit it is unclear whether the absence of a third, high-energy, component may have boosted the inferred temperature. Similarly as for NGC 1313 X-2, most of the luminosity variability is again seen in the hard component (the soft component varies by a factor of $\sim$ 2 and the hard component by a factor of $\sim$ 5). However, for NGC 7793 P13 and M51 ULX-7, the soft component does seem to brighten with source luminosity.   

The lack of HR variability might indicate that anisotropic emission caused by the wind funnel is not important in these sources. As stated before, in the presence of a strong magnetic field, the disk might be truncated before radiation pressure starts to be significant to inflate the disk and drive the strong outflow. Assuming the geometry of the accretion flow outside \rmag\ is similar to that of supercritically accreting black holes, our soft \diskbb\ could represent the emission from the outer regions of the accretion disk, with partial reprocessing by the wind if \rmag\ $>$ \rsph\ \citep{kitaki_theoretical_2017}. If we consider that this model component can give us a rough estimate of the size of this emitting region, a larger emitting area compared to NGC 1313 X-2, for which we have argued that outflows are important, could indicate that the disk is being truncated further from the accretor in the case of NGC 7793 P13 and M51 ULX-7. To illustrate this, we compute the mean radius of the inner disk given by the soft \diskbb\ normalisation ($N$) from all epochs. Using:
\begin{equation}
    R_\text{in} = \sqrt{N / \cos i} \times D_{10} \times f_\text{col}^2
\end{equation}
where $i$ is the inclination of the system, $D_{10}$ is its distance in units of 10 kpc and \fcol\ is the colour correction factor \citep{shimura_spectral_1995} which we take as 1.8 \citep[e.g.][]{gierlinski_black_2004} throughout this paper. This gives 1637\errors{106}{66}${(\cos i)}^{-1/2}$ km, 1975\errors{197}{134} and 2257\errors{125}{104}${(\cos i)}^{-1/2}$ km for NGC 1313 X-2, M51 ULX-7 and NGC 7793 P13, respectively. Naively assuming this radius gives a rough estimate of the size of the magnetospheric radius (\rmag), it may indicate a higher-mass accretion rate for NGC 1313 X-2 (and/or a lower magnetic field strength) and thus a scenario in which the disk becomes thick and outflows cause anisotropic emission. Instead, the larger radius of NGC 7793 P13 and M51 ULX-7 may suggest that either the mass-accretion rate is lower or the magnetic field strength is higher, which can result in the disk remaining geometrically thin \citep[see also][]{chashkina_super-eddington_2017} and therefore in reduced anisotropy. We note that these would support previous studies by \cite{koliopanos_ulx_2017}, where the same relationships for \rmag\ and \rsph\ were found for NGC 7793 P13 and NGC 1313 X-2 (see their Table 2 and 4).

Assuming the mass-accretion rate varies within a similar range in the three sources, a stronger magnetic field in NGC 7793 P13 and M51 ULX-7 is also supported by the fact both sources undergo periods of inactivity to $\lesssim$ 10$^{38}$ erg/s, likely associated with the propeller regime \citep[e.g.][]{furst_discovery_2016, Vas2021}, indicating that the condition \rmag $>$ \rco\ is more easily achieved. Furthermore, the lower pulse-fraction of NGC 1313 X-2 compared to that of NGC 7793 P13 and M51 ULX-7 is consistent with less material reaching the magnetosphere and thus the accretion column, as a result of the mass loss in the disk. Therefore, the emission at high-energies is not only intrinsically diminished but also down-scattered in the outflow \citep{mushtukov_pulsating_2021}, which might explain why NGC 1313 X-2 is generally softer than NGC 7793 P13 and M51 ULX-7.

We thus suggest that the magnetic field in NGC 7793 P13 and M51 ULX-7 is likely to be higher than that of NGC 1313 X-2 so that \rmag $>$ \rsph. The low degree of beaming and high-magnetic field implied by this solution is in agreement with previous magnetic field estimates \citep{vasilopoulos_m51_2020, rodriguez-castillo_discovery_2020}, that suggested a magnetic field of $\sim$ 10$^{13}$ G in M51 ULX-7. 

\textit{NGC 5907 ULX1}: \rmag $\sim$ \rsph\ -- Contrary to the rest of PULXs in our sample, the luminosity of NGC 5907 ULX1 clearly exceeds 10$^{40}$ erg/s. The variability between the observations clustered at L$_\text{X}$ $\sim$ (6--8) $\times$ 10$^{40}$ erg/s (epochs 2003 and 2014) and epoch 2012 was shown to be associated with different phases of the 78-day super-orbital period \citep{walton_78_2016} of the source by \cite{furst_spectral_2017}. This suggests that these changes are not due to a change in the mass-accretion rate and suggests instead changes in the viewing angle as the sources precesses. It has been speculated that the extreme luminosity of the source could be due to a high-degree of beaming \citep[e.g.][]{king_pulsing_2017, king_no_2019}, in which the super-orbital modulation was due to a conical outflow beaming the emission into and out of our line of sight \citep[e.g.][]{dauser_modelling_2017}. However, our Figure \ref{fig:individual_hl} shows that the source is harder when it becomes dimmer in epoch 2012-02-09 \cite[see also Figure 3 from][]{sutton_ultraluminous_2013} compared to epochs 2003/2014, and thus these changes associated with the super-orbital period may be hard to reconcile with changes imprinted by the precession of an outflowing cone, as we would expect a softer emission at low luminosities. This might imply that \rsph\ $\leq$ \rmag\ is likely in the case of NGC 5907 ULX1.

We note that other mechanisms could also cause the luminosity to be overestimated under the assumption of isotropic emission. In fact, the emission from the accretion column is not expected to be emitted isotropically. Instead, radiation-hydrodynamic simulations of super-Eddington accretion onto magnetised NSs by \cite{kawashima_radiation-hydrodynamics_2016} show that the accretion column is expected to have a flat emission profile along its sides, as the emission is only able to escape through the lateral sides of the confined material in it. This naturally creates highly anisotropic emission and can cause the observed emission to be greatly in excess of the Eddington limit.

A high $\dot{m}_\text{0}$ is still likely required to produce the observed luminosities. This may require the presence of a strong magnetic field so that the disk is truncated roughly at the point where it becomes supercritically. Thus, as suggested previously \citep{walton_evidence_2018, king_no_2019}
\rsph\ $\sim$ \rmag\ seems a plausible condition to explain the high-luminosity of NGC 5907 ULX1. 

\subsection{The non-pulsating NS: M51 ULX8} \label{sub:m51_ulx8}
This source was identified as a NS through the identification of a possible cyclotron resonance feature \citep{brightman_magnetic_2018} \footnote{\chandra\ obs id 13813, where the putative line was identified by \cite{brightman_magnetic_2018} was not considered in this work due to a certain degree of pile-up affecting the observation.}. We note that its position in the HLD (see Figure \ref{fig:hid_diagram}) may be consistent with a lack of pulsations \citep{brightman_magnetic_2018}, as this source is markedly dimmer and softer than the overall PULX sample, and sources with higher pulse fractions tend to sit in the harder end of the diagram, as stated before. However, given the apparent lack of variability, it is hard to give a comparison between this source and other PULXs in our sample. Strong variability is only observed in epoch 2018-05-25, where the hard component increased by a factor of 2 in luminosity. Its behaviour is somewhat similar to NGC 1313 X-2 and M81 X-6, that we argue is a good PULX candidate (see Section \ref{sub:sub_pulxcandidates}), although we lack enough observations of the source at higher luminosities to confirm this similarity. The soft component shows also little variability and no clear correlation, suggesting again a link between these three sources, which would favour a weakly magnetised NS in M51 ULX8 in agreement with the study presented by \cite{middleton_magnetic_2019}. Should this be the case, then long-term monitoring of the source will be crucial to attempt to identify any quasi-periodicity similar to those seen in NGC 1313 X-2 and M81 X-6.  

\subsection{PULX candidates} \label{sub:sub_pulxcandidates}
\textit{M81 X-6} -- As stated above, M81 X-6 constitutes our best NS-ULX candidate. The spectral evolution and the track in the HLD for this source is strikingly similar to that of NGC 1313 X-2 (Figure \ref{fig:individual_hl}). Both sources transit back and forth from a soft (HR $\lesssim$ 1.5) and low luminous state ($\lesssim$ 4 $\times$ 10$^{39}$ erg/s) to a hard (HR $\sim$ 3) and brighter state. The similar temperature, luminosity range and variability (or lack of it) of the soft component also suggests a link between these two sources. Instead, the variability is driven mostly by the hard component (the soft \diskbb\ changes by a factor $\lesssim$ 2 in luminosity while the hard component luminosity varies by a factor $\lesssim$ 5 in both sources). Furthermore, the variability of M81 X-6 is also likely associated with a 115 days quasi-periodicity \citep{weng_evidence_2018}, compatible with the one seen in NGC 1313 X-2. Given the similarity of these two sources, we suggest that it also harbours a weakly magnetised NS, so that the presence of strong outflows along with source precession may account for the source spectral variability. 

Prompted by this similarity, we searched for coherent pulsations in M81 X-6 using the code \hendrics\ \citep{bachetti_hendrics_2018} which is based on the publicly available \python\ library \stingray\ \citep{huppenkothen_stingray:_2019}. Unfortunately, all of the observations except one have less than 5000 counts in pn, and typically 10000 counts seem to be required in order to detect pulsations \citep{rodriguez-castillo_discovery_2020}. We thus searched in the observation with the longest exposure ($\sim$ 73 ks in pn, see Table \ref{tab:observations_xmm} and Figure \ref{fig:individual_hl} epoch 2001-04-23, HR $\sim$ 2, $L$ $\sim$ 4 $\times$ 10$^{39}$ erg/s) suitable for pulse searches, assuming M81 X-6 has a period and period derivative similar to the other PULXs. We used the \chandra\ coordinates given by \cite{swartz_chandra_2003} to extract barycentred corrected events in the 0.2 -- 12 keV band. We ran \hendrics\ on the unbinned event file searching for coherent pulsations in the 0.2 -- 8 Hz range, based on previous PULX detections, using the $Z^2$ statistic \citep{buccheri_search_1983} suitable for sinusoidal pulses. The search was performed using the option \textit{fast}, that optimises the search in the $f$--$\dot{f}$ space and reduces the computational time 10--fold compared to a classical search. We found no detection above the 3$\sigma$ level. We ran the same search in the 4 longest GTI intervals, ranging from 20 ks to 40 ks, as pulsations in PULXs have been shown to vary during the course of an observation \citep[e.g.][]{bachetti_all_2020}, but again we did not find any significant detections. Overall there is no peak that can be robustly identified as several peaks with similar $Z^2$ power are found, well below the 3$\sigma$ level. We also looked in the pn data of the individual observations with shorter exposures, but found no significant detections. We performed a last search looking into the 0.02 --  0.2 Hz range to look for longer periods as those seen in NGC 300 ULX1 \citep[see e.g.][]{vasilopoulos_ngc_2018}, with similar results. This could indicate that pulsations in this source are as elusive and faint as those found in NGC 1313 X-2 \citep{sathyaprakash_discovery_2019} and that deep exposures with the source on-axis or deeper searches correcting for the orbital parameters will be required to detect pulsations. 

Alternatively, the source spectral state may play a role in the detectability of the pulsations. Considering the case of NGC 1313 X-2, the epochs when pulsations were detected by \cite{sathyaprakash_discovery_2019} are 2017-09-02 and 2017-12-09 (i.e. the last two epochs in Figure \ref{fig:individual_hl}). The authors also found that the pulse fraction (and hardness, as shown in this work, Figure \ref{fig:hid_diagram}) decreases with the source luminosity. As argued before, we understand the hardening of the source as a decrease in the viewing angle as the system precesses. Considering the pulse-fraction calculations for super-critical accretion columns proposed by \cite{inoue_pulsed_2020}, this might imply that the angle between the rotational axis and the magnetic field axis ($\Theta_B$) is greater than the angle between the observer's line of sight and the rotational axis ($\Theta_\text{obs}$) (e.g. $\Theta_B$ $>$ 30$^\circ$ and $\Theta_\text{obs}$ $<$ 30$^\circ$, see Figure 5 from \cite{inoue_pulsed_2020}). Assuming the same applies to M81-X6, this could imply that pulsations are more likely to be found in softer and dimmer states (HR $\sim$ 1.5, L $\sim$ 2.5 $\times$ 10$^{39}$ erg/s) where we expect the pulse fraction to be higher.

If instead, the dilution of the pulsed emission is mainly due to a stochastic process such as multiple scatterings through the wind, then it may be possible to find a PULX in similar spectral states, with and without pulsations. Nevertheless, studying the dependence of the appearance of pulsations on the source spectral states seems a promising tool to put constraints on the accretion flow geometry in PULXs.

\subsection{Geometrical effects of a supercritical funnel} \label{sub:all_other_ulxs}
Several of the softer sources for which the accretor is unknown show a common pattern in their long-term evolution: three distinct spectral states, two of them at similar low luminosities but distinct hardness and a third one at a higher luminosity (see for instance NGC 1313 X-1, Circinus ULX5 and IC342 X-1 in Figure \ref{fig:individual_hl}). Two other sources that show also three marked spectral states are Holmberg II X-1 and NGC 5204 X-1, albeit the luminosity of the dimmer states differ in this case by a factor of $\sim$ 2--4. The difference in luminosity between one of the two dim states and the bright state might be naturally explained by changes in $\dot{m}_\text{0}$. However, the presence of an additional dim state requires another explanation. A super-critical funnel, as we discuss below, may offer an explanation to these three states either through obscuration as $\dot{m}_\text{0}$ increases (sources in Section \ref{sub:opticall_thick}) or due to changes in the inclination of the system (sources in Section \ref{sub:inclination_effects}). Given some of the common transitions and other spectral properties as we show below, we discuss together NGC 1313 X-1, Holmberg IX X-1, NGC 55 ULX1, Circinus ULX5 and IC 342 X-1 in Section \ref{sub:opticall_thick} and Holmberg II X-1 and NGC 5204 X-1 in Section \ref{sub:inclination_effects}.  

For the discussion, we will use NGC 1313 X-1 as our benchmark to discuss some of the transitions observed to the soft and dim states. In some cases, the timescale between these transitions and the duration of each state are poorly constrained due to the sparsity of our data. However, in a few cases, like for NGC 5204 X-1 and NGC 1313 X-1, the sampling rate is high enough so that we do observe the source switching from one state to another and thus we refer to these changes as transitions. 

\subsubsection{Optically thick funnels} \label{sub:opticall_thick}
\textit{NGC 1313 X-1} and \textit{Holmberg IX X-1} -- As we show later, NGC 1313 X-1 undergoes a transition similar to that observed in the super-soft ULXs in NGC 247 \cite{feng_nature_2016} and M101 \citep{soria_revisiting_2016}, which are seen to transit from ULSs to a soft ULXs spectra. A similar transition is also seen in NGC 55 ULX1 \citep{pinto_ultraluminous_2017} (and in this work as we argue later), although the source would still classify as a ULX when in this dim-state. These transitions are all marked by an increase in the size of the emitting region of the soft component and a decrease in temperature, interpreted as an expansion of the wind photosphere as the spherisation radius increases with the corresponding decrease in temperature \citep{poutanen_supercritically_2007}. While the ULXs in M101 and NGC 247 are frequently thought to be viewed at high inclinations \citep[e.g.][]{ogawa_radiation_2017}, NGC 1313 X-1 is likely viewed down the optically thin funnel \citep[e.g.][]{poutanen_supercritically_2007, narayan_spectra_2017}, so that the hard component dominates the emission (epoch 2012-12-16 in Figure \ref{fig:individual_hl}).

The spectral transitions of NGC 1313 X-1 between the low state (L$\sim$ 8 $\times$ 10$^{39}$ erg/s, epoch 2012-12-06) and the high-state (L$\sim$ 18 $\times$ 10$^{39}$ erg/s, epoch 2004-06-05) have been frequently interpreted as the wind entering our line due to a narrowing of the funnel as the mass-accretion rate increases \citep[e.g.][]{sutton_ultraluminous_2013}. However, the fact that the high-energy emission ($\gtrsim$ 10 keV) remains relatively stable may be at odds with this interpretation, as we should expect the wind to down-scatter \citep{kawashima_comptonized_2012, middleton_spectral-timing_2015} or even absorb \citep{abolmasov_optically_2009} the high-energy emission from the inner parts of the accretion flow. We should also expect the increase in the mass-transfer rate to lead to an increase in the Thomson scattering optical depth of the funnel \citep{kawashima_comptonized_2012}, also causing the high-energy emission to drop.

The physical processes at play to produce these high-energy photons are still poorly understood \citep[e.g.][]{walton_unusual_2020} but it is generally accepted that this emission is produced in the vicinity of the accretor \citep[e.g.][]{kawashima_comptonized_2012, takahashi_formation_2016, walton_unusual_2020}. For the remainder of this part, we assume that this high-energy component is indeed produced in the inner regions of the accretion flow and focus on the influence of the wind/funnel structure on the spectra, rather than on the origin of this emission, which will be discussed in Section \ref{sub:hard_component}.

The fact that the high-energy component remains stable, could therefore imply that the gas within the funnel has remained optically thin over a certain range of mass-transfer rate, and that the inclination of the system ($i$) remains well below the half-opening angle of the funnel ($\theta_\text{f}$). The second condition is required so that the higher degree of beaming caused by the reduction of $\theta_f$, will only result in an increase in the amount of photons that are down-scattered off the wind walls into the observer's line of sight, with the wind remaining out of the line of sight. Since the optical depth of the wind is lowest near the rotational axis of the compact object \citep{poutanen_supercritically_2007}, the emission from the innermost regions are more likely to reach the observer without suffering severe energy loses \citep{kawashima_comptonized_2012}. This may support previous works suggesting that NGC 1313 X-1 is seen at low viewing angles \citep[e.g.][]{middleton_spectral-timing_2015}.

The lack of obscuration of the high-energy emission suggests that the mass-accretion rate has to be moderate ($\dot{M}$ $\lesssim$ 10 $\dot{M}_\text{Edd}$) \footnote{Here we adopt the definition of $\dot{M}$ of \citep{narayan_spectra_2017} where $\dot{M}$ = $\frac{L_\text{Edd}}{\eta c^2}$ where $\eta$ depends on the black hole spin.} as for higher mass-transfer rates the gas within the funnel is expected to become optically thick \citep{narayan_spectra_2017}. Therefore, regardless of the exact nature of this high-energy powerlaw tail, we argue that albeit the increase in mass-accretion rate in NGC 1313 X-1 up to epoch 2004-06-05, the physical conditions within the funnel have remained stable. The similar persistent high-energy emission seen in Holmberg IX X--1 and the similar L--T positive correlation, suggests a similar evolution in both sources, albeit NGC 1313 X--1 is seen in an obscured state (see below) not seen in Holmberg IX X--1. We discuss in more detail the possible differences between these two sources focussing on the nature of the high-energy tail in Section \ref{sub:hard_component}.

Interestingly, after NGC 1313 X--1 reaches its maximum luminosity (epoch 2004-06-05), it becomes extremely soft and its luminosity decreases (epoch 2004-08-23 -- \textit{obscured state}). This spectral transition (note that these two observations are just two months apart) can be understood if a further increase in the mass-accretion rate leads to a narrowing of the opening angle of the funnel as the wind becomes more mass-loaded, to the point where the gas within the funnel becomes optically thick to the high-energy radiation. This implies that now $\theta_f$ $<$ $i$ and thus the wind effectively enters the line of sight and starts obscuring the inner accretion flow. The optical depth of the wind in the direction parallel to the disk is also expected to be an order of magnitude higher than in the perpendicular direction \citep{poutanen_supercritically_2007}. High-energy photons are now heavily down-scattered or absorbed and therefore we mostly observe the soft emission from the expanded wind photosphere, as supported by the increase in the normalisation of the soft component (from $\sim$ 5 to $\sim$ 50 before and after the obscuration respectively).

This transition is also in good agreement with the GRRMHD simulations presented by \cite{narayan_spectra_2017} of super-Eddington accretion onto black holes. The authors observe a transition from a hard spectrum to a very soft one, as the mass accretion rate increases ($\dot{M}$ $\sim$ 23 $\dot{M}_\text{Edd}$ from their simulations) and the gas within the funnel becomes optically thick (see their Figure 9). Furthermore, their simulations also show that the luminosity for an observer with a line of sight close to the rotational axis of the accretor ($i$ $\sim$ 10$^\circ$) is capped at around 2 $\times$ 10$^{40}$ erg/s, which is in very good agreement with the maximum luminosity observed in NGC 1313 X-1. Their simulations also predict an increase in the spectral emission at low energies ($\lesssim$ 0.6 keV), which seem at odds with our observations, where the luminosity of the soft component has remained stable compared to the brightest state. We note however, that \cite{kawashima_comptonized_2012}, who found qualitatively the same results as \cite{narayan_spectra_2017} for high-mass transfer rates ($\gtrsim$ 23 $\dot{M}_\text{Edd}$), do not observe an increase in luminosity at low energies. It is also possible that due to the expansion of the photosphere, the soft component peaks now in the extreme UV and thus given our limited bandpass, we cannot reliable assess whether the luminosity of the soft component has increased.

Numerical simulations by \cite{ogawa_radiation_2017} also predict a steep decline of the high-energy emission as the outflow photosphere enters the line of sight. We find that the temperature of the hard \diskbb\ diminishes from 1.6$\pm$0.1 keV in epoch 2004-06-05 to 0.92\errors{0.10}{0.08} keV in epoch 2004-08-23 and its unabsorbed bolometric luminosity decays from (11.2$\pm$0.6) $\times$ 10$^{39}$ erg/s to (2.9$\pm$0.4) $\times$ 10$^{39}$ erg/s, which seems to support this interpretation.

\textit{NGC 55 ULX1} -- Similarly, we argue that the soft and dim spectral state observed in NGC 55 ULX1 is analogue to the obscured state observed in NGC 1313 X-1. In both cases, we observe a possible increase in the neutral absorption column (see Section \ref{sub:absorption_column}). Albeit this might be model dependent (this is discussed further below), it suggests that their spectrum has evolved in a similar manner. The ratio of unabsorbed bolometric fluxes in the obscured state are F$_\text{harddiskbb}$/F$_\text{softdiskbb}$ = 0.34\errors{0.12}{0.08} and 0.29\errors{0.01}{0.02} in NGC 1313 X-1 (epoch 2004-08-23) and NGC 55 ULX1 (epoch 2010-05-24), respectively. This is a factor of $\sim$ 3 times lower in both cases compared to when the sources were at their brightest and indicates that the hard component is responsible for the drop in luminosity. We see again an increase in the normalisation of the soft \diskbb\ \citep[see also][]{pintore_spectral_2015}, akin to the transitions seen in the ULXs in M101 and NGC 247 \citep{feng_nature_2016, soria_revisiting_2016}. The increase in the neutral absorption column may indicate that now we see parts of the wind less exposed to the central source \citep{pinto_thermal_2020}, where self-absorption could start to be important, albeit more physically motivated models are needed to address this. The presence of outflows is supported by studies using high-resolution spectroscopy \citep{pinto_ultraluminous_2017, pinto_xmmnewton_2020} which might have revealed the presence of soft residuals associated with outflowing winds in NGC 1313 X-1 and NGC 55 ULX1. Their similar transitions highlighted here support therefore the unification scenario proposed by \citep{middleton_spectral-timing_2015, pinto_thermal_2020}. 

\textit{IC 342 X-1} and \textit{Circinus ULX5} -- As shown in Figure \ref{fig:individual_hl}, IC 342 X-1 and Circinus ULX5, not only share a very similar evolution in the HLD, but are also found in a soft and dim state (epochs 2012-10-29 and 2016-08-23 for IC 342 X-1 and Circinus ULX5 respectively), reminiscent again of the obscured state seen in NGC 1313 X-1. When both sources are hard and dim (e.g. epochs 2012-08-07 and 2001-08-06 for IC 342 X-1 and Circinus ULX5 respectively) the mass-transfer rate is likely to be low, similar to NGC 1313 X-1 in epoch 2012-12-16. The brighter and harder states (epochs 2005-02-10 and 2013-02-30 for IC 342 X-1 and Circinus ULX5 respectively) might correspond to an increase in the mass-transfer rate while the softest and dimmest states are likely again due to the central source being obscured by the funnel becoming optically thick at high-transfer rates. This is supported by the diminishing of the hard component in both temperature and luminosity. In this case, we do not see an increase in the local \nh-value as for NGC 1313 X-1 or NGC 55 ULX1, that could strengthen the similarities of these obscured states, but we note that these are the two sources with the largest \nh$_\text{Gal}$-values ($\gtrsim$ 30 $\times$ 10$^{20}$ cm$^{-2}$) in our sample and that we noted some calibration uncertainties at low energies (see Section \ref{sub:data_reduction}). Nevertheless, these sources show that both archetypal soft ULXs (e.g. NGC 55 ULX1) and hard ULXs (e.g. IC 342 X-1) undergo similar type of transitions.

\subsubsection{Inclination effects}\label{sub:inclination_effects}
\textit{Holmberg II X-1} and \textit{NGC 5204 X-1} -- The similarities between Holmberg II X-1 and NGC 5204 X-1 are clear when looking at their long-term evolution in the HLD (see also the similarity between the three spectra shown in Figure \ref{fig:individual_hl}) and are further supported by the similar L--T correlations found for the soft \diskbb\ ($\alpha_\text{NGC5204X-1}$ = 3.6$\pm$0.4, $\alpha_\text{HolmbergIIX-1}$ = 3.3$\pm$0.4). Therefore, regardless of the physical processes powering these two sources, these similarities strongly suggest that we are witnessing the same type of source and/or accretion flow (see also G\'urpide et al. in \textit{prep}).

In epochs 2004-04-15 and 2006-11-16 of Holmberg II X-1 and NGC 5204 X-1 respectively, both sources are found with a hard spectrum and an intermediate luminosity -- \textit{hard/intermediate} state. Again, we favour a low viewing angle as for NGC 1313 X-1 as the hard component dominates the emission, although the inclination in this case may be higher than for NGC 1313 X-1, given their softer spectra. As both these sources move in the HLD (see 2003 \xmm\ epochs for NGC 5204 X-1 for the transition) from these epochs to softer spectra and brightest luminosities -- \textit{bright/soft state} (e.g. epochs 2004-04-15 and 2006-11-16 for Holmberg II X-1 and NGC 5204 X-1 respectively), the temperature of the hard component decreases as seen in NGC 1313 X-1\footnote{For NGC 5204 X-1, see those epochs of higher quality as the epochs when \simpl\ was not included tend to appear with artificially hotter temperatures for the hard \diskbb.}, while the soft component increases in temperature and luminosity (see Table \ref{tab:individual_fits}). Similarly, most of the variability between these two epochs is seen at mid to soft energies ($\sim$ 0.3 -- 5 keV), whereas the high-energy emission ($\gtrsim$ 5 keV) shows little variability. We argue that these transitions are due once again to an increase in the mass-accretion rate as the slight softening of both sources at high-energy and the increase in the soft component seems to match the evolution presented by \cite{kawashima_comptonized_2012} (see their Figure 2). In this case, the small dimming seen at high-energies ($\gtrsim$ 10 keV, see especially spectra for Holmberg II X-1) may be due to the fact that our line of sight may be now grazing the optically thick walls of the wind, and thus some of the high-energy photons from the inner parts of the accretion flow are now being Compton down-scattered by the wind. The transitions from hard/intermediate to soft/bright in NGC 5204 X-1 were also reported by \cite{sutton_ultraluminous_2013} in terms of hard and soft ULX transitions.

Our study shows that these sources are also seen to transit from bright/soft to another state that we term \textit{dim/soft} (epochs 2002-09-18 and 2001-05-02 for Holmberg II X-1 and NGC 5204 X-1 respectively) and vice-versa. For NGC 5204 X-1, this is also seen in the \chandra\ observations of 2003 (see also the full set of \chandra\ observations presented by \cite{roberts_chandra_2006}\footnote{Albeit these transitions were well sampled by \chandra\ at the end of 2003, we were not able to use the short exposure ($\sim$ 5ks) observations when the source was caught repeatedly in the low state, given the limited number of counts registered.}). For Holmberg II X-1, this is also seen in 2010 (see also the 2009/2010 \swift-XRT monitoring of Holmberg II X-1 in \cite{grise_x-ray_2010}). 

While there are certain similarities between these transitions and that seen in NGC 1313 X-1 in the obscured state, namely a softening and dimming of the source, we found also certain differences that may indicate that these transitions are not due to the same phenomenon as for NGC 1313 X-1. From the bright/soft state to the dim/soft state, the entire spectrum seems to have diminished in luminosity in the case of Holmberg II X-1 and NGC 5204 X-1 (see the spectra from Figure \ref{fig:individual_hl}), while for NGC 1313 X-1 we showed that it was the hard component that was mostly responsible for the dimming. Indeed, the soft \diskbb\ is the faintest in the dim/soft state, while the temperature of the hard \diskbb\ remains consistent within errors with respect to the bright/soft state (for both Holmberg II X-1 and NGC 5204 X-1). If the funnel has become optically thick due to an increase in the mass-transfer rate and is now obscuring the hard emission, we should expect the soft component to be relatively stable as seen in NGC 1313 X-1. Therefore, the dimming of this component may be at odds with this interpretation.

It is worth noting that the spectral evolution of NGC 5204 X-1 and Holmberg II X-1 from bright/soft to dim/soft bears some resemblance with how the inclination affects the spectral shape \citep[see for example][]{kawashima_comptonized_2012, kitaki_theoretical_2017, narayan_spectra_2017}, which could suggest that changes in the inclination are responsible for these spectral changes. Indeed, numerical simulations by \cite{narayan_spectra_2017} predict a decrease in about one order of magnitude in luminosity between a source with a face-on aspect and a source viewed at high inclinations ($i$ $>$ 30$^\circ$), which is consistent with our observations of the transitions from bright/soft to dim/soft. However, if the inclination of the system is indeed changing due to precession of the supercritical funnel, we should expect these changes to be associated with some periodicity. Thus, the fact that these transitions do not seem to be periodic \citep{grise_x-ray_2010} may be at odds with the effects of a precessing funnel (albeit see G\'urpide et al. in \textit{prep}).  

These transitions were also shown to occur rapidly, in some cases in timescales shorter than half a day \citep{grise_x-ray_2010}. This short-term variability may be expected if dense clumps of the wind are intersecting our line of sight \citep{takeuchi_clumpy_2013, middleton_spectral-timing_2015} or if our line of sight is rapidly changing between seeing down the funnel and seeing through the wind walls, expected if our line of sight grazes the wind as argued before. More information and monitoring is needed about the timescale of these transitions to address this issue and thus this will be further studied in a forthcoming publication.

\subsection{Origin of the high-energy tail} \label{sub:hard_component}
As stated above, NGC 1313 X-1 and Holmberg IX X-1 show a remarkably stable emission above $\sim$8 keV, albeit the sources are clearly varying at lower energies. Several physical processes have been proposed in order to explain the high-energy emission in ULXs. Here we attempt to discuss its nature based on this observed stability.

Crucially, the absence of a hard surface in black holes naturally offers an explanation for the presence of a stable emission component: advection and photon trapping effects \citep{abramowicz_slim_1988, ohsuga_why_2007}. In the super-Eddington regime, the diffusion photon time in the vicinity of the black hole is expected to be greater than the accretion timescale, while part of the excess energy will go into powering the outflow, so that the luminosity only increases logarithmically with the mass-transfer rate \citep{poutanen_supercritically_2007}:
\begin{equation}
    L \sim L_\text{Edd} (1 + x\ln \dot{m}_\text{0})
\end{equation}
where $x$ = 1 if only advection is considered or $x$ = 0.6 if all the energy goes into powering the wind. Theoretical works show that this can lead to saturation of the continuum in the hard X-ray band for stellar-mass black holes \citep{feng_global_2019} and thus this stability at high-energies may represent the smoking gun in differentiating BH- from NS-ULXs, as NS have no means of swallowing any excess energy. Therefore, if we assume the high-energy emission is rather insensitive to $\dot{m}_\text{0}$ as suggested by the stability of the high-energy component in Holmberg IX X-1 and NGC 1313 X-1, we suggest that the former might harbour a heavier black hole than the latter given its brighter high-energy component (around a factor $\sim$ 1.6).

Alternatively, \cite{kawashima_comptonized_2012} and \cite{kitaki_theoretical_2017} based on numerical radiation hydrodynamic (RHD) simulations of super-Eddington accretion onto black holes, showed that an overheated region (T $\sim$ 8 keV) is formed in the vicinity of the black hole, where a shock is produced in the region where the outflow collides with the inflow. Photons from the disk entering this overheated region will gain energy through multiple Compton up-scatterings prior to escaping through the funnel or the wind itself (where they will undergo Compton down-scattering). Photons escaping mostly through the funnel will be less affected by Compton down-scattering and will be observed as a high-energy powerlaw tail in the spectrum. \cite{kitaki_theoretical_2017} argued that the temperature of this region is independent on the black hole mass, resulting in a similar spectral shape for the high-energy tail regardless of the black hole mass. Therefore, provided that the mass-transfer rate between two given sources is similar, the luminosity of this high-energy component may provide means to estimate the black hole mass ratio between two given sources. The two sources for which this stability is best observed, NGC 1313 X-1 and Holmberg IX X-1, have indeed similar slope of the hard tail $\Gamma_\text{NGC1313X-1}$=2.90\errors{0.05}{0.04} and $\Gamma_\text{1HolmbergIXX-1}$=2.9\errors{0.2}{0.3} and $\Gamma_\text{2HolmbergIXX-1}$=3.4\errors{0.2}{0.3} where 1 and 2 indicate the two different $\Gamma$ values we have used for the low and high flux epochs respectively. If we instead refit all the high data quality sets of Holmberg IX X-1 assuming one single value for $\Gamma$ tied between all epochs, we obtain $\Gamma$=3.1$\pm$0.2 with $\chi^2_r$ = 1.06 for 3722 degrees of freedom, consistent with the slope found in NGC 1313 X-1. Therefore, the brighter luminosity of the high-energy component in Holmberg IX X-1, would again suggest that Holmberg IX X-1 harbours a heavier black hole compared to NGC 1313 X-1.

It is however not clear yet how the physical properties of this overheated region change with the mass-transfer rate. Assuming the overheated region is indeed responsible for the stable high-energy tail, then its physical properties should also be rather insensitive to the mass-transfer rate. Given that the overheated region is formed due to the shock of the outflows colliding with the inflow, we may also expect it to form around weakly magnetised NSs too. However, it is to be seen if the physical conditions within this overheated region and the funnel are expected to lead to the formation of a similar high-energy powerlaw tail as predicted for super-critically accreting BHs.

Lastly, in the case of accretion onto a NS magnetic poles, the accretion column is expected to be responsible for the emission at high-energies \citep{walton_evidence_2018}. Unfortunately, simulated spectra from super-critically accreting NS from numerical simulations are still under study \citep[e.g.][]{takahashi_general_2017, takahashi_supercritical_2018} and it is currently hard to know how the emission from the accretion column reacts to changes in the mass-transfer rate and magnetic field strength. Nevertheless, if the accretion column is responsible for the stable emission observed in these sources, it is still unclear why we observe this difference in terms of stability between PULXs and sources like NGC 1313 X-1 and Holmberg IX X-1. 

\subsection{M83 ULX1: a stellar-mass black hole?} \label{sub:m83}
This source sits at the lower end of the ULX luminosity distribution. We found the maximum unabsorbed luminosity of M83 ULX1 to be $\sim$ 3.5 $\times$ 10$^{39}$ erg/s, close to the maximum observed luminosity of $\sim$ 4.5 $\times$ 10$^{39}$ erg/s by \cite{soria_slim-disk_2015}, although this was computed using a powerlaw which could have boosted the unabsorbed luminosity. This maximum luminosity, using Eddington mass scaling, suggests an accretor of $\sim$ 25 \msun. We also found that the hard \diskbb\ component follows L$_\text{hard}$ $\propto$ T$_\text{hard}^{4.4\pm0.7}$, which as suggested by Figure \ref{fig:degeneracy_hard}, does not seem to be spuriously created by existing degeneracies. This may suggest that this component arises from an accretion disk with constant inner radius. This fact together with its low maximum luminosity, may suggest that we are witnessing a black hole accreting close to the Eddington-limit, as binary synthesis population studies predict that BH-ULXs tend to emit isotropically, since super-Eddington mass-transfer rates (a factor $\sim$ 8 above the classical Eddington limit) are harder to obtain from binary population evolution in the case of BHs \citep{wiktorowicz_observed_2019}. 

We thus consider the possibility that the source could be accreting close to the Eddington-limit. If so, then we could expect the accretion disk to deviate from the standard thin accretion disk, as radiation pressure inflates the disk making it geometrically slim or thick. This may lead to a departure of the radial temperature index ($p$) of --0.75 for a standard thin accretion disk \citep{shakura_black_1973}, implicitly assumed in the \diskbb\ model. In order to explore such deviations, we refitted all our data with an absorbed broadened disk model (\diskpbb\ in \xspec). As in Section \ref{sec:spectral_fitting}, we assumed again constant absorption column and we jointly fitted our data tying \nh\ across all datasets with the \diskpbb\ model. We obtained an excellent fit with $\chi^2$ $\sim$ 1.02 for 1493 degrees of freedom (parameters are listed in Table \ref{tab:m83_diskpbb}). 

\begin{figure}
    \centering
    \includegraphics[width=0.45\textwidth]{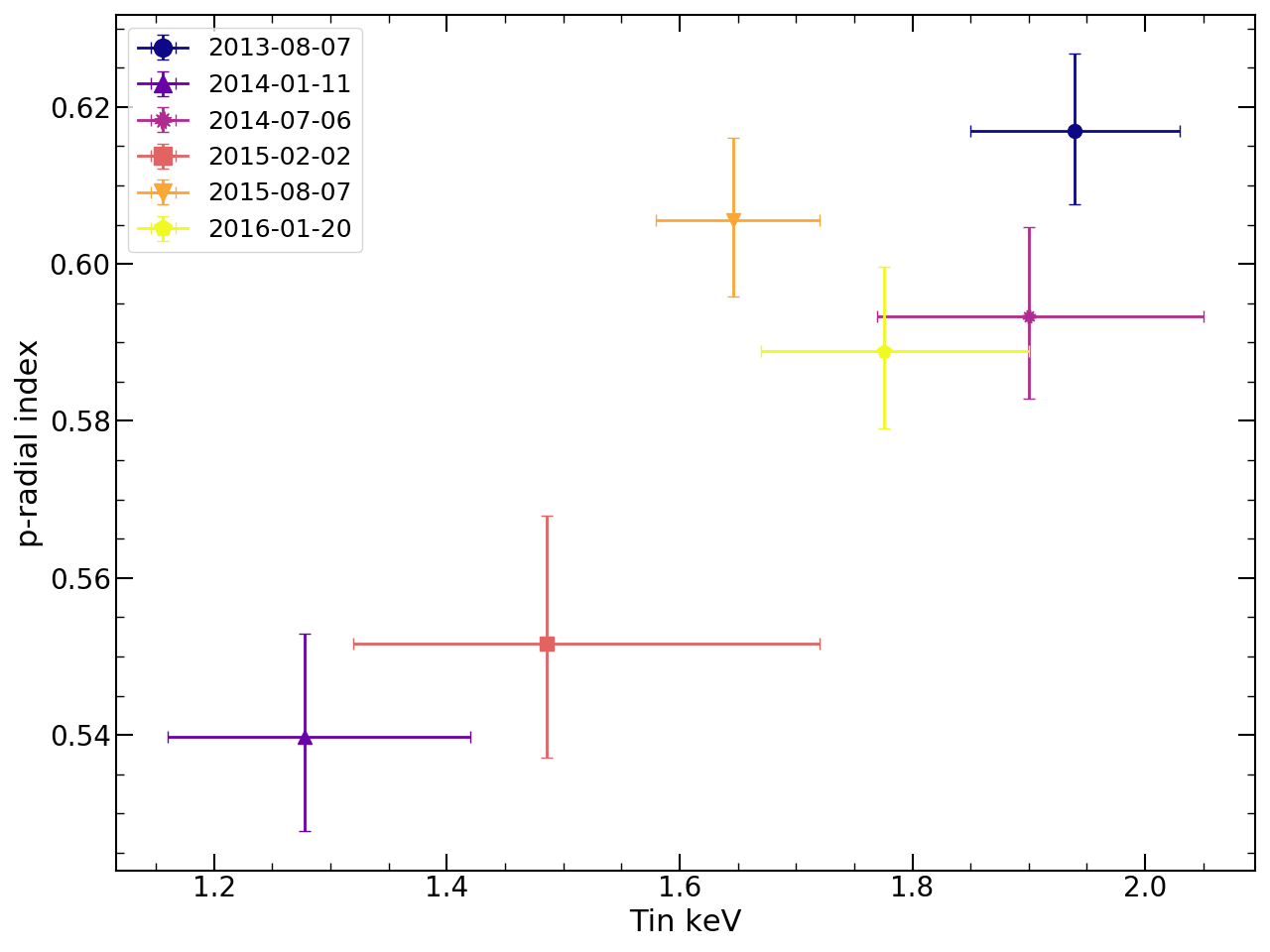}
    \caption{Dependency of the $p$ radial index of the \diskpbb\ with its temperature for M83 ULX1 (see text for details). This dependency is similar to that seen in stellar-mass black holes in the standard regime \citep{kubota_three_2004}. Symbols are as per Figure \ref{fig:positive_sources}.}
    \label{fig:m83_pindex}
\end{figure}

   \begin{table}
      \caption{Results from jointly fitting all datasets of M83 ULX1 with an absorbed \diskpbb\ model.}
         \label{tab:m83_diskpbb}
         \begin{tabular}{lcccc}
           \hline 
            \noalign{\smallskip}
            Epoch  & \nh\ & T$_\text{in}$ & $p$ & norm \\
                  & 10$^{20}$ cm$^{-2}$ & keV& &10$^{-4}$\\
            \noalign{\smallskip}
            \hline 
     \noalign{\smallskip}
1X & \multirow{6}{*}{3.2$\pm$0.7}& 1.94$\pm$0.09& 0.62$\pm$0.01& 24\errors{6}{5} \\
2X & & 1.3$\pm$0.1& 0.54$\pm$0.01& 16\errors{9}{6}\\
3X & & 1.9\errors{0.2}{0.1}& 0.59$\pm$0.01& 17\errors{7}{5} \\
4X & & 1.5$\pm$0.2& 0.55\errors{0.02}{0.01}& 14\errors{11}{7} \\
5X & & 1.65$\pm$0.07& 0.61$\pm$0.01& 30\errors{7}{6}\\
6X &  &1.8$\pm$0.1& 0.59$\pm$0.01& 17\errors{6}{5}\\
\noalign{\smallskip}
\hline
\noalign{\smallskip}
$\chi^2_r$ & & & & 1.02 \\
dof& & & & 1493 \\ 
            \noalign{\smallskip}
            \hline
         \end{tabular}
   \end{table}

The overall evolution of M83 ULX1 is very similar to that seen in the $\sim$ 10 \msun\ black hole XTE J1550--564 \citep{kubota_three_2004}. Our hard \diskbb\ when using the dual-thermal component shows constant inner-disk radius, closely following the L$\propto$ T$^4$ relationship as XTE J1550--564 in the \textit{standard} regime (i.e. when $L_\text{disk}$ $\propto$ T$^4$) \citep[e.g. period 3 in][]{kubota_three_2004}\footnote{Note that \cite{kubota_three_2004} uses RXTE that covers the 3--20 keV range and thus their soft component corresponds roughly to our hard component.}. Additionally, when using the \diskpbb\ model, the temperature increases with the radial index $p$ of the \diskpbb\ (Figure \ref{fig:m83_pindex}) as seen in XTE J1550--564 and LMC X-3 when fitted with the same model (see their Figure 9). \cite{kubota_three_2004} argued that this increase in $p$ with temperature is an artefact caused by the limited bandpass and the fact that the radial dependency is flatter near the innermost disk radius than the --0.75 given by the \diskbb\ approximation. Our \diskpbb\ L--T relationship follows $\alpha$ = 3.0 $\pm$ 0.6 (90\% confidence level) with Spearman's test coefficient of 0.94, again with roughly constant normalisation, suggesting we are witnessing the inner-most stable orbit as in XTE J1550--564 and LMC X-3 in the standard regime. A typical value of the normalisation $N$ $\sim$ 0.002, corresponds to R$_\text{in}$ $\sim$ 95 ${(\cos i)}^{-1/2}$ km, assuming \fcol\ = 1.8 as previously. If we assume that the constant radius we observe corresponds to the innermost stable orbit for a non-spinning black hole, then R$_\text{in}$ = \risco\ = 3\rschw, and we obtain a BH mass estimate of $\sim$ 10 $(\cos i)^{-1/2}$ \msun. This mass estimate would be a factor 6 larger if we consider instead a maximally spinning Kerr black hole. 

A further constraint on the mass of the black hole comes from the fact that \cite{kubota_three_2004} argued that a source enters the \textit{anomalous regime} when L$_\text{diskb}$ / L$_\text{Edd}$ $\sim$ 0.4. Therefore, given that L$_\text{disk}$ $\sim$ 4 $\times$ 10$^{39}$ erg/s, then the mass of the M83 ULX1 could be of the order of 60 \msun, which could easily be accounted for with reasonable values of inclination and spin of the black hole. We note that it is unlikely that the source is in the anomalous state as we should expect $p$ to \textit{decrease} with temperature, as this parameter starts to deviate from standard value for a thin disk of --0.75 \citep{kubota_three_2004, shakura_black_1973}, at odds with our observations (Figure \ref{fig:m83_pindex}). 

Albeit the exact mass estimate is rather uncertain, we conclude that the behaviour of this source is consistent with a massive stellar-mass black hole accreting close to the Eddington limit, in the high/soft state, given its similar evolution with accreting black holes in the standard regime. This conclusion is in agreement with previous studies by \cite{soria_slim-disk_2015} and other studies suggesting that sources below 3 $\times$ 10$^{39}$ erg/s could be consistent with massive black holes accreting close to the Eddington limit \citep{middleton_bright_2013, sutton_ultraluminous_2013}. Such massive black holes might not be rare given the black hole masses estimated from gravitational wave events \citep{abbott_binary_2019}.

\section{Conclusions} \label{sec:conclusions}
We have presented a thorough study of the long-term spectral evolution of a representative sample of ULXs and PULXs using data from \xmm, \chandra, and \nustar. By studying their spectral states and transitions, we have been able to explain the main sources of variability in these sources which can be summarised as: changes in the mass-transfer rate, changes in the degree of beaming, precession and obscuration by the optically thick parts of the wind as this becomes more mass-loaded.

We have shown that PULXs are among the hardest sources in the sample and discussed their evolution in terms of the interplay between the magnetospheric radius and the spherisation radius. We favour a scenario in which the softest PULX, NGC 1313 X-2, is consistent with being a weakly magnetised NS so that \rsph\ $>$ \rmag\ and the wind/funnel structure is responsible for imprinting highly anisotropic emission as the source precesses, given the wide HR variability the source spans. This interpretation can explain the significantly softer spectra of NGC 1313 X-2 and its lower pulsed-fraction, as the primary emission from the accretion column is expected to be downscattered in the cool electrons of the outflow. Additionally, the lack of transitions to the propeller regime in NGC 1313 X-2 supports this interpretation, as the weak magnetic field (or high-mass transfer rate) implied by the condition \rsph $>$ \rmag, will naturally lead to smaller magnetospheric radii and thus transitions to the propeller regime are less likely to occur. Instead, the hardest PULXs, NGC 7793 P13 and M51-ULX7, are consistent with being strongly magnetised, so that \rmag\ $>$ \rsph\ given the lack of HR variability which we interpret as lack of strong anisotropy. In this scenario, the accretion disk is being truncated before it becomes supercritical, suppressing the anisotropy that the funnel/wind structure would otherwise imprint. For NGC 5907 ULX1, we have shown that in those epochs associated with the super-orbital variability, the source appears harder when dimmer. This is hard to reconcile with the anisotropy expected from the funnel/wind structure in which we expect the source to become harder when brighter (as for NGC 1313 X-2). Still, a high-mass transfer rate is required to explain its high luminosity and therefore we conclude that \rsph $\sim$ \rmag\ is a plausible condition to explain the source variability. 

By comparing the evolution of PULXs with the sources in our sample, we have been able to identify a strong NS candidate with very similar evolution to that seen in NGC 1313 X-2: M81 X-6. Albeit we were not able to detect pulsations in the source, it is possible that longer exposures sampling the source in different spectral states may be needed to detect pulsations. Additionally, deeper pulse searches taking into account orbital parameters corrections may be needed. 

Most of the softer sources for which the accretor is unknown, show three markedly different spectral states: one at highest luminosity and two at similar low luminosities but different hardness ratio. A super-critical funnel can offer an explanation of such degeneracy between luminosity and hardness, because a source is expected to be dim at both low mass-transfer rates and when the gas within the funnel becomes optically thick at high-mass transfer rates, so that the hard radiation from the inner regions of the accretion flow becomes abruptly obscured. This could explain the evolution seen in NGC 1313 X-1, NGC 55 ULX1, IC 342 X-1 and Circinus ULX5. For Holmberg II X-1 and NGC 5204 X-1, these transitions may be better explained if our line of sight is grazing the half-opening angle of the funnel, so that our view of the accretion flow rapidly transits between seeing down the funnel and seeing through the optically thick wind walls as the source precesses. Future higher cadence monitoring of these transitions will be key in order to determine the exact nature of these transitions, by studying both their timescale and the source evolution prior and after them. Nevertheless, these transitions are suggestive of strong winds in these sources, which together with their softer appearance compared to most PULXs supports a scenario in which the sources considered here are powered by weakly magnetised NSs or BHs.

Finally we have reported on the stability of the high-energy emission ($\gtrsim$ 10 keV) in some of the sources in our sample. Notably, none of the PULXs show such stability, albeit further high-quality \nustar\ observations are needed to probe the different spectral states of both PULXs and those sources for which the accretor is unknown. Black holes are favoured candidates to explain this stability, as they naturally offer means to swallow any excess radiation, stabilising the output radiation even as the mass-transfer rate increases. Should this be the case, this high-energy emission may be the smoking gun to identify BH-ULXs. On the other hand, should some of this sources host NS, then this stability may offer interesting clues about the accretion flow geometry around NSs in the super-Eddington regime. Nevertheless, we stress the importance of obtaining future \nustar\ observations as we may expect to see most of the observational differences between BH- and NS-ULXs at high-energies, where the mechanism responsible for the emission is expected to differ (i.e. an accretion column compared to the case of the inner regions of the accretion disk around a black hole).

\longtab{
\begin{longtable}{cccccc}
 \caption{\label{tab:ftest} Results of the F-test between \tbabs$\otimes$\tbabs$\otimes$(\diskbb + \diskbb) (1) and \tbabs$\otimes$\tbabs$\otimes$(\diskbb + \simpl$\otimes$\diskbb) (2) models. Epochs are sorted chronologically and the nomenclature indicate the number \xmm, \chandra\ and \nustar\ observations fit together. (see Table \ref{tab:observations_xmm}). } \\ 
 \hline
 \noalign{\smallskip}
Epoch & $\chi_{r1}$/dof$_1$ \tablefootmark{a} & $\chi_{r2}$/dof$_2$ \tablefootmark{b} & $\Delta\chi$ & [1-Prob(F-test)] $\times$ 100 (\%)& \simpl\ ?\\
\noalign{\smallskip}
\hline 
\noalign{\smallskip}
\endfirsthead
\caption{Continued.}\\
\hline
 Epoch & $\chi_{r1}$/dof$_1$ \tablefootmark{a} & $\chi_{r2}$/dof$_2$ \tablefootmark{b} & $\Delta\chi$& [1-Prob(F-test)] $\times$ 100 (\%) & \simpl\ ?\\
 \noalign{\smallskip}
\hline 
\noalign{\smallskip}
\endhead
\endfoot
\noalign{\smallskip}
\hline 
\endlastfoot 
\multicolumn{6}{c}{Holmberg II X--1} \\ 
 \noalign{\smallskip} 
1X&1.01/281 &0.90/279 &32.8&100.0&yes\\
2X&1.06/246 &1.02/244 &10.7&99.40&yes\\ 
3X&1.07/124 &1.08/122 &1.3&45.02&no\\ 
4X&1.32/392 &1.15/390 &70.9&100.0&yes\\
5X&1.12/316 &1.06/314 &22.6&100.0&yes\\
6XNN&1.31/490 &0.99/488 &159.2&100.0&yes\\ 
7XN&1.17/467 &0.91/465 &125.5&100.0&yes\\ 
\noalign{\smallskip} 
 \hline 
 \noalign{\smallskip}
\multicolumn{6}{c}{Holmberg IX X--1}\\
 \noalign{\smallskip}
1X&1.19/198 &1.20/196 &0.9&31.29&no\\ 
2X&1.08/342 &1.04/340 &13.2&99.80&yes\\ 
3X&1.05/353 &0.98/351 &25.4&100.0&yes\\
4X&1.25/502 &1.20/500 &30.0&100.0&yes\\
5X&0.82/81 &0.80/79 &3.1&85.00&no\\
6X&1.19/398 &1.09/396 &40.3&100.0&yes\\
7X&0.97/372 &0.97/370 &3.6&84.24&no\\ 
8X&1.05/301 &1.05/299 &2.6&70.80&no\\ 
9XXNN&1.24/1211 &1.07/1209 &208.0&100.0&yes\\ 
10XNN&1.22/956 &1.04/954 &175.8&100.0&yes\\ 
11XXNN&1.18/1055 &0.98/1053 &211.0&100.0&yes\\ 
\noalign{\smallskip}
\hline
\noalign{\smallskip}
\multicolumn{6}{c}{IC 342 X--1} \\
\noalign{\smallskip}
1X&0.89/156 &0.90/154 &-&0.00&no\\ 
1C&0.80/84 &0.82/82 &0.0&0.00&no\\ 
2C&1.07/77 &1.10/75 &0.0&0.00&no\\
2X&1.00/336 &1.00/334 &1.9&61.32&no\\ 
3X&1.11/286 &1.11/284 &1.2&41.54&no\\ 
4X&0.98/288 &0.97/286 &4.0&87.06&no\\
5XNN&1.09/775 &0.95/773 &111.0&100.0&yes\\ 
6XN&1.29/802 &1.10/800 &154.5&100.0&yes\\ 
3C&1.04/74 &1.01/72 &4.4&87.94&no\\ 
\noalign{\smallskip}
\hline
\noalign{\smallskip}
\multicolumn{6}{c}{NGC 5204 X--1} \\ 
 \noalign{\smallskip}
1C&0.86/50 &0.89/48 &0.1&5.46&no\\ 
1X&1.04/249 &1.04/247 &2.34&67.45&no\\ 
2X&1.22/157 &1.21/155 &4.0&80.66&no\\ 
3X&1.08/180 &1.08/178 &2.0&60.09&no\\
2C&1.27/151 &1.27/149 &1.7&48.59&no\\ 
3CC&0.94/152 &0.95/150 &0.1&5.12&no\\ 
4C&0.97/47 &1.01/45 &0.0&0.00&no\\ 
5C&1.06/78 &1.09/76 &0.4&16.77&no\\ 
6C&1.08/47 &1.04/45 &3.9&83.55&no\\ 
7C&0.66/68 &0.68/66 &0.0&0.00&no\\
4X&0.97/185 &0.94/183 &7.8&98.28&yes\\ 
5X&1.01/339 &0.97/337 &15.3&99.96&yes\\ 
6X&1.03/307 &1.00/305 &12.1&99.74&yes\\ 
7XXNN&1.13/625 &1.02/623 &72.5&100.0&yes\\
8X&1.07/260 &1.06/258 &4.7&88.84&no\\ 
\noalign{\smallskip}
\hline
\noalign{\smallskip}
\multicolumn{6}{c}{NGC5408X--1} \\ 
 \noalign{\smallskip} 
 1X&1.18/138 &1.16/136 &4.7&86.41& no\\
2X&1.01/139 &1.01/137 &1.2&44.48& no\\ 
3X&1.13/97 &1.12/95 &2.9&71.99& no\\ 
4X&0.97/106 &0.97/104 &2.4&70.65& no\\ 
5X&1.86/352 &1.57/350 &106.4&100.0&yes\\
6X&1.56/314 &1.41/312 &48.9&100.0&yes\\
1CC&1.15/201 &1.15/199 &3.9&81.52& no\\
2C&1.07/99 &1.04/97 &5.5&92.42&no\\
7X&1.65/379 &1.45/377 &79.9&100.0&yes\\
8X&1.85/382 &1.58/380 &105.4&100.0&yes\\
9X&1.79/366 &1.40/364 &145.8&100.0&yes\\
10X&1.75/377 &1.42/375 &129.1&100.0&yes\\ 
11XX&1.51/568 &1.36/566 &86.3&100.0&yes\\
\noalign{\smallskip}
\hline
\noalign{\smallskip}
\multicolumn{6}{c}{NGC 1313 X--1}\\
\noalign{\smallskip}
1X&1.16/338 &1.15/336 &4.8&87.49&no\\ 
2XX&0.85/150 &0.84/148 &3.7&88.67&no\\ 
3XX&1.00/343 &0.98/341 &6.9&99.05&yes$^c$\\ 
4X&0.91/109 &0.86/107 &7.8&98.73&no\\ 
5X&1.01/286 &1.01/284 &0.7&29.23&no\\ 
6X&1.18/166 &1.19/164 &-&0.00&no\\ 
7X&1.01/157 &1.02/155 &-&0.00&no\\ 
8X&1.09/218 &1.06/216 &9.3&98.67&yes$^c$\\ 
9X&1.27/457 &1.21/455 &30.0&100.0&yes\\
10XN&1.43/649 &1.20/647 &157.3&100.0&yes\\ 
11XN&1.43/678 &1.15/676 &189.5&100.0&yes\\ 
12XN&1.61/558 &1.18/556 &244.2&100.0&yes\\
13X&0.98/256 &0.98/254 &1.8&60.07&no\\ 
14XN&1.38/512 &1.11/510 &142.6&100.0&yes\\ 
15XN&1.71/759 &1.16/757 &415.8&100.0&yes\\  
16X&1.58/487 &1.30/485 &139.2&100.0&yes\\ 
17XXN&1.39/983 &1.24/981 &148.8&100.0&yes\\  
18X&1.31/457 &1.12/455 &87.3&100.0&yes\\
19XN&1.82/655 &1.26/653 &368.1&100.0&yes\\ 
\noalign{\smallskip}
\hline
\noalign{\smallskip}
\multicolumn{6}{c}{Circinus ULX5} \\
\noalign{\smallskip}
1X&0.78/119 &0.80/117& 0.0&64.12&no\\ 
1C&0.78/84 &0.77/82 &1.9&70.14&no\\ 
2XNN&1.30/678 &1.21/676 &67.9&100.0&yes\\ 
3X&1.04/271 &1.01/269 &12.2&99.73&no$^d$\\ 
4XN&0.99/305 &0.98/303 &3.6&83.73&no\\ 
5X&0.99/201 &0.99/199 &2.0&63.55&no\\  
6X&1.30/486 &1.30/484 &5.4&87.40&no\\ 
\noalign{\smallskip}
\hline
\noalign{\smallskip}
\multicolumn{6}{c}{NGC 55 ULX1}\\
\noalign{\smallskip}
1X&1.11/293 &1.06/291 &17.5&99.97&yes\\ 
2X&1.36/178 &1.32/176 &11.2&98.44&yes$^c$\\ 
3X&1.43/317 &1.36/315 &26.9&99.99&yes\\ 
\hline
\noalign{\smallskip}
\multicolumn{6}{c}{NGC 6946 X--1} \\
\noalign{\smallskip}
1C&1.1/162 &1.0/160 &19.1&100.0&yes\\
1X&1.1/130 &1.0/128 &8.6&98.7&yes$^c$\\
2C&1.1/97 &1.0/95 &11.5&99.5&yes\\ 
3CC&1.0/103 &0.9/101 &6.2&95.9&no\\
2XX&1.1/329 &1.1/327 &19.0&100.0&yes\\ 
3X&1.4/357 &1.2/355 &72.0&100.0&yes\\ 
4C&1.0/102 &0.9/100 &3.6&84.82&no\\
4XNN&1.3/328 &1.2/326 &29.6&100.0&yes\\  \noalign{\smallskip}
\hline
\noalign{\smallskip}
\multicolumn{6}{c}{NGC 1313 X--2}\\
\noalign{\smallskip}
1X&0.98/135 &0.97/133 &3.3&81.43&no\\ 
2X&0.89/279 &0.89/277 &0.0&0.00&no\\ 
3X&1.06/199 &1.06/197 &2.7&71.91&no\\ 
4X&0.91/170 &0.92/168 &0.0&0.00&no\\ 
5X&0.91/175 &0.90/173 &3.8&87.53&no\\ 
6X&1.00/110 &0.97/108 &5.6&93.99&no\\ 
7X&0.78/293 &0.79/291 &0.4&22.49&no\\ 
8X&1.22/95 &1.21/93 &3.6&76.86&no\\ 
9X&1.18/181 &1.18/179 &2.1&58.75&no\\ 
10X&0.93/302 &0.93/300 &1.1&44.53&no\\ 
11X&1.05/118 &1.05/116 &2.0&61.00&no\\  
12X&1.10/154 &0.92/152 &29.7&100.0&yes\\
13XN&1.11/484 &1.09/482 &8.3&97.69&yes\\
14XN&1.15/393 &1.15/391 &4.8&87.56&no\\ 
15X&1.09/59 &1.12/57 &0.0&0.00&no\\ 
16X&1.30/127 &1.29/125 &3.1&69.48&no\\ 
17X&1.11/85 &1.14/83 &0.0&0.00&no\\ 
18X&0.88/335 &0.88/333 &2.0&67.96&no\\ 
19X&1.11/232 &1.12/230 &0.0&0.00&no\\ 
20X&1.09/433 &1.03/431 &26.2&100.0&yes\\
21XN&1.14/160 &1.04/158 &17.4&99.96&yes\\ 
22X&1.06/246 &1.07/244 &0.0&0.00&no\\ 
23X&1.51/135 &1.48/133 &0.0&87.02&no\\
24X&1.02/100 &1.01/98 &2.8&74.54&no\\  
25X&0.90/108 &0.91/106 &1.1&45.26&no\\  
26XXN&1.23/625 &1.23/623 &1.3&66&no\\ 
\noalign{\smallskip}
\hline
\noalign{\smallskip}
\multicolumn{6}{c}{NGC 300 ULX1} \\
\noalign{\smallskip}
1XN&1.59/857 &1.34/855 &221.3&100.0&yes\\ 
2X&1.1/450 &1.1/448 &12.1&99.6  & yes\\
1CC&1.05/165 &1.06/163 &-&0.00&no\\
\noalign{\smallskip}
\hline
\noalign{\smallskip}
\multicolumn{6}{c}{NGC 5907 ULX-1} \\
\noalign{\smallskip}
1X&0.80/215 &0.81/213 &0.0&0.00&no\\ 
2X&1.11/187 &1.12/185 &0.0&0.00&no\\ 
3XXCC&0.91/239 &0.92/237&0.0&-&no\\ 
4XN&1.12/232 &1.11/230 &5.2&90.13&no\\ 
5XNN&1.03/491 &0.94/489 &47.7&100.0&yes\\ 
6X&1.06/151 &1.03/149 &6.4&95.19&no\\ 
7X&0.93/122 &0.94/120 &0.0&0.00&no\\ 
8XX&1.14/125 &1.15/123 &2.2&61.46&no\\
\noalign{\smallskip}
\hline
\noalign{\smallskip}
\multicolumn{6}{c}{NGC 7793 P13} \\ 
\noalign{\smallskip}
1C&1.05/221 &1.06/219 &0.2&8.98&no\\
1X&1.07/336 &1.08/334 &0.0&0.00&no\\ 
2X&1.25/401 &1.25/399 &1.3&40.56&no\\  
3XN&1.13/812 &1.12/810 &6.8&95.10&no\\ 
4X&0.90/168 &0.91/166 &0.6&28.11&no\\
5XN&1.17/579 &1.18/577 &-0.5&-&no\\ 
6X&1.23/403 &1.23/401 &4.6&84.56&no\\ 
7X&1.04/423 &1.05/421 &-3.0&-&no\\ 
8X&1.09/422 &1.09/420 &3.6&80.70&no\\ 
9XN&1.10/661 &1.10/659 &2.9&73.12&no\\  
 \end{longtable}
\textbf{Notes:}$^a\chi_{r}$ and degrees of freedom of the \tbabs$\otimes$\tbabs$\otimes$(\diskbb + \diskbb) model. $^b$ Similarly for model  \tbabs$\otimes$\tbabs$\otimes$(\diskbb + \simpl$\otimes$\diskbb). $^c$\simpl\ model included as we could constrain it with another observations where the source was similar in flux and hardness-ratio. $^d$\simpl\ model not included as we could not constrain its parameters with another observations where the source was similar in flux and hardness-ratio.
}
\begin{acknowledgements}
      The authors would like to thank the anonymous referee for his comments and suggestions that helped improve the quality of the manuscript. A. G\'urpide would like to thank M. Bachetti for his help and assistance during the search of pulsations and to I. Pastor-Marazuela for the computational resources provided. NW acknowledges support by the CNES. This work made used of the free software Veusz developed by J. Sanders to produce some of the plots.
\end{acknowledgements}

%
%
\bibliographystyle{aa}
\bibliography{aanda}
\end{document}